\pdfoutput=1

\documentclass[a4paper,11pt]{article}

\usepackage{jheppub} 
\usepackage{bbm}
\usepackage{graphicx}
\usepackage{slashed}
\usepackage{subfigure}
\usepackage[utf8]{inputenc}
\graphicspath{ {./images/} }
\usepackage{mathtools}

\usepackage{tikz}
\usetikzlibrary{arrows,decorations.markings}
\tikzset{myptr/.style={decoration={markings,mark=at position 1 with %
    {\arrow[scale=2]{>}}},postaction={decorate}}}
\definecolor{ublue}{HTML}{0063A6}
\definecolor{ured}{HTML}{DD4814}
\definecolor{uviolet}{HTML}{A71C49}

\newcommand{\comment}[1]{}

\newcommand{\Tr}{\mathrm{Tr}}

\newcommand{\td}{\mathrm{d}}

\newcommand{\TT}[1]{\mathrm{#1}}

\renewcommand{\v}[1]{\mathbf{#1}}

\renewcommand{\>}{\right\rangle}
\newcommand{\<}{\left\langle}
\newcommand{\lkl}{\left|}
\newcommand{\rkl}{\right|}
\newcommand{\Pole}{\mathop{\TT{Pole}}}
\newcommand{\qqquad}{\qquad\qquad}

\newcommand{\Nc}{N_{\mathrm{c}}}
\newcommand{\As}{\alpha_{\mathrm{s}}}
\newcommand{\bcdot}{\boldsymbol{\cdot}}
\usepackage{breakurl} 

\hypersetup{urlcolor=black}

\title{\boldmath Rings and strings: a basis for understanding subleading colour and QCD coherence beyond the two-jet limit}

\preprint{\begin{flushright}
MCNET-21-13
\end{flushright}}    

\author[a]{Jeffrey R. Forshaw}
\author[a,b]{Jack Holguin}
\author[c,d,e]{Simon Pl\"atzer}

\affiliation[a]{Consortium for Fundamental Physics, Department of Physics \& Astronomy, \\ University of Manchester, Manchester M13 9PL, United Kingdom}
\affiliation[b]{CPHT, CNRS, Ecole polytechnique, IP Paris, F-91128 Palaiseau, France}
\affiliation[c]{Institute of Physics, NAWI Graz, University of Graz, Universit\"atsplatz 5, A-8010 Graz, Austria}
\affiliation[d]{Particle Physics, Faculty of Physics, University of Vienna, Boltzmanngasse 5, A-1090 Wien, Austria}
\affiliation[e]{Erwin Schr\"odinger Institute for Mathematics and
  Physics, University of Vienna, Boltzmanngasse 9, A-1090 Wien}
\emailAdd{jack.holguin@polytechnique.edu}
\emailAdd{jeffrey.forshaw@manchester.ac.uk}
\emailAdd{simon.plaetzer@uni-graz.at}

\date{\today}

\abstract{Guided by the colour-diagonal structure of collinear singularities, we identify a set of kinematic basis functions that are well suited to the simplification of soft gluon emission amplitudes. In particular, these basis functions, which emerge naturally in the colour flow basis, isolate the subleading colour contributions and improve the statistical convergence of the \texttt{CVolver} amplitude-evolution code. They also allow us to extend current angular-ordered parton showers beyond the azimuthally-averaged, two-jet limit.}

\begin{document} 
\maketitle
\flushbottom

\section{Introduction}
\label{sec:intro}

Resummation of logarithmically enhanced contributions in perturbation theory, and the closely related design and analysis of parton shower algorithms, which reside at the core of event generator simulation programs \cite{Bahr:2008pv,Bellm:2015jjp,Herwig_shower,Pythia8,VINCIA,Gleisberg:2008ta}, are crucial to predicting the details of final states at collider experiments. Recent developments of parton shower algorithms have been driven by attempts to achieve higher precision. Notably, a significant amount of work has been performed to make decisive statements about the perturbative accuracy of existing algorithms \cite{Dasgupta:2018nvj,Bewick:2019rbu,Nagy:2020gjv} and to propose modifications that improve their formal accuracy \cite{Dasgupta:2020fwr,Forshaw:2020wrq,Holguin:2020joq,Nagy:2020rmk,Nagy:2020gjv,Nagy:2020dvz,Bewick:2019rbu,Bewick:2021nhc,Hamilton:2020rcu,Karlberg:2021kwr,Hoeche:2020nsx,Gellersen:2021eci}. 

The baseline accuracy of an angular-ordered shower \cite{Gieseke:2003rz} has long been established, drawing from the seminal work on QCD coherence \cite{CATANI1991635,CATANI1992419,Resum_large_logs_ee}, which guarantees next-to-leading-log (NLL) accuracy for inclusive, global, two-jet observables with a colour singlet initial state and lower accuracy otherwise. Recent improvements include the addition of spin correlations \cite{Richardson:2018pvo}, higher-order DGLAP emission kernels \cite{Hoche:2017hno} and modifications to the evolution variable so as to better populate the emission phase-space \cite{Bewick:2019rbu,Bewick:2021nhc}. However, at their core, modern angular-ordered parton showers are still based on the early implementations of QCD coherence and their accuracy has not been systematically improved beyond the two-jet limit. Today, it is known that QCD coherence can be used to achieve NLL accuracy across a broad class of exponentiating observables beyond the two-jet limit (the rIRC safe observables) \cite{Banfi:2004yd,Banfi:2000si,Banfi:2010xy,Banfi:2014sua,Arpino:2019ozn}. However, these insights have yet to be formulated as an algorithm that can be incorporated into a general purpose angular-ordered parton shower.

In contrast to angular-ordering, early work on dipole showers focused on the large $\Nc$ limit of amplitudes with soft gluons \cite{Gustafson:1986db,Gustafson:1987rq,Lonnblad:1992tz}. Their wide-ranging applicability to inclusive and exclusive observables at LL and leading-colour (LC) accuracy was demonstrated without needing arguments of QCD coherence. Only in recent years has the accuracy of dipole showers been systematically extended beyond LL \cite{Dasgupta:2020fwr,Forshaw:2020wrq,Holguin:2020joq,Hamilton:2020rcu,Karlberg:2021kwr} and LC NLL accuracy can now be achieved across a range of inclusive and exclusive observables. 

In parallel to the recent efforts to improve existing cross-section-level showers, significant amounts of work have also been devoted to exploring a new paradigm on which parton branching algorithms can eventually be based \cite{Nagy:2015hwa,SoftEvolutionAlgorithm,Forshaw:2019ver,Platzer:2020lbr}. By working directly with the full quantum mechanical amplitudes, these new approaches naturally achieve greater accuracy but are also burdened by significantly increased complexity. Consequently, their direct application to phenomenology has so far seen limited success \cite{DeAngelis:2020rvq,Nagy:2019bsj,Nagy:2020dvz}.  In what follows, we use this amplitude-level formalism to explore the structure of subleading colour corrections from soft (eikonal) gluons. These efforts lead us to a deeper understanding of the kinematic structure of `off-diagonal' subleading colour, and to a proposed improvement of existing angular-ordered showers.

The rest of the paper is structured as follows. In Section~\ref{sec:amplitudes} we present an analysis of soft gluon amplitudes, and introduce our basic objects: the `ring' and `string' functions. Using these functions, we will isolate the kinematic dependence of colour diagonal and colour suppressed contributions using the known fact that collinear poles are diagonal in colour space, and then also by explicit construction in a colour flow basis. In Section~\ref{sec:simulation} and Section~\ref{sec:example}, we illustrate the benefits of using these functions when evaluating full-colour amplitudes: the functions are economical for expressing subleading colour and they improve the numerical convergence of the parton branching algorithm implemented into the \texttt{CVolver} code \cite{Platzer:2013fha,DeAngelis:2020rvq}. In Section~\ref{sec:coherence} we study the kinematic limits of rings and strings, which enables us to identify how the coherent branching formalism can be generalized beyond the azimuthally-averaged two-jet limit, where colour correlations from the hard process and the azimuthal dependence of eikonal gluons play a significant role. The bottom line is a simple modification to the emission kernels in an angular-ordered shower: 
\begin{align}
    \frac{\td(\cos \theta_{in} )\td \phi_{n}}{1 - \cos \theta_{in}} \Theta(\theta_{in} < \theta)  \mapsto  \, {}^{(n-1)}\TT{S}^{j,k}_{i} \, E^{2}_{n} \, \Theta(\theta_{in} < \theta) \, \td(\cos \theta_{in} )\td \phi_{n}, \label{eq:key}
\end{align}
where $\, {}^{(n-1)}\TT{S}^{j,k}_{i}$ is a string function (introduced in the next section):
\begin{align}
    \, {}^{(n-1)}\TT{S}^{j,k}_{i} = \frac{s_{ij} s_{nk} + s_{ik} s_{nj} - s_{jk} s_{ni}}{2 s_{ni} s_{nj} s_{nk}},
\end{align}
for $s_{ij} = q_{i}\cdot q_{j}$. The emitted parton is labelled $n$, $i$ is its parent, $j,k$ are spectators (defined later), and $\theta$ the previous angular scale in the shower. This modification extends the angular-ordered framework to the three-jet limit. It includes the azimuthal correlations from an eikonal gluon dressing any number of collinear partons in the three jets, and when one of the three jets is soft it encodes the azimuthal correlations between two eikonal gluons dressing a two-jet system. When more than three coloured hard particles are involved, the modification offers no formal improvement over existing angular-ordered showers \cite{Gieseke:2003rz,Herwig_shower}\footnote{Though some improvement might be possible by reducing hard process colour flows to tripoles rather than dipoles.}. In this case, the large-$\Nc$ colour flows of the hard process are used to initiate the shower, and each colour flow is individually showered with a weight given by the sub-amplitude for the particular colour structure. The showering is performed by treating each colour dipole as if it were a two-jet system.
This procedure loses the interference between the hard process colour flows, thus restricting the angular-ordered shower to leading-colour accuracy. 

\section{Soft amplitudes, collinear poles and colour structures}
\label{sec:amplitudes}

\subsection{Recap of density matrices for soft gluons}
\label{sec:background}

We assume that an $n$-particle QCD amplitude, for which one of the $n$ particles is a soft gluon, factorises:\footnote{This factorisation is violated at two loops \cite{Catani:2011st,factorisationBreaking} but that is beyond the scope of this paper.}
\begin{align}
    \lkl M_{n}(q_{1},\dots , q_{n}) \> = \v{J}_{n}(q_{n}) \lkl M_{n-1} (q_{1},\dots , q_{n-1}) \> + \mathcal{O}(\lambda^{0}),
\end{align}
where $\v{J}_{n}$ is the eikonal current,  $q_{n} \sim \lambda q_{i}$ for $\lambda \ll 1$ and the other momenta are $q_{i} \in \{q_{1},\dots , q_{n-1}\}$. In what follows, we are interested in $n>3$ so that the colour dynamics is non-trivial. Differential cross sections are computed using
\begin{align}
    \td \sigma_{n} = \< M_{n} | M_{n} \> \td \Phi_{n} = \< M_{n-1}  \rkl \v{J}^{\dagger}_{n}\v{J}_{n} \lkl M_{n-1}  \> \td \Phi_{n-1} \; \frac{(2 \pi\mu)^{2\epsilon}}{2 \pi} \, \frac{\td^{3-2\epsilon}\vec{q}_{n}}{2 E_{n} } + \mathcal{O}(\lambda^{1-\epsilon}),
\end{align}
where $\td \Phi_{n}$ is the $n$-particle phase-space measure. At lowest order in $\As$, the eikonal current is
\begin{align}
    \v{J}^{(0)}_{n}(q_{n}) = \left(\frac{\As}{\pi}\right)^{\frac{1}{2}} \sum_{i,\lambda_i}\v{T}_{i} \frac{\varepsilon_{\lambda_i}(q_{n}) \cdot q_{i}}{q_{n} \cdot q_{i}}.
\end{align}
Similarly, a soft loop factorises as
\begin{align}
    \lkl M^{(1)}_{n}\> = \v{I}^{(1)}_{n} \lkl M^{(0)}_{n} \> + \mathcal{O}(\lambda^{1-\epsilon}),
\end{align}
where the bracketed superscript indicates the number of loops. $\v{I}^{(1)}_{n}$ is related to $\v{J}^{(0)}_{n}$ via unitarity:
\begin{align}
\label{eqs:unitarity}
    \v{I}_{n} + \v{I}_{n}^{\dagger} = \frac{(2 \pi\mu)^{2\epsilon}}{2 \pi}\int \frac{\td^{3-2\epsilon}\vec{q}_{n}}{2 E_{n}} \; \v{J}^{\dagger}_{n}\v{J}_{n}.
\end{align}

It is useful to express the soft gluon amplitude and its conjugate as a density matrix, $\v{A}_{n}$, such that $\Tr \v{A}_{n}= \< M_{n} | M_{n} \>$. Written in this form 
\begin{align}
    \v{A}_{n} = \v{J}^{(0)}_{n} \lkl M_{n-1}  \> \< M_{n-1}  \rkl \v{J}^{(0) \, \dagger}_{n} =- \frac{\As}{\pi} \sum_{i \neq j} \omega_{ij}(q_{n})\v{T}_{i} \lkl M_{n-1}  \> \< M_{n-1}  \rkl \v{T}^{\dagger}_{j},
\end{align}
where $\omega_{ij}(q_{n})$ is an antenna function,
\begin{align}
 \omega_{ij}(q_{n}) = \frac{q_{i} \cdot q_{j}}{q_{n} \cdot q_{i}\; q_{n} \cdot q_{j}}.
\end{align}
We will make use of an abridged notation:
\begin{align}
    [i \bcdot j] = \v{T}_{i} \lkl M_{n-1}  \> \< M_{n-1}  \rkl \v{T}^{\dagger}_{j}.
\end{align}
Crucially, both the emission of a soft gluon and the real part of a soft loop, when expressed as operators on density matrices, only depend on antenna functions.

Each antenna function is accompanied by two collinear poles. Let us now extract the functional form of a pole in $\omega_{ij}$ arising when $q_{n} || q_{i}$. We define a dipole momentum fraction and dipole transverse momentum as
\begin{align}
    z_{ij} = \frac{q_{n} \cdot q_{j}}{q_{i} \cdot q_{j}}, \qquad (k^{ij}_{\bot})^{2} = \frac{2 \, q_{i}\cdot q_{n} \, q_{j} \cdot q_{n}}{q_{i} \cdot q_{j}},
\end{align}
and use them to write
\begin{align}
    \omega_{ij}(q_{n}) \frac{\td^{3-2\epsilon}\vec{q}_{n}}{2 E_{n}} = \frac{\td k^{ij}_{\bot} }{(k^{ij}_{\bot})^{1+ 2\epsilon}}\frac{\td z_{ij} }{(z_{ij})^{1+\epsilon}} \td \phi.
\end{align}
Working in the lab  frame (or equivalently any frame where $q_{1}, \dots, q_{n-1}$ are hard relative to $q_{n}$), we can use the small angle approximation to isolate the collinear poles. In this approximation $k^{ij}_{\bot} \approx E_{n}\theta_{ni}$ where $\theta_{ni}$ is the angle  between $i$ and $n$, and $z_{ij} \approx E_{n} /E_{i}$. Thus
\begin{align}
    \Pole_{q_{n} || q_{i}}\left(\omega_{ij}(q_{n}) \frac{\td^{3-2\epsilon}\vec{q}_{n}}{2 E_{n}}\right) = \frac{\td \theta_{ni}}{(\theta_{ni})^{1+2\epsilon}}  \frac{E_{i}^{\epsilon}\td E_{n} }{(E_{n})^{1+ 3\epsilon}} \td \phi, \label{eq:smallangleantenna}
\end{align}
which is independent of $j$. Consequently, colour conservation ($\sum_{j}\v{T}_{j} = 0$) can be used so that for a fixed $i$
\begin{align}
    \Pole_{q_{n} || q_{i}}\left(\v{A}_{n} \frac{\td^{3-2\epsilon}\vec{q}_{n}}{2 E_{n}} \right) = \frac{\As}{\pi} [i \bcdot i] \frac{\td \theta_{ni}}{(\theta_{ni})^{1+2\epsilon}}  \frac{E_{i}^{\epsilon}\td E_{n} }{(E_{n})^{1+ 3\epsilon}} \td \phi.
\end{align}
This highlights a general feature: the colour structures accompanying a single collinear pole are diagonal\footnote{Strictly speaking the colour operator $[i \cdot i]$, is only diagonal after using the cyclicity of the trace, $\Tr[i \cdot i] = \< M_{n-1}  \rkl \v{T}^{2}_{i} \lkl M_{n-1}  \>$. It is the quadratic Casimir operator, $\v{T}^{2}_{i}$, which is diagonal in colour space. This subtlety is not of relevance in this paper since, with tree-level soft currents, there is a one-to-one relationship between a structure $[i \cdot i]$ and a quadratic Casimir.}. Equivalently, the off-diagonal colour density matrix elements for a soft emission $n$ are independent of collinear poles associated with $q_{n}$. QCD coherence can be viewed as a consequence of this pole structure in the amplitude density matrix. Consequently, the tree-level eikonal current for the emission of a soft gluon from a bunch of collinear particles couples only to the combined colour charge of the collinear particles.

\subsection{Isolating collinear poles in density matrices}
\label{sec:rings}

The absence of collinear poles in off-diagonal colour density matrix elements is well known. However, it is also somewhat striking given that the antenna functions, from which we can express any leading soft amplitude, each contain two collinear poles. Necessarily, collinear poles cancel for some linear combinations of antenna functions. We refer to these linear combinations as ring functions, it will be clear why shortly. Linear combinations of antenna functions with one or more collinear poles we refer to as string functions. Whilst diagonal elements of a colour density matrix can depend on both rings and strings, the collinear finiteness of off-diagonal elements ensures they only ever depend on rings. This motivates us to look for a set of basis functions spanning all rings and strings for a given particle multiplicity. Expressing $\v{A}_{n}$ in this basis will partition collinear poles from the off-diagonal elements in the amplitude density matrix.  

In Appendix \ref{app:rings} we give a detailed account of the construction of ring functions, here we collect the important features. The simplest ring is\footnote{From here on we use all roman indices other than $n$ to index a particle other than $n$.}
\begin{align}
    \, {}^{(n-1)}\TT{R}^{i,j}_{k,l} = \omega_{ij}(q_{n}) - \omega_{ik}(q_{n}) - \omega_{jl}(q_{n}) + \omega_{kl}(q_{n}),
\end{align}
where $i,j,k,l \in \{1,2,\dots, n-1\}$ and the $n$th gluon is soft. It is necessary that none of $i,j,k,l,n$ are equal. Consequently, $\v{A}_{n}$ has no non-zero, off-diagonal elements for $n<5$ at tree-level (or $n<4$ if $q_{n}$ is a loop momentum). From this point on we will assume that $\v{A}_{n}$ is a tree-level density matrix. One-loop density matrices can be found through the unitarity relation, Eq.~\eqref{eqs:unitarity}.

We can represent antenna functions diagrammatically as sources and sinks,
\begin{align}
    \omega_{ij}(q_{n}) \equiv 
    \begin{array}{c}
    \begin{tikzpicture}
    \draw[black] [myptr] (0,0) -- (0.25,0.75);
    \draw[black] (0.25,0.75) -- (0.5,1.5);
    \draw[black] [myptr] (0,0) -- (0.25,-0.75);
    \draw[black] (0.25,-0.75) -- (0.5,-1.5);
    \filldraw [black] (0,0) circle (1pt);
    \node at (0,0.8) {$i$};
    \node at (0,-0.8) {$j$};
    \end{tikzpicture} \end{array} 
    \qquad \TT{and} \qquad -\omega_{ij}(q_{n}) =  
    \begin{array}{c}
    \begin{tikzpicture}
    \draw[black]  (0,0) -- (0.25,0.75);
    \draw[black] [myptr] (0.5,1.5) -- (0.25,0.75);
    \draw[black]  (0,0) -- (0.25,-0.75);
    \draw[black] [myptr] (0.5,-1.5) -- (0.25,-0.75);
    \filldraw [black] (0,0) circle (1pt);
    \node at (0,0.8) {$i$};
    \node at (0,-0.8) {$j$};
    \end{tikzpicture}
    \end{array},
\end{align}
from which we can construct a ring:
\begin{align}
    \, {}^{(n-1)}\TT{R}^{i,j}_{k,l} \equiv 
    \begin{array}{c}
    \begin{tikzpicture}
    \draw[black] [myptr] (1.5,0) -- (1.5,0.75);
    \draw[black] (1.5,0.75) -- (1.5,1.5);
    \draw[black] [myptr] (0,1.5) -- (0,0.75);
    \draw[black] (0,0.75) -- (0,0);
    \draw[black] [myptr] (0,1.5) -- (0.75,1.5);
    \draw[black] (0.75, 1.5) -- (1.5,1.5);
    \draw[black] [myptr] (1.5,0) -- (0.75,0);
    \draw[black] (0,0) -- (0.75,0);
    \filldraw [black] (0,0) circle (1pt);
    \filldraw [black] (1.5,0) circle (1pt);
    \filldraw [black] (0,1.5) circle (1pt);
    \filldraw [black] (1.5,1.5) circle (1pt);
    \node at (-0.3,0.8) {$i$};
    \node at (0.7,1.8) {$j$};
    \node at (1.8,0.8) {$l$};
    \node at (0.7,-0.3) {$k$};
    \end{tikzpicture}
    \end{array}~.
\end{align}
All rings can be expressed as a sum over even sided polygons. For example
\begin{align}
    \omega_{13} - \omega_{12} + \omega_{25} - \omega_{56} + \omega_{46} - \omega_{34} \equiv \begin{array}{c}
    \begin{tikzpicture}
    \draw[black] [myptr] (0,0) -- (0.29,0.5);
    \draw[black] (0.29,0.5) -- (0.58,1);
    \draw[black] [myptr] (0,0) -- (0.29,-0.5);
    \draw[black] (0.58,-1) -- (0.29,-0.5);
    \draw[black]  (0.58*4,0) -- (0.58*4-0.29,0.5);
    \draw[black] [myptr] (0.58*3,1) -- (0.58*4-0.29,0.5);
    \draw[black]  (0.58*4,0) -- (0.58*4-0.29,-0.5);
    \draw[black] [myptr](0.58*3,-1) -- (0.58*4-0.29,-0.5);
    \draw[black]  (0.58,1) -- (0.58*2,1);
    \draw[black] [myptr] (0.58*3,1) -- (0.58*2,1);
    \draw[black]  (0.58,-1) -- (0.58*2,-1);
    \draw[black] [myptr] (0.58*3,-1) -- (0.58*2,-1);
    \filldraw [black] (0,0) circle (1pt);
    \filldraw [black] (0.58,1) circle (1pt);
    \filldraw [black] (0.58*3,1) circle (1pt);
    \filldraw [black] (0.58*3,-1) circle (1pt);
    \filldraw [black] (0.58*4,0) circle (1pt);
    \filldraw [black] (0.58,-1) circle (1pt);
    \node at (0,0.6) {$1$};
    \node at (0,-0.6) {$3$};
    \node at (0.58*4,0.6) {$5$};
    \node at (0.58*4,-0.6) {$6$};
     \node at (0.58*2,1.3) {$2$};
    \node at (0.58*2,-1.3) {$4$};
    \end{tikzpicture} \end{array},
\end{align}
where we have dropped the $(q_{n})$ argument of the antenna functions. For a second example,
\begin{align}
    &\omega_{18} - \omega_{17} + \omega_{57} - \omega_{56} + \omega_{46} - \omega_{48} + \omega_{23} - \omega_{36} + \omega_{46} - \omega_{24} \nonumber \\
    &\qqquad \equiv \begin{array}{c}
    \begin{tikzpicture}
    \draw[black] [myptr] (0,0) -- (0.29,0.5);
    \draw[black] (0.29,0.5) -- (0.58,1);
    \draw[black] [myptr] (0,0) -- (0.29,-0.5);
    \draw[black] (0.58,-1) -- (0.29,-0.5);
    \draw[black]  (0.58*4,0) -- (0.58*4-0.29,0.5);
    \draw[black] [myptr] (0.58*3,1) -- (0.58*4-0.29,0.5);
    \draw[black]  (0.58*4,0) -- (0.58*4-0.29,-0.5);
    \draw[black] [myptr](0.58*3,-1) -- (0.58*4-0.29,-0.5);
    \draw[black]  (0.58,1) -- (0.58*2,1);
    \draw[black] [myptr] (0.58*3,1) -- (0.58*2,1);
    \draw[black]  (0.58,-1) -- (0.58*2,-1);
    \draw[black] [myptr] (0.58*3,-1) -- (0.58*2,-1);
    \filldraw [black] (0,0) circle (1pt);
    \filldraw [black] (0.58,1) circle (1pt);
    \filldraw [black] (0.58*3,1) circle (1pt);
    \filldraw [black] (0.58*3,-1) circle (1pt);
    \filldraw [black] (0.58*4,0) circle (1pt);
    \filldraw [black] (0.58,-1) circle (1pt);
    \node at (0,0.6) {$1$};
    \node at (0,-0.6) {$8$};
    \node at (0.58*4,0.6) {$5$};
    \node at (0.58*4,-0.6) {$6$};
     \node at (0.58*2,1.3) {$7$};
    \node at (0.58*2,-1.3) {$4$};
    \end{tikzpicture} \end{array} ~~~+~~ \begin{array}{c}
    \begin{tikzpicture}
    \draw[black] [myptr] (1.5,0) -- (1.5,0.75);
    \draw[black] (1.5,0.75) -- (1.5,1.5);
    \draw[black] [myptr] (0,1.5) -- (0,0.75);
    \draw[black] (0,0.75) -- (0,0);
    \draw[black] [myptr] (0,1.5) -- (0.75,1.5);
    \draw[black] (0.75, 1.5) -- (1.5,1.5);
    \draw[black] [myptr] (1.5,0) -- (0.75,0);
    \draw[black] (0,0) -- (0.75,0);
    \filldraw [black] (0,0) circle (1pt);
    \filldraw [black] (1.5,0) circle (1pt);
    \filldraw [black] (0,1.5) circle (1pt);
    \filldraw [black] (1.5,1.5) circle (1pt);
    \node at (-0.3,0.8) {$4$};
    \node at (0.7,1.8) {$2$};
    \node at (1.8,0.8) {$3$};
    \node at (0.7,-0.3) {$6$};
    \end{tikzpicture}
    \end{array}~.
\end{align}
A string is a sequence of sources and sinks which does not form a polygon. It necessarily has at least one collinear pole.

Rings can be manipulated with the following two rules. We have a cut rule for splitting an $n$ sided polygon into two smaller polygons:
\begin{align}
    \begin{array}{c }
    \begin{tikzpicture}
    \draw[black] [myptr] (0,0) -- (0.2,0.5);
    \draw[black] (0.2,0.5) -- (0.4,1);
    \draw[black] [myptr] (0,0) -- (0.2,-0.5);
    \draw[black] (0.4,-1) -- (0.2,-0.5);
    \draw[black]  (0.4*6,0) -- (0.4*6-0.2,0.5);
    \draw[black] [myptr] (0.4*5,1) -- (0.4*6-0.2,0.5);
    \draw[black]  (0.4*6,0) -- (0.4*6-0.2,-0.5);
    \draw[black] [myptr](0.4*5,-1) -- (0.4*6-0.2,-0.5);
    \filldraw [black] (0,0) circle (1pt);
    \filldraw [black] (0.4*6,0) circle (1pt);
    \node at (0,0.6) {$i$};
    \node at (0,-0.6) {$j$};
    \node at (0.4*6,0.6) {$k$};
    \node at (0.4*6,-0.6) {$l$};
     \node at (0.4*3,1) {$\cdots\cdots$};
    \node at (0.4*3,-1) {$\cdots\cdots$};
    \end{tikzpicture} 
    \end{array} = 
    \begin{array}{c}
    \begin{array}{c}\begin{tikzpicture}
    \draw[white] [myptr] (0,0+0.15) -- (0.2,0.5+0.15);
    \draw[white] (0.2,0.5+0.15) -- (0.4,1+0.15);
    \draw[black] [myptr] (0,0-0.15) -- (0.2,-0.5-0.15);
    \draw[black] (0.4,-1-0.15) -- (0.2,-0.5-0.15);
    \draw[white]  (0.4*6,0+0.15) -- (0.4*6-0.2,0.5+0.15);
    \draw[white] [myptr] (0.4*5,1+0.15) -- (0.4*6-0.2,0.5+0.15);
    \draw[black]  (0.4*6,0-0.15) -- (0.4*6-0.2,-0.5-0.15);
    \draw[black] [myptr](0.4*5,-1-0.15) -- (0.4*6-0.2,-0.5-0.15);
    \draw[white] [myptr] (0,0+0.15) -- (0.4*3,0+0.15);
    \draw[white] (0.4*6,0+0.15) -- (0.4*3,0+0.15);
    \draw[black] [myptr] (0,-0.15) -- (0.4*3,-0.15);
    \draw[black] (0.4*6,-0.15) -- (0.4*3,-0.15);
    \filldraw [white] (0,0.15) circle (1pt);
    \filldraw [white] (0.4*6,0.15) circle (1pt);
    \filldraw [black] (0,-0.15) circle (1pt);
    \filldraw [black] (0.4*6,-0.15) circle (1pt);
    \node at (0,-0.6-0.15) {$j$};
    \node at (0.4*6,-0.6-0.15) {$l$};
    \node at (0.4*3,-0.3-0.15) {$i$};
    \node at (0.4*3,-1-0.15) {$\cdots\cdots$};
    \end{tikzpicture} \end{array} ~~ + ~~ \begin{array}{c}\begin{tikzpicture}
    \draw[black] [myptr] (0,0+0.15) -- (0.2,0.5+0.15);
    \draw[black] (0.2,0.5+0.15) -- (0.4,1+0.15);
    \draw[white] [myptr] (0,0-0.15) -- (0.2,-0.5-0.15);
    \draw[white] (0.4,-1-0.15) -- (0.2,-0.5-0.15);
    \draw[black]  (0.4*6,0+0.15) -- (0.4*6-0.2,0.5+0.15);
    \draw[black] [myptr] (0.4*5,1+0.15) -- (0.4*6-0.2,0.5+0.15);
    \draw[white]  (0.4*6,0-0.15) -- (0.4*6-0.2,-0.5-0.15);
    \draw[white] [myptr](0.4*5,-1-0.15) -- (0.4*6-0.2,-0.5-0.15);
    \draw[black] [myptr] (0,0+0.15) -- (0.4*3,0+0.15);
    \draw[black] (0.4*6,0+0.15) -- (0.4*3,0+0.15);
    \draw[white] [myptr] (0,-0.15) -- (0.4*3,-0.15);
    \draw[white] (0.4*6,-0.15) -- (0.4*3,-0.15);
    \filldraw [black] (0,0.15) circle (1pt);
    \filldraw [black] (0.4*6,0.15) circle (1pt);
    \filldraw [white] (0,-0.15) circle (1pt);
    \filldraw [white] (0.4*6,-0.15) circle (1pt);
    \node at (0,0.6+0.15) {$i$};
    \node at (0.4*6,0.6+0.15) {$k$};
    \node at (0.4*3,0.3+0.15) {$l$};
    \node at (0.4*3,1+0.15) {$\cdots\cdots$};
    \end{tikzpicture} \end{array} \TT{for} \; i \neq k, i \neq l, j \neq l.\label{eq:cutrule}
    \end{array}
\end{align}
There is also a rule for repeated edges, which can be used to reduce an $n$-sided polygon to an $(n-2)$-sided polygon:
\begin{align}
    \begin{array}{c}\begin{tikzpicture}
    \draw[black] [myptr] (1,3) -- (0.7,2.6);
    \draw[black] (0.7,2.6) -- (0.4,2.2);
    \draw[black] [myptr] (0,1.2) -- (0.2,1.7);
    \draw[black] (0.2,1.7) -- (0.4,2.2);
    \draw[black] [myptr] (0,1.2) -- (0,0.6);
    \draw[black] (0,0.6) -- (0,0);
    \draw[black] [myptr] (0.4,-1) -- (0.2,-0.5);
    \draw[black] (0.2,-0.5) -- (0,0);
    \draw[black] [myptr] (0.4,-1) -- (0.7,-1.4);
    \draw[black] (0.7,-1.4) -- (1,-1.8);
    \filldraw [black] (0,0) circle (1pt);
    \filldraw [black] (0.4,2.2) circle (1pt);
    \filldraw [black] (0,1.2) circle (1pt);
    \filldraw [black] (0.4,-1) circle (1pt);
    \node at (-0.3,0.7) {$j$};
    \node at (-0.1,2) {$i$};
    \node at (0.5,2.8) {$k$};
    \node at (-0.1,-0.7) {$i$};
    \node at (0.5,-1.6) {$l$};
    \end{tikzpicture} \end{array} = \begin{array}{c}\begin{tikzpicture}
    \draw[black] [myptr] (1,3) -- (0.7,2.6);
    \draw[black] (0.7,2.6) -- (0.4,2.2);
    \draw[black] [myptr] (0.4,-1) -- (0.4,0.6);
    \draw[black] (0.4,0.6) -- (0.4,2.2);
    \draw[black] [myptr] (0.4,-1) -- (0.7,-1.4);
    \draw[black] (0.7,-1.4) -- (1,-1.8);
    \filldraw [black] (0.4,2.2) circle (1pt);
    \filldraw [black] (0.4,-1) circle (1pt);
    \node at (0.1,0.7) {$i$};
    \node at (0.5,2.8) {$k$};
    \node at (0.5,-1.6) {$l$};
    \end{tikzpicture} \end{array} .\label{eq:repeatedgerule}
\end{align}
Using these rules, it is always possible to reduce a ring function to a sum over the functions $\, {}^{(n-1)}\TT{R}^{i,j}_{k,l}$. Therefore we can use the functions $\, {}^{(n-1)}\TT{R}^{i,j}_{k,l}$ to form a basis over rings. The functions ${}^{(n-1)}\TT{R}^{i,j}_{k,l}$ have the following symmetries and anti-symmetries:
\begin{align}
    \, {}^{(n-1)}\TT{R}^{i,j}_{k,l} = \, {}^{(n-1)}\TT{R}^{j,i}_{l,k} \, , \qquad \, {}^{(n-1)}\TT{R}^{i,j}_{k,l} = -\, {}^{(n-1)}\TT{R}^{l,j}_{k,i} \, , \qquad \, {}^{(n-1)}\TT{R}^{i,j}_{k,l} = -\, {}^{(n-1)}\TT{R}^{i,k}_{j,l} \, . \label{eq:basis}
\end{align}
Diagrammatically (anti-)symmetries of the rings are generated by axes of reflection of their polygons. Reflections with axes bisecting vertices generate the symmetries and axes bisecting edges generate anti-symmetries.

In Appendix \ref{app:rings} we show that for a particle multiplicity of $(n-1)$ there are $$_{n-1}C_{2}-n+1= \frac{(n-1)(n-4)}{2}$$ linearly independent rings. For example, when $(n-1)=4$ there are $2$ linearly independent basis rings: 
\begin{align}
    \, {}^{(4)}\TT{R}^{1,2}_{3,4} \equiv \begin{array}{c}\begin{tikzpicture}
    \draw[black] [myptr] (1.5,0) -- (1.5,0.75);
    \draw[black] (1.5,0.75) -- (1.5,1.5);
    \draw[black] [myptr] (0,1.5) -- (0,0.75);
    \draw[black] (0,0.75) -- (0,0);
    \draw[black] [myptr] (0,1.5) -- (0.75,1.5);
    \draw[black] (0.75, 1.5) -- (1.5,1.5);
    \draw[black] [myptr] (1.5,0) -- (0.75,0);
    \draw[black] (0,0) -- (0.75,0);
    \filldraw [black] (0,0) circle (1pt);
    \filldraw [black] (1.5,0) circle (1pt);
    \filldraw [black] (0,1.5) circle (1pt);
    \filldraw [black] (1.5,1.5) circle (1pt);
    \node at (-0.3,0.8) {$1$};
    \node at (0.7,1.8) {$2$};
    \node at (1.8,0.8) {$4$};
    \node at (0.7,-0.3) {$3$};
    \end{tikzpicture} \end{array} , \quad \, {}^{(4)}\TT{R}^{1,2}_{4,3} \equiv \begin{array}{c}\begin{tikzpicture}
    \draw[black] [myptr] (1.5,0) -- (1.5,0.75);
    \draw[black] (1.5,0.75) -- (1.5,1.5);
    \draw[black] [myptr] (0,1.5) -- (0,0.75);
    \draw[black] (0,0.75) -- (0,0);
    \draw[black] [myptr] (0,1.5) -- (0.75,1.5);
    \draw[black] (0.75, 1.5) -- (1.5,1.5);
    \draw[black] [myptr] (1.5,0) -- (0.75,0);
    \draw[black] (0,0) -- (0.75,0);
    \filldraw [black] (0,0) circle (1pt);
    \filldraw [black] (1.5,0) circle (1pt);
    \filldraw [black] (0,1.5) circle (1pt);
    \filldraw [black] (1.5,1.5) circle (1pt);
    \node at (-0.3,0.8) {$1$};
    \node at (0.7,1.8) {$2$};
    \node at (1.8,0.8) {$3$};
    \node at (0.7,-0.3) {$4$};
    \end{tikzpicture} \end{array}.
\end{align}
Neither square can be transformed into the other by application of the cutting or repeated edge rules or symmetries. When $n-1=5$ there are $_{5}C_{2}-5=5$ linearly independent rings: $$\, {}^{(5)}\TT{R}^{1,2}_{3,4}, ~ \, {}^{(5)}\TT{R}^{1,2}_{4,3}, ~ \, {}^{(5)}\TT{R}^{1,2}_{3,5}, ~ \, {}^{(5)}\TT{R}^{1,2}_{5,3}, ~ \TT{and} ~ \, {}^{(5)}\TT{R}^{1,2}_{4,5}.$$
Finding a complete basis over ring functions for the amplitude $\v{A}_{n}$ is achieved by finding a set $\{{}^{(n-1)}\TT{R}^{i,j}_{k,l}\}$ of dimension $(n-1)(n-4)/2$ where each ${}^{(n-1)}\TT{R}^{i,j}_{k,l}$ is linearly independent. One such complete basis is
\begin{align}
    \left\{{}^{(n-1)}\TT{R}^{1,2}_{3,i},{}^{(n-1)}\TT{R}^{1,2}_{i,3},{}^{(n-1)}\TT{R}^{1,2}_{i,j}\; \Big| \; (i>3) \wedge (j>i)\right\}.
\end{align}

We have found a basis over rings, let us now do the same for strings. Our goal is to disentangle the collinear poles in $\v{A}_{n}$ and so to construct a set of strings that forms a complete basis over the collinear poles in $\v{A}_{n}$ (which we refer to as our string basis\footnote{Note that this string basis will not be a basis over all possible strings but rather a basis over the collinear poles in $\v{A}_{n}$. This is because a string with a pole in the direction of $i$ can be transformed into another string of the same length and with the same pole by the addition of a ring.}). This is mostly easily achieved by constructing a set of strings for which each string only contains a collinear pole in a single direction. A string with a collinear pole only in the direction of particle $i$ has the form
\begin{align}
    \begin{array}{c}\begin{tikzpicture}
    \draw[black]  (1,3) -- (0.7,2.6);
    \draw[black] [myptr] (0.4,2.2) -- (0.7,2.6);
    \draw[black] (0,1.2) -- (0.2,1.7);
    \draw[black] [myptr] (0.4,2.2) -- (0.2,1.7);
    \draw[black] [myptr] (0.4,-1) -- (0.2,-0.5);
    \draw[black] (0.2,-0.5) -- (0,0);
    \draw[black] [myptr]  (0.4,-1) -- (0.7,-1.4);
    \draw[black] (1,-1.8) -- (0.7,-1.4);
    \filldraw [black] (0.4,2.2) circle (1pt);
    \filldraw [black] (0.4,-1) circle (1pt);
    \filldraw [black] (0.1,0.3) circle (0.5pt);
    \filldraw [black] (0.1,0.6) circle (0.5pt);
    \filldraw [black] (0.1,0.9) circle (0.5pt);
    \node at (0.5,2.8) {$i$};
    \node at (0.5,-1.6) {$i$};
    \end{tikzpicture} \end{array},
\end{align}
where the dots represent the insertion of sources and sinks. There are $(n-1)$ collinear poles in $\v{A}_{n}$ and with them $(n-1)$ basis string functions. We will use the shortest string basis:
\begin{align}
   {}^{(n-1)}\TT{S}^{j,k}_{i} = \frac{1}{2}\left(\omega_{ij} + \omega_{ik} - \omega_{jk} \right) \equiv \frac{1}{2} \times\begin{array}{c}\begin{tikzpicture}
    \draw[black] [myptr] (0.2,1) -- (0.5,1.4);
    \draw[black] (0.5,1.4) -- (0.8,1.8);
    \draw[black] [myptr] (0.2,1) -- (0.1,0.5);
    \draw[black] (0.1,0.5) -- (0,0);
    \draw[black] [myptr] (0.2,-1) -- (0.1,-0.5);
    \draw[black] (0.1,-0.5) -- (0,0);
    \draw[black] [myptr] (0.2,-1) -- (0.5,-1.4);
    \draw[black] (0.5,-1.4) -- (0.8,-1.8);
    \filldraw [black] (0,0) circle (1pt);
    \filldraw [black] (0.2,1) circle (1pt);
    \filldraw [black] (0.2,-1) circle (1pt);
    \node at (-0.1,0.7) {$j$};
    \node at (0.3,1.6) {$i$};
    \node at (-0.1,-0.7) {$k$};
    \node at (0.3,-1.6) {$i$};
    \end{tikzpicture} \end{array},
\end{align}
where $j,k$ can be chosen freely except that $ j,k\neq i$. The factor of a half is included so that
\begin{align}
   \Pole_{q_{n} || q_{i}}\left({}^{(n-1)}\TT{S}^{j,k}_{i} \frac{\td^{3-2\epsilon}\vec{q}_{n}}{2 E_{n}}\right)   = \Pole_{q_{n} || q_{i}}\left(\omega_{ij}(q_{n}) \frac{\td^{3-2\epsilon}\vec{q}_{n}}{2 E_{n}}\right) = \Pole_{q_{n} || q_{i}}\left(\omega_{ik}(q_{n}) \frac{\td^{3-2\epsilon}\vec{q}_{n}}{2 E_{n}}\right).
\end{align}
The set of functions $\{{}^{(n-1)}\TT{S}^{j,k}_{i} \}$ forms a complete basis over the collinear poles in the amplitude density matrix $\v{A}_{n}$. Combining a ring basis and this set of strings gives a set of $_{n-1}C_{2}$ linearly independent functions, equal to the total number of antenna functions in $\v{A}_{n}$.

Now we have constructed a basis of rings and strings, we can use it to separate the collinear poles from the colour off-diagonal physics in $\v{A}_{n}$. An antenna function can be expressed as
\begin{align}
    \omega_{ij}(q_{n}) \equiv \begin{array}{c}\begin{tikzpicture}
    \draw[black]  (1,2) -- (0.7,1.6);
    \draw[black] [myptr] (0.4,1.2) -- (0.7,1.6);
    \draw[black] [myptr]  (0.4,0) -- (0.7,-0.4);
    \draw[black] (1,-0.8) -- (0.7,-0.4);
    \filldraw [black] (0.5,0.3) circle (0.5pt);
    \filldraw [black] (0.5,0.6) circle (0.5pt);
    \filldraw [black] (0.5,0.9) circle (0.5pt);
    \node at (0.5,1.8) {$i$};
    \node at (0.5,-0.6) {$i$};
    \end{tikzpicture} \end{array} +
    \begin{array}{c}\begin{tikzpicture}
    \draw[black]  (1,2) -- (0.7,1.6);
    \draw[black] [myptr] (0.4,1.2) -- (0.7,1.6);
    \draw[black] [myptr]  (0.4,0) -- (0.7,-0.4);
    \draw[black] (1,-0.8) -- (0.7,-0.4);
    \filldraw [black] (0.5,0.3) circle (0.5pt);
    \filldraw [black] (0.5,0.6) circle (0.5pt);
    \filldraw [black] (0.5,0.9) circle (0.5pt);
    \node at (0.5,1.8) {$j$};
    \node at (0.5,-0.6) {$j$};
    \end{tikzpicture} \end{array}
    + \TT{Ring}
\end{align}
where the two strings take care of the pole regions and the ring (which will depend on the string basis chosen) takes care of any residual finite pieces. Using the string basis $\{{}^{(n-1)}\TT{S}^{j,k}_{i} \}$, and reducing the ring to a sum over basis rings, we have the following relation\footnote{Note this is not unique. The rings in this relation come from just one combination of cuts on an octagon with two repeated edges: \begin{align}
    \omega_{ij}(q_{n}) = \begin{array}{c}\begin{tikzpicture}
    \draw[black] [myptr] (0.2,0.9) -- (0.5,1.3);
    \draw[black] (0.5,1.3) -- (0.8,1.7);
    \draw[black] [myptr] (0.2,0.9) -- (0.1,0.4);
    \draw[black] (0.1,0.4) -- (0,0);
    \draw[black] [myptr] (0.2,-0.9) -- (0.5,-1.3);
    \draw[black] (0.5,-1.3) -- (0.8,-1.7);
    \draw[black] [myptr] (0.2,-0.9) -- (0.1,-0.4);
    \draw[black] (0.1,-0.4) -- (0,0);
    \filldraw [black] (0,0) circle (1pt);
    \filldraw [black] (0.2,0.9) circle (1pt);
    \filldraw [black] (0.2,-0.9) circle (1pt);
    \node at (-0.15,0.7) {$k$};
    \node at (0.3,1.6) {$i$};
    \node at (-0.15,-0.6) {$l$};
    \node at (0.3,-1.6) {$i$};
    \end{tikzpicture} \end{array} +
    \begin{array}{c}\begin{tikzpicture}
    \draw[black] [myptr] (0.2,0.9) -- (0.5,1.3);
    \draw[black] (0.5,1.3) -- (0.8,1.7);
    \draw[black] [myptr] (0.2,0.9) -- (0.1,0.4);
    \draw[black] (0.1,0.4) -- (0,0);
    \draw[black] [myptr] (0.2,-0.9) -- (0.5,-1.3);
    \draw[black] (0.5,-1.3) -- (0.8,-1.7);
    \draw[black] [myptr] (0.2,-0.9) -- (0.1,-0.4);
    \draw[black] (0.1,-0.4) -- (0,0);
    \filldraw [black] (0,0) circle (1pt);
    \filldraw [black] (0.2,0.9) circle (1pt);
    \filldraw [black] (0.2,-0.9) circle (1pt);
    \node at (-0.15,0.7) {$m$};
    \node at (0.3,1.6) {$j$};
    \node at (-0.15,-0.6) {$n$};
    \node at (0.3,-1.6) {$j$};
    \end{tikzpicture} \end{array} + \frac{1}{2}
    \begin{array}{c}\begin{tikzpicture}
    \draw[black] [myptr] (0.3,0.8) -- (0.15,0.4);
    \draw[black] (0.15,0.4) -- (0,0);
    \draw[black] [myptr] (0.3,0.8) -- (0.7,0.95);
    \draw[black] (0.7,0.95) -- (1.1,1.1);
    \draw[black] [myptr] (1.9,0.8) -- (1.5,0.95);
    \draw[black] (1.5,0.95) -- (1.1,1.1);
    \draw[black] [myptr] (1.9,0.8) -- (2.1,0.4);
    \draw[black] (2.1,0.4) -- (2.3,0);\
    \draw[black] [myptr] (0.3,-0.8) -- (0.15,-0.4);
    \draw[black] (0.15,-0.4) -- (0,0);
    \draw[black] [myptr] (0.3,-0.8) -- (0.7,-0.95);
    \draw[black] (0.7,-0.95) -- (1.1,-1.1);
    \draw[black] [myptr] (1.9,-0.8) -- (1.5,-0.95);
    \draw[black] (1.5,-0.95) -- (1.1,-1.1);
    \draw[black] [myptr] (1.9,-0.8) -- (2.1,-0.4);
    \draw[black] (2.1,-0.4) -- (2.3,0);
    \filldraw [black] (0,0) circle (1pt);
    \filldraw [black] (0.3,0.8) circle (1pt);
    \filldraw [black] (1.1,1.1) circle (1pt);
    \filldraw [black] (1.9,0.8) circle (1pt);
    \filldraw [black] (2.3,0) circle (1pt);
    \filldraw [black] (0.3,-0.8) circle (1pt);
    \filldraw [black] (1.1,-1.1) circle (1pt);
    \filldraw [black] (1.9,-0.8) circle (1pt);
    \node at (-0.15,0.5) {$i$};
    \node at (-0.15,-0.5) {$k$};
    \node at (2.5,-0.5) {$j$};
    \node at (2.5,0.5) {$n$};
    \node at (0.7,1.2) {$j$};
    \node at (1.5,1.2) {$m$};
    \node at (0.7,-1.2) {$l$};
    \node at (1.5,-1.2) {$i$};
    \end{tikzpicture} \end{array}.
\end{align} }:
\begin{align}
    \omega_{ij}(q_{n}) = \, {}^{(n-1)}\TT{S}^{k,l}_{i} + \, {}^{(n-1)}\TT{S}^{m,n}_{j} + \frac{1}{2} \left(\, {}^{(n-1)}\TT{R}^{i,j}_{k,l} + \, {}^{(n-1)}\TT{R}^{j,l}_{m,i} + \, {}^{(n-1)}\TT{R}^{i,j}_{m,n} \right). \label{eq:2.28}
\end{align}
Thus we can write
\begin{align}
\label{eq:recursion}
    \v{A}_{n} =&- \frac{\As}{\pi} \sum_{i \neq j} \omega_{ij}(q_{n}) [i \bcdot j]  \nonumber \\
    = &- \frac{\As}{\pi} \sum_{i \neq j} \bigg(\, {}^{(n-1)}\TT{S}^{k,l}_{i} + \, {}^{(n-1)}\TT{S}^{m,n}_{j}  +\frac{1}{2} \left(\, {}^{(n-1)}\TT{R}^{i,j}_{k,l} + \, {}^{(n-1)}\TT{R}^{j,l}_{m,i} + \, {}^{(n-1)}\TT{R}^{i,j}_{m,n} \right) \bigg) [i \bcdot j]  , \nonumber \\
    = & \frac{2\As}{\pi} \sum_{i} \, {}^{(n-1)}\TT{S}^{k,l}_{i}  [i \bcdot i] - \frac{\As}{2\pi} \sum_{i \neq j} \left(\, {}^{(n-1)}\TT{R}^{i,j}_{k,l} + \, {}^{(n-1)}\TT{R}^{j,l}_{m,i} + \, {}^{(n-1)}\TT{R}^{i,j}_{m,n} \right) [i \bcdot j].
\end{align}
We immediately observe that all terms with collinear poles are colour diagonal, and this property emerges without the small-angle approximation. Note, there are only $_{n-1}C_{2}-n+1$ linearly independent rings whilst there are $3(_{n-1}C_{2})$ terms in the summation on the final line. Consequently, the ring terms must simplify. We can simplify matters by a choice of basis, $\{\, {}^{(n-1)}\TT{S}^{k,l}_{i}\}$.  Specifically, for all $i \neq 1,2$, we choose $k=1$ and $l=2$, for $i = 1$ we choose $k=2$ and $l=3$, and for $i = 2$, $k=1$ and $l=3$, i.e. 
$$\{\, {}^{(n-1)}\TT{S}^{k,l}_{i}\} \rightarrow \{\, {}^{(n-1)}\TT{S}^{2,3}_{1}, \, {}^{(n-1)}\TT{S}^{1,3}_{2}, \, {}^{(n-1)}\TT{S}^{1,2}_{3}, \, {}^{(n-1)}\TT{S}^{1,2}_{4},...,\, {}^{(n-1)}\TT{S}^{1,2}_{n-1}\}.$$ 
We have chosen this basis, which singles out partons 1,2 and 3, with a view to Section~\ref{sec:coherence} where it is a natural choice. This basis can be substituted into Eq.~\eqref{eq:recursion} in which $k,l,m,n$ appear arbitrary. $k,l,m,n$ are specified through Eq.~\eqref{eq:2.28}, wherein the first string fixes $k,l$ for a given $i$ and similarly the second fixes $j,m,n$. After some simplification using the relations given in Appendix \ref{app:rings}, we find
\begin{align}
    & \v{A}_{n} = \frac{2\As}{\pi} \left(\, {}^{(n-1)}\TT{S}^{2,3}_{1}  [1\bcdot 1] + \, {}^{(n-1)}\TT{S}^{1,3}_{2}  [2\bcdot 2] + \sum_{i  \neq 1,2} \, {}^{(n-1)}\TT{S}^{1,2}_{i}  [i\bcdot i] \right) \nonumber \\ 
    & -  \frac{\As}{2\pi} \sum_{i \neq 1,2,3} \, {}^{(n-1)}\TT{R}^{i,1}_{2,3} ([i\bcdot 1-2] + [1-2\bcdot i]) - \frac{\As}{2\pi} \sum_{i \neq 1,2}\sum_{j \neq i,1,2} \left(\, {}^{(n-1)}\TT{R}^{i,j}_{1,2} + \, {}^{(n-1)}\TT{R}^{j,i}_{1,2}  \right) [i\bcdot j],
\end{align}
where $[1-2\bcdot i]= (\v{T}_{1} - \v{T}_{2})\lkl M_{n-1}  \> \< M_{n-1}  \rkl \v{T}^{\dagger}_{i}$. This can be written as\footnote{We have $2(_{n-3}C_{2}) + (n-4)$ rings in this expression, which means $_{n-3}C_{2}$ of them are not linearly independent. For example,
$$
\, {}^{(n-1)}\TT{R}^{5,4}_{1,2} = \, {}^{(n-1)}\TT{R}^{4,5}_{1,2} + \, {}^{(n-1)}\TT{R}^{5,3}_{1,2} + \, {}^{(n-1)}\TT{R}^{3,4}_{1,2} - \, {}^{(n-1)}\TT{R}^{4,3}_{1,2} - \, {}^{(n-1)}\TT{R}^{3,5}_{1,2}.
$$
Moreover, $\, {}^{(n-1)}\TT{R}^{3,4}_{1,2}-\, {}^{(n-1)}\TT{R}^{4,3}_{1,2} = \, {}^{(n-1)}\TT{R}^{4,1}_{2,3}$ and so rings in the first sum are not independent from rings in the second. Therefore further simplifications can be made if desired.}
\begin{align}
    \v{A}_{n} = & \frac{2\As}{\pi} \left(\, {}^{(n-1)}\TT{S}^{2,3}_{1}  [1 \bcdot 1] + \, {}^{(n-1)}\TT{S}^{1,3}_{2}  [2\bcdot 2] + \sum_{i  \neq 1,2} \, {}^{(n-1)}\TT{S}^{1,2}_{i}  [i\bcdot i] \right) \nonumber \\ 
    & -  \frac{\As}{2\pi} \sum_{i \neq 1,2,3} \, {}^{(n-1)}\TT{R}^{i,1}_{2,3} ([i\bcdot 1-2] + [1-2\bcdot i]) - \frac{\As}{2\pi} \sum_{i \neq 1,2}\sum_{j \neq i,1,2} \, {}^{(n-1)}\TT{R}^{i,j}_{1,2} ([i\bcdot j] + [j\bcdot i]). \label{eq:amplitudeinringsandstrings}
\end{align}
Thus we have disentangled the colour diagonal collinear poles from the rest of the soft physics. We have done so without introducing subtraction terms (which introduce non-antenna function dependence into the colour density matrix elements \cite{Forshaw:2019ver}), without introducing a scale to separate collinear and soft modes (as in \cite{Magnea:2021fvy}), and without approximating the eikonal current in the collinear limit (as one typically does when considering zero bins or double counting between the soft and collinear limits). 

At this point, it is worth noting the integrated form of the ring functions we have introduced. Though the rings are collinear finite, the integral is most easily performed analytically whilst using a regulator:
\begin{align}
    \int \frac{\td \Omega_{n}^{3-2\varepsilon}}{4\pi } E^{2}_{n}\;{}^{(n-1)}\TT{R}^{i,j}_{k,l} &= \frac{1}{-2\varepsilon} \frac{\pi^{\frac{1-2\varepsilon}{2}}}{\Gamma(\frac{1-2\varepsilon}{2})} \left( f_{ij} - f_{ik} - f_{jl} + f_{kl} \right), \nonumber \\
    &= 2 \ln \left(\frac{n_{i} \cdot n_{j} \, n_{k} \cdot n_{l}}{n_{i} \cdot n_{k} \, n_{j} \cdot n_{l}}\right) + \mathcal{O}(\epsilon) \equiv 2 \ln \left(\rho_{ijkl}\right) + \mathcal{O}(\epsilon), \label{eq:dimvirtuals}
\end{align}
where $\Omega^d = 2 \pi^{d/2}/\Gamma(d/2)$, $f_{ij} = n_{i} \cdot n_{j} \, F_{1,2}(1,1,1-\varepsilon,\tfrac{n_{i}\cdot n_{j}}{2}),$ $n_{i}=q_{i}/E_{i}$ and $\rho_{ijkl}$ is a conformal cross-ratio \cite{Gardi:2009qi}. Using this integral, one can compare Eq.~\eqref{eq:amplitudeinringsandstrings} for $n=5$ with Eq.~(B.6) in \cite{Gardi:2009qi} and where it can be checked that, upon setting $\v{A}_{4} = 1$, the two are related term-by-term via the unitarity relation, Eq.~\eqref{eqs:unitarity}.

\subsection{Rings emerge from the colour flow basis}
In this section, we explore how rings emerge in the context of the amplitude-evolution framework in the colour flow basis. It is our hope that the key ideas can be appreciated without the need to descend into the details of the formalism, for which we refer to \cite{Platzer:2013fha,SoftEvolutionAlgorithm}. Let us first discuss the virtual evolution, which is controlled by 
the soft anomalous dimension:
\begin{equation}
    { [\sigma|{\mathbf \Gamma}|\tau\rangle} \propto \delta_{\tau\sigma} \left(\Gamma_\sigma + \frac{1}{\Nc^2}\rho\right) + \frac{1}{\Nc} \Sigma_{\tau\sigma} \ , \label{eq:GammaCF}
\end{equation}
where $\tau$ and $\sigma$ label colour flows (expressed as a permutation of the integers 1 to $n$, where $n$ is the number of flows). Since the colour flow basis is not orthogonal we need to take a little care defining the basis states: $|\sigma\rangle$ denotes a colour flow obtained from a string of Kronecker symbols $\delta^{i_1}_{j_{\sigma (1)}}\cdots \delta^{i_n}_{j_{\sigma(n)}}$ for $n$ colour lines, and the dual basis state is defined by $|\sigma]\equiv \sum_{\tau} (\underline{S}^{-1})_{\tau\sigma}|\sigma\rangle$ (for $S$ the scalar product matrix defined below) such that $\sum_\alpha |\alpha\rangle [\alpha|=1$. We refer the reader to \cite{Platzer:2013fha,SoftEvolutionAlgorithm} for more details. The colour diagonal, leading-$\Nc$, contribution $\Gamma_\sigma$ can be expressed as a sum over dipoles connecting colour charge and anti-charge. The $1/\Nc^2$ suppressed contribution, $\rho$, is also proportional to the unit operator in colour space, and involves a sum over conventional dipoles as well as a sum over colour connected colour/colour and anti-colour/anti-colour pairs:
\begin{equation}
    \Gamma_\sigma = \sum_{i,j \text{ c.c. in } \sigma} \Omega_{ij},\qquad
    \rho = -\sum_{i,j}\lambda_i \lambda_j\Omega_{ij}, \qquad \Omega_{ij} = \frac{(2 \pi\mu)^{2\epsilon}}{2 \pi}\int_{\TT{P.S.}} \frac{\td^{3-2\epsilon}\vec{q}}{2 E} \omega_{ij}(q) ,
\end{equation}
where $\lambda_i=\pm 1$ if $i$ is coloured or anti-coloured (with a gluon carrying both colour and anti-colour). Here `$i,\,j$ c.c. in $\sigma$' means that particles $i$ and $j$ are colour connected in the flow $\sigma$. 
Notably, the colour-suppressed, off-diagonal contributions involve certain combinations of eikonal factors \cite{Platzer:2013fha,non_global_logs}:
\begin{equation}
  \Sigma_{\tau\sigma} =\sum_{i,l \text{ c.c. in } \sigma}\sum_{~~j,k \text{ c.c. in } \tau}
  \Omega_{ij}+\Omega_{kl}-\Omega_{ik} - \Omega_{jl} \ , \label{eq:SigmaRing}
\end{equation}
which we recognise as a ring.

Now we turn our attention to the case of real emissions. What follows is closely related to the implementation in \texttt{CVolver} \cite{Platzer:2013fha,DeAngelis:2020rvq}.
In the colour flow basis we can write
\begin{equation}
    {\rm Tr}[{\mathbf A}_n] \equiv {\rm Tr} \left[\underline{A}_n \underline{S}_n\right],
\end{equation}
where $\underline{A}_n=([\tau|{\mathbf A}_n|\sigma])$ is the matrix representation of the colour space operator ${\mathbf A}_n$, and $\underline{S}_n=(\langle\sigma|\tau\rangle)$ is the matrix of scalar products of colour flows in the $n$-parton colour space, which satisfies 
\begin{equation}
   \langle\tau|\sigma\rangle = \Nc^{m - \#(\tau,\sigma)} \ ,
\end{equation}
where $m$ is the number of colour flows involved, and $\#(\tau,\sigma)$ denotes the minimal number of transpositions between the permutations $\tau$ and $\sigma$. Perturbative unitarity arises because of the cyclicity of the trace:
\begin{equation}
  {\rm Tr} [{\mathbf T}_i {\mathbf A}_n {\mathbf T}_j^\dagger] = {\rm Tr}[{\mathbf T}_j^\dagger {\mathbf T}_i {\mathbf A}_n], 
\end{equation}
and in the colour flow basis this is
\begin{equation}
    \underline{T_j}^\dagger \underline{S}_{n+1} \underline{T_i} = \underline{S}_n  \underline{T}_{ij} \ . \label{eq:colourflowunitarity}
\end{equation}
Our aim is to consider a sequence of real emissions, for which we need
\begin{equation}
   {\rm Tr}\left[\underline{T}_{i_m}^\dagger\cdots \underline{T}_{i_1}^\dagger \underline{S}_{n+m}\underline{T}_{j_1}\cdots \underline{T}_{j_m}\underline{A}_{n}\right] = 
   {\rm Tr}\left[\underline{S}_n\underline{T}_{i_m,...,i_1,j_1,...,j_m}\underline{A}_n\right]\ ,
\end{equation}
which we have written in terms of a correlator of many colour charges,
\begin{equation}
\label{eq:higherpoint}
    \underline{T}_{i_m,...,i_1,j_1,...,j_m} \equiv
    ([\tau|{\mathbf T}_{i_m}^\dagger \cdots {\mathbf T}_{i_1}^\dagger {\mathbf T}_{j_1}\cdots {\mathbf T}_{j_m}|\bar{\tau}\rangle)\ .
\end{equation}
In order to analyze the $\Nc$ dependence of the emission contributions alone, we recursively evaluate the trace until we hit the level of the hard density operator $\underline{A}_n$ and the according scalar product matrix $\underline{S}_n$.
Using the completeness relation
\begin{equation}
  {\mathbf 1} = \sum_{\lambda} |\lambda\rangle [\lambda| \ ,
\end{equation}
we can reduce the multi-colour correlator to product of matrix elements of single colour charges acting in the amplitude and conjugate amplitude, i.e. the quantity
\begin{equation}
\xi_{ij}^{(\tau\sigma\bar{\sigma}\bar{\tau})}=
   \frac{\langle\sigma|\bar{\sigma}\rangle}{\langle \tau|\bar{\tau}\rangle}[\sigma|{\mathbf T}_i|\tau\rangle[ \bar{\sigma}| {\mathbf T}_j |\bar{\tau}\rangle \ .
\end{equation}
This object is used to sample colour flows for real emissions in \texttt{CVolver} \cite{DeAngelis:2020rvq}, and using it we can evaluate the multi-colour correlator (Eq.~\eqref{eq:higherpoint}) recursively:
\begin{equation}
\label{eq:multirecursion}
\frac{
    \left(\underline{T}_{i_m,...,i_1,j_1,...,j_m} \right)_{\sigma\bar{\sigma}}}{\langle\sigma|\bar{\sigma}\rangle}=
    \sum_{\tau,\bar{\tau}}  \xi_{i_1,j_1}^{(\sigma\tau\bar{\tau}\bar{\sigma})} \times
    \frac{\left(\underline{T}_{i_m,...,i_2,j_2,...,j_m}\right)_{\tau\bar{\tau}}}{\langle\tau|\bar{\tau}\rangle} \ ,
\end{equation}
with initial condition $(\underline{T}_{ij})_{\sigma\bar{\sigma}}=\sum_{\lambda}[\lambda|{\mathbf T}_i|\bar{\sigma}\rangle [\lambda|{\mathbf T}_j|\sigma\rangle$.

\begin{figure}
    \centering
    \includegraphics[scale=0.8]{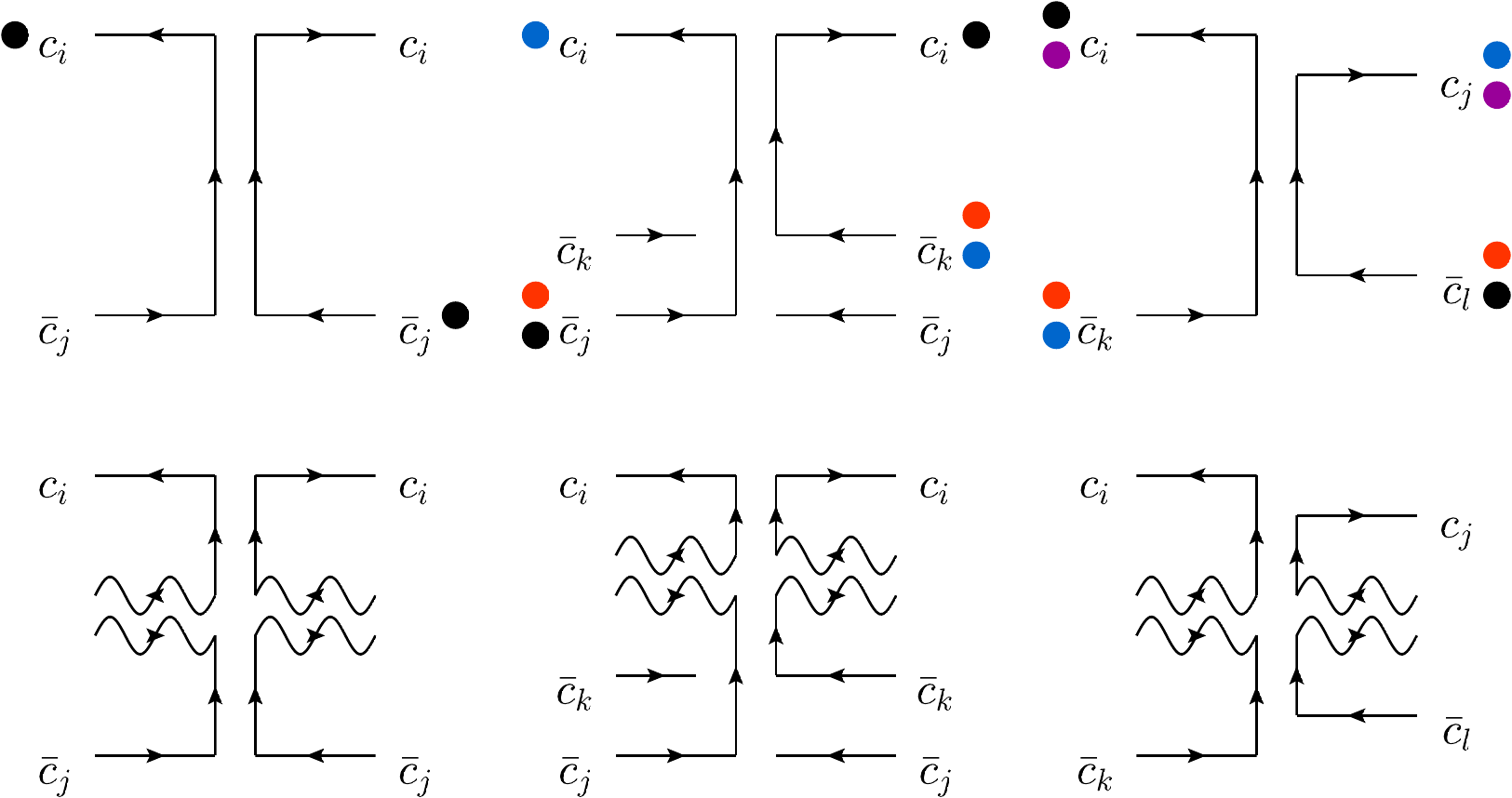}
    \caption{The colour flows in the density operator relevant for the emission of a gluon (the amplitude is drawn to the left and the conjugate amplitude to the right). Top row is before the emission and the bottom row is after (the emitted gluon's colour lines are indicated as wavy lines). See text for more details.}
    \label{fig:flows}
\end{figure}

The emission operator can be decomposed into a sum over operators acting on individual colour lines ($\mathbf{t}$ and $\mathbf{\bar{t}}$), and a `singlet' gluon operator ($\mathbf{s}$):
\begin{equation}
{\mathbf T}_i = \frac{1}{\sqrt{2}} \left(\lambda_i{\mathbf t}_{c_i} - \bar{\lambda}_i \bar{\mathbf t}_{\bar{c}_i} + (\lambda_i - \bar{\lambda}_i){\mathbf s}\right) \ ,
\label{eq:emdecomp}
\end{equation}
where $c_i$ is the colour line index assigned to parton $i$ (if this carries colour like a quark or a gluon), $\bar{c}_j$ the anti-colour index for parton $j$ (if this carries anti-colour like an anti-quark or a gluon) \cite{SoftEvolutionAlgorithm}. In order to analyze matrix elements of the density operator, we need to consider definite colour states in the amplitude and conjugate amplitude after the emission, as depicted in the lower row of Figure~\ref{fig:flows}. The three scenarios depicted cover all possible colour states after an emission (for a completely general set of prior colour connections). We have only drawn the colour lines relevant to the emission and the three cases differ by the number of colour lines that are common in the amplitude and its conjugate (two, one and zero from left to right). The wavy lines are drawn as such in order to clearly identify the emitted gluon.
The top row of the figure indicates the colour flows prior to the emission and the coloured circles indicate the two lines from which the gluon is emitted that deliver the colour flow in the bottom row. The point is that there can be several combinations that give rise to the same final colour flow. The left-most transition will only receive contributions from one dipole pair $i,j$, originating as either $-\omega_{ij}\lambda_i\bar{\lambda}_j t_{c_i}\bar{t}_{\bar{c}_j}^\dagger$, as indicated by the black dots, or with $i$ and $j$ interchanged. This contribution will only be non-vanishing if $i$ and $j$ carry colour and anti-colour for which $\lambda_i\bar{\lambda}_j=-1$. The middle flow topology allows for more pairs to contribute, and includes contributions from two colour/anti-colour pairs (black and blue) as well as one anti-colour/anti-colour pair, for which $\bar{\lambda}_i\bar{\lambda}_j=+1$. These transitions are all associated with a real emission contribution $\omega_{ij}+\textcolor{ublue}{\omega_{ik}}-\textcolor{uviolet}{\omega_{jk}}$, which is a string. The rightmost topology can receive contributions from up to four partons, and is accordingly weighted by a ring $\omega_{il}+\textcolor{ublue}{\omega_{kj}}-\textcolor{uviolet}{\omega_{ij}}-\textcolor{ured}{\omega_{kl}}$. 

In the presence of quarks a complication arises due to the action of the $\mathbf{s}$ operator. In this case we need to sum over the quark or anti-quark lines as emitters, leaving us with a sum over dipoles, $\sum_{j \in q} \omega_{ij} - \sum_{j \in \bar{q}} \omega_{ij} + (i\leftrightarrow k)$ if we consider a colour flow connecting partons $i$ and $k$ in the amplitude, and a `singlet' emission in the conjugate amplitude (and also a contribution corresponding to emitting a `singlet' in the amplitude and conjugate amplitude). This is a superficial complication since the resulting contribution to the radiation pattern again collapses into a sum over dipoles, strings and rings (though with a slightly different structure to those in the second row of Fig.~\ref{fig:flows}).

While the above observation explains the emergence of strings and rings directly from the colour flow basis, it does not yet explain the $1/N_c$ suppression. For this we must analyze $\xi$. One can show that $\xi_{ij}^{(\tau\sigma\bar{\sigma}\bar{\tau})} = \Nc$ whenever $i$ and $j$, or their colour connected partners, share at least one common colour line in the amplitude and conjugate. This applies to both the dipole and string transitions. More interesting is the case where no colour lines are common. This is the rightmost case in Figure~\ref{fig:flows}, which we have identified as a ring. In this case $\xi_{ij}^{(\tau\sigma\bar{\sigma}\bar{\tau})} = 1/\Nc\text{ or } \Nc$. For the former case, we get an explicit $1/\Nc$ suppression, for the latter the subsequent recursion needs to contribute a $1/\Nc$ suppressed contribution since the powers of $\Nc$ need to match on either side of Eq.~\eqref{eq:multirecursion}: this is because the matrix elements of the correlator $(T_{i_m,...,i_1;j_1,...,j_m})_{\sigma\bar{\sigma}}$ in the recursion evaluate to powers of $\pm 1$. The entire $\Nc$ dependence is hence carried by the choice of normalization $1/\langle\sigma|\bar{\sigma}\rangle$ that we associate with each step in the recursion, i.e. in Eq.~\eqref{eq:multirecursion}. 

\subsection{Impact on Numerical Simulations in CVolver}
\label{sec:simulation}

The emission contribution can be expressed as a sum over dipoles and a simple Monte Carlo implementation selects one such dipole to set up the kinematics of the emission, as well as to select one of the possible pairs of final colour flows. However, as we have seen in Figure~\ref{fig:flows},  two (dipoles), three (strings) or four (rings) terms can contribute to the same transition between the colour flows we have selected. Using dipoles does not present any problems in the large-$\Nc$ limit, but the account of subleading colour contributions relies on large collinear cancellations within a string or ring to take place on a statistical level, with severe consequences on the convergence of the Monte Carlo simulation. Clearly a better way to proceed is to use rings and strings directly.

We illustrate this with a result from \texttt{CVolver} \cite{Platzer:2013fha,DeAngelis:2020rvq}, in which we exploit a sampling over dipoles in comparison to a sampling which explicitly implements strings and rings according to the discussion we have presented in Section~\ref{sec:rings}. This is implemented by identifying which dipole factors contribute to the same colour flow transition (as in the previous section). We select one pair $i,l$ to importance sample a direction such that the collinear enhancement with respect to that particular dipole factor $\omega_{il}$ is mapped out, {\it i.e.} with a probability density proportional to $\omega_{il}$. Such a pair uniquely identifies a pair of colour flows we intend to emit from in the amplitude and the conjugate amplitude. This pair of colour flows fixes four parton indices out of which two will drive the kinematic sampling: two connecting to the colour line ends, say $i$ and $k$, and two connecting to the anti-colour line ends, say $j$ and $l$. We then apply a weight
\begin{equation}
   \frac{\omega_{il} + \omega_{jk} - \omega_{ik}-\omega_{jl}}{\omega_{il} + \omega_{jk} + \omega_{ik}+\omega_{jl}}
\end{equation}
which makes the cancellations in the string or ring structures manifest. This procedure removes the ambiguity in defining the string basis used in Eq.~\eqref{eq:recursion}, since we use the colour flow information to fix the accompanying indices of the emitter $i$ or pair $i,j$ picking up exactly either an $^{(n)}S_i^{k,l}$ or an $^{(n)}R^{i,j}_{k,l}$ contribution.

\begin{figure}
    \centering
    \includegraphics[scale=0.8]{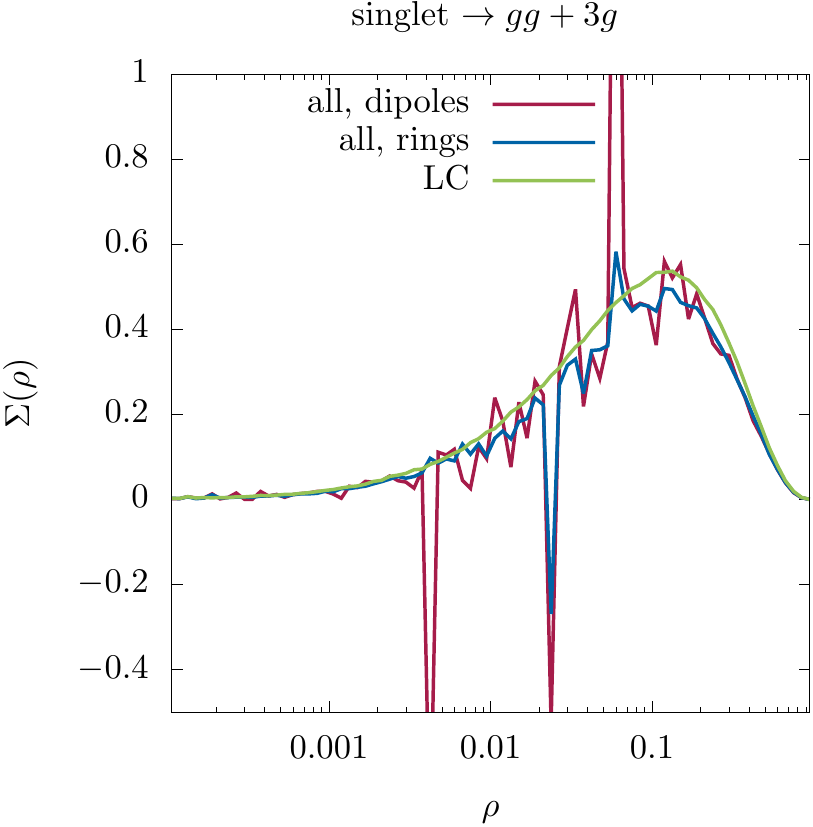}\hspace{0.5cm}
    \includegraphics[scale=0.8]{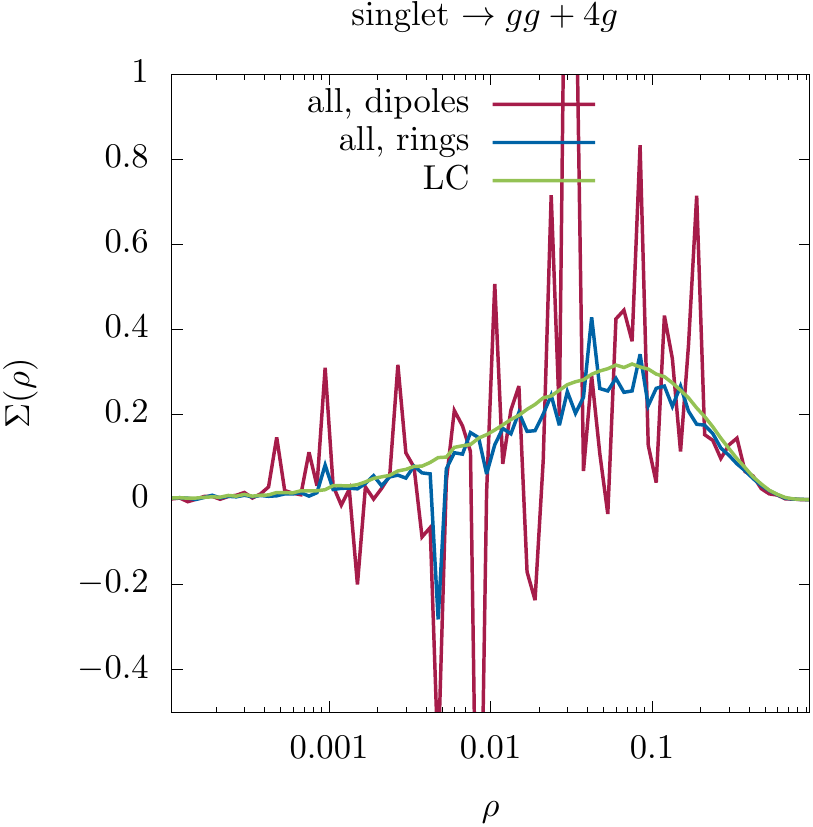}
    \caption{The three and four emission contributions to the soft gluon evolution of $h\to gg$ in the gaps-between-jets observable (including virtual corrections). The green line indicates the leading colour contribution, which agrees between the two sampling strategies. The violet curve uses only dipoles to generate each emission, while the blue curve corresponds to summing dipoles to form rings where appropriate. All runs have the same (deliberately low) statistics to illustrate the improvement.}
    \label{fig:DipolesAndRings}
\end{figure}

In Figure~\ref{fig:DipolesAndRings} we show the three and four gluon emission contributions to $h\to gg$ decay for the gaps-between-jets observable, where $\rho$ is the ratio of the veto scale to the hard scale and the veto region corresponds to $-\pi/4 < \theta < \pi/4$.  \texttt{CVolver} is able to separate the contributions by the powers of $N_c$ with which they contribute from the same Monte Carlo run, and we can thus explicitly observe that the leading colour curve (green) is not affected by the new implementation. The subleading colour results show a significant improvement over the version which only draws from dipoles. In this case our leading colour result takes the hard process normalization to be $\Nc^2$, rather than $\Nc^2-1$.

\subsection{An example: computing subleading colour at $\As^4$ in $H\rightarrow gg$}
\label{sec:example}
In this short section we will demonstrate how the use of rings and strings allows us to isolate and simplify the subleading colour contribution to $H\rightarrow g^{(1)}g^{(2)} (g^{(3)}g^{(4)}g^{(5)}g^{(6)})$ at $\mathcal{O}(\As^4)$, where $(g^{(3)}g^{(4)}g^{(5)}g^{(6)})$ are four energy-ordered, soft gluons. The cross section is given by
\begin{align}
    \frac{\td \Sigma^{\TT{RRRR}}_{4}}{\td \Phi_{4}} = \frac{\As^{4}}{\pi^{4}}\sum_{i_{4}\neq j_{4}}\sum_{i_{3}\neq j_{3}}\sum_{i_{2}\neq j_{2}}\sum_{i_{1}\neq j_{1}}\omega^{(6)}_{i_{4}j_{4}}\omega^{(5)}_{i_{3}j_{3}}\omega^{(4)}_{i_{2}j_{2}}\omega^{(3)}_{i_{1}j_{1}}\Tr( \v{T}_{i_{4}}\v{T}_{i_{3}}\v{T}_{i_{2}}\v{T}_{i_{1}}\v{H} \v{T}^{\dagger}_{j_{1}}\v{T}^{\dagger}_{j_{2}}\v{T}^{\dagger}_{j_{3}}\v{T}^{\dagger}_{j_{4}}),
\end{align}
where $\omega^{(n)}_{ij} = \omega_{ij}(q_{n})$ and $\sigma_{0} = \Tr \, \v{H} \propto (\Nc^{2}- 1)$, where $\sigma_{0}$ is the Born $H\rightarrow gg$ cross section. We find, after rotating the amplitude density matrix to rings and strings and computing the colour trace, 
\begin{align}
    \frac{\td \Sigma^{\TT{RRRR}}_{4}}{\td \Phi_{4}} &= \frac{\As^{4}\Nc^{4}\sigma_{0}}{\pi^{4}}\omega^{(3)}_{12}\sum_{i_{4} \, \TT{c.c.} \, j_{4}}\sum_{i_{3}\, \TT{c.c.} \, j_{3}}\sum_{i_{2}\, \TT{c.c.} \, j_{2}} \omega^{(6)}_{i_{4}j_{4}}\omega^{(5)}_{i_{3}j_{3}}\omega^{(4)}_{i_{2}j_{2}} + \nonumber \\ & 24 \frac{\As^{4}\Nc^{2}\sigma_{0}}{\pi^{4}} \omega^{(3)}_{12} \left(\, {}^{(3)}\TT{S}^{2,3}_{1}\, {}^{(4)}\TT{R}^{1,2}_{3,4}\, {}^{(5)}\TT{R}^{1,2}_{3,4} + \, {}^{(3)}\TT{S}^{1,2}_{3}\, {}^{(4)}\TT{R}^{1,2}_{3,4}\, {}^{(5)}\TT{R}^{1,4}_{3,2} \right) + (1 \leftrightarrow 2). \label{eq:H4grings}
\end{align}
The first line is the leading colour piece, computable by summing over colour connected dipoles (the sums over `$i\, \TT{c.c.} \, j$' are sums over gluons $i,j$ that are colour connected in planar topologies). Note that the structure of Eq.~\eqref{eq:H4grings} is predictable from Eq.~\eqref{eq:amplitudeinringsandstrings} since, for $H\rightarrow gg(ggg\cdots)$, we know that after taking the trace terms are either proportional to the 't Hooft coupling (as all Casimirs accompanying collinear poles will be in the adjoint representation) or must contain the product of at least two rings (so that colour has chance to leave the diagonal then return)\footnote{See Appendix \ref{app:LargeNc} for a more complete discussion on why for this process rings do not contribute to the leading colour diagonal.}. This is what we see in Eq.~\eqref{eq:H4grings} and it explains the absence of subleading colour in $H\rightarrow g^{(1)}g^{(2)} (g^{(3)}g^{(4)}g^{(5)})$. 

Eq.~\eqref{eq:H4grings} can be compared with the same cross-section computed in terms of antenna functions:
\begin{align}
   &\frac{\td \Sigma^{\TT{RRRR}}_{4}}{\td \Phi_{4}} = \nonumber \\ 
   &\mathcal{O}(\Nc^{4}) + 12  \frac{\As^{4}\Nc^{2}\sigma_{0}}{\pi^{4}} \omega ^{(3)}_{1 2} \bigg(\omega ^{(4)}_{2 3} \Big(-\omega ^{(5)}_{1 4} \omega ^{(6)}_{1 2}+\omega ^{(5)}_{2 4} \omega ^{(6)}_{1 2}+\omega ^{(5)}_{1 2} \omega ^{(6)}_{1 3}-\omega ^{(5)}_{1 4} \omega ^{(6)}_{1 3}+\omega ^{(5)}_{3 4} \omega ^{(6)}_{1 3}
   \nonumber \\ 
   &-\omega ^{(5)}_{1 2} \omega ^{(6)}_{2 3}+2 \omega ^{(5)}_{1 4} \omega ^{(6)}_{2 3}-\omega ^{(5)}_{2 4} \omega ^{(6)}_{2 3}-\omega ^{(5)}_{3 4} \omega ^{(6)}_{2 3}-\omega ^{(5)}_{1 2} \omega ^{(6)}_{1 4}+2 \omega ^{(5)}_{1 4} \omega ^{(6)}_{1 4}-\omega ^{(5)}_{2 4} \omega ^{(6)}_{1 4}-\omega ^{(5)}_{3 4} \omega ^{(6)}_{1 4}
   \nonumber \\ 
   &+\omega ^{(5)}_{1 2} \omega ^{(6)}_{2 4}-\omega ^{(5)}_{1 4} \omega ^{(6)}_{2 4}+\omega ^{(5)}_{3 4} \omega ^{(6)}_{2 4}-\omega ^{(5)}_{1 4} \omega ^{(6)}_{3 4}+\omega ^{(5)}_{2 4} \omega ^{(6)}_{3 4}+\omega ^{(5)}_{1 3} (\omega ^{(6)}_{1 2}-\omega ^{(6)}_{2 3}-\omega ^{(6)}_{1 4}+\omega ^{(6)}_{3 4})
   \nonumber \\ 
   &-\omega ^{(5)}_{2 3} (\omega ^{(6)}_{1 2}+\omega ^{(6)}_{1 3}-2 \omega ^{(6)}_{2 3}-2 \omega ^{(6)}_{1 4}+\omega ^{(6)}_{2 4}+\omega ^{(6)}_{3 4})\Big)
   \nonumber \\ 
   &-\omega ^{(4)}_{1 3} \Big(-\omega ^{(5)}_{1 4} \omega ^{(6)}_{1 2}+\omega ^{(5)}_{2 4} \omega ^{(6)}_{1 2}+\omega ^{(5)}_{1 2} \omega ^{(6)}_{1 3}+\omega ^{(5)}_{1 4} \omega ^{(6)}_{1 3}-2 \omega ^{(5)}_{2 4} \omega ^{(6)}_{1 3}+\omega ^{(5)}_{3 4} \omega ^{(6)}_{1 3}-\omega ^{(5)}_{1 2} \omega ^{(6)}_{2 3}
   \nonumber \\ 
   &+\omega ^{(5)}_{2 4} \omega ^{(6)}_{2 3}-\omega ^{(5)}_{3 4} \omega ^{(6)}_{2 3}-\omega ^{(5)}_{1 2} \omega ^{(6)}_{1 4}+\omega ^{(5)}_{2 4} \omega ^{(6)}_{1 4}-\omega ^{(5)}_{3 4} \omega ^{(6)}_{1 4}+\omega ^{(5)}_{1 2} \omega ^{(6)}_{2 4}+\omega ^{(5)}_{1 4} \omega ^{(6)}_{2 4}
   \nonumber \\ 
   &-2 \omega ^{(5)}_{2 4} \omega ^{(6)}_{2 4}+\omega ^{(5)}_{3 4} \omega ^{(6)}_{2 4}+\omega ^{(5)}_{2 3} (-\omega ^{(6)}_{1 2}+\omega ^{(6)}_{1 3}+\omega ^{(6)}_{2 4}-\omega ^{(6)}_{3 4})
   \nonumber \\ 
   &-\omega ^{(5)}_{1 4} \omega ^{(6)}_{3 4}+\omega ^{(5)}_{2 4} \omega ^{(6)}_{3 4}+\omega ^{(5)}_{1 3} (\omega ^{(6)}_{1 2}-2 \omega ^{(6)}_{1 3}+\omega ^{(6)}_{2 3}+\omega ^{(6)}_{1 4}-2 \omega ^{(6)}_{2 4}+\omega ^{(6)}_{3 4})\Big)
   \nonumber \\ 
   &+\omega ^{(4)}_{1 2} \Big(-\omega ^{(5)}_{2 3} \omega ^{(6)}_{1 2}-\omega ^{(5)}_{1 4} \omega ^{(6)}_{1 2}-\omega ^{(5)}_{2 4} \omega ^{(6)}_{1 2}+2 \omega ^{(5)}_{3 4} \omega ^{(6)}_{1 2}+\omega ^{(5)}_{2 3} \omega ^{(6)}_{1 3}+\omega ^{(5)}_{1 4} \omega ^{(6)}_{1 3}-\omega ^{(5)}_{3 4} \omega ^{(6)}_{1 3}
   \nonumber \\ 
   &+\omega ^{(5)}_{2 4} \omega ^{(6)}_{2 3}-\omega ^{(5)}_{3 4} \omega ^{(6)}_{2 3}+\omega ^{(5)}_{2 4} \omega ^{(6)}_{1 4}-\omega ^{(5)}_{3 4} \omega ^{(6)}_{1 4}+\omega ^{(5)}_{2 3} \omega ^{(6)}_{2 4}+\omega ^{(5)}_{1 4} \omega ^{(6)}_{2 4}-\omega ^{(5)}_{3 4} \omega ^{(6)}_{2 4}
   \nonumber \\ 
   &+\omega ^{(5)}_{1 3} (-\omega ^{(6)}_{1 2}+\omega ^{(6)}_{2 3}+\omega ^{(6)}_{1 4}-\omega ^{(6)}_{3 4})-\omega ^{(5)}_{2 3} \omega ^{(6)}_{3 4}-\omega ^{(5)}_{1 4} \omega ^{(6)}_{3 4}-\omega ^{(5)}_{2 4} \omega ^{(6)}_{3 4}+2 \omega ^{(5)}_{3 4} \omega ^{(6)}_{3 4}
   \nonumber \\ 
   &+\omega ^{(5)}_{1 2} (2 \omega ^{(6)}_{1 2}-\omega ^{(6)}_{1 3}-\omega ^{(6)}_{2 3}-\omega ^{(6)}_{1 4}-\omega ^{(6)}_{2 4}+2 \omega ^{(6)}_{3 4})\Big)\bigg).
\end{align}
The virtue of using rings and strings for expressing subleading colour is apparent.

\section{QCD coherence}

In this section we use rings and strings to study QCD coherence. We start by re-deriving the classic results for two-jet coherence \cite{CATANI1991635} and then extend these to include three-jet processes. The traditional approach to coherence in the two-jet limit exploits the azimuthal symmetry of the process (in a frame where the jets are back-to-back) in order to average azimuthal degrees of freedom. There is no equivalent symmetry for a general three-jet process and so we are required to discuss the subtleties of taking a collinear limit without azimuthal averaging. This is the topic of Section~\ref{sec:collinearlimit}. In Section~\ref{sec:revisit}, we use the results of Section~\ref{sec:collinearlimit} to derive two-jet coherence without any azimuthal averaging. In Section~\ref{sec:coherencegenerally} we perform an all-orders calculation of coherence in the three-jet limit. This calculation is supported by the analysis in Appendix \ref{sec:fullderivation}. In Section~\ref{sec:extensions} we discuss the addition of collinear physics and finally, in Section~\ref{sec:partonshower}, we discuss the modification to an angular-order parton shower that is motivated by our analysis.

\label{sec:coherence}
\subsection{An elegant re-derivation of two-jet coherence}
\label{sec:rederivation}

\begin{figure}[h]
	\centering
	\includegraphics[width=0.3\textwidth]{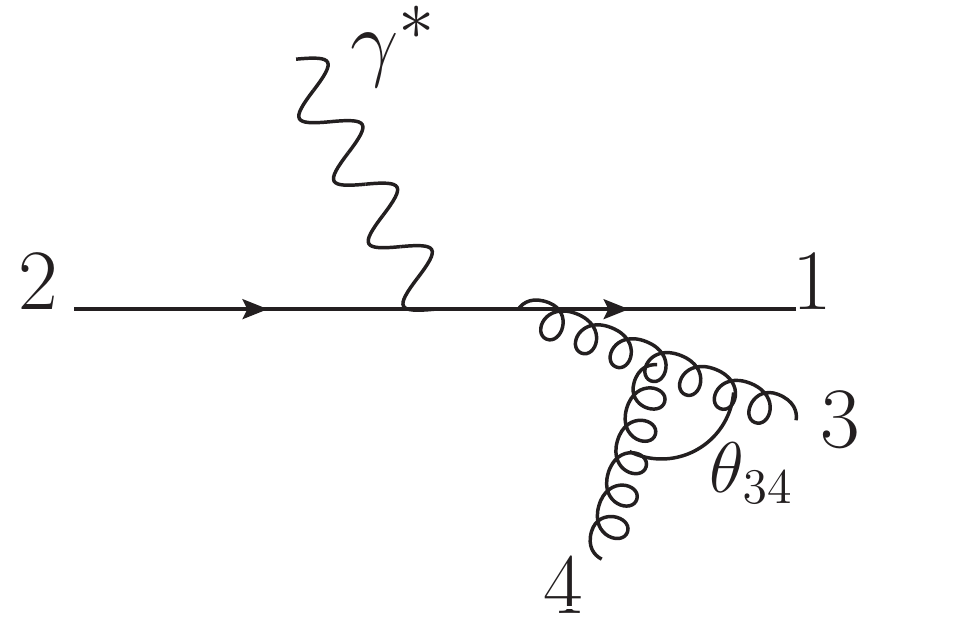}
	\caption{One of the $\mathcal{O}(\As^{2})$ $\gamma^{*}\rightarrow q \bar{q} gg$ Feynman diagrams that contribute to the amplitude density matrix $\v{A}_{4}$.}	\label{fig:matrixlement}
\end{figure}

\begin{figure}[t]
	\centering
	\subfigure[$\propto P_{13}^{[1]}$]{\includegraphics[width=0.2\textwidth]{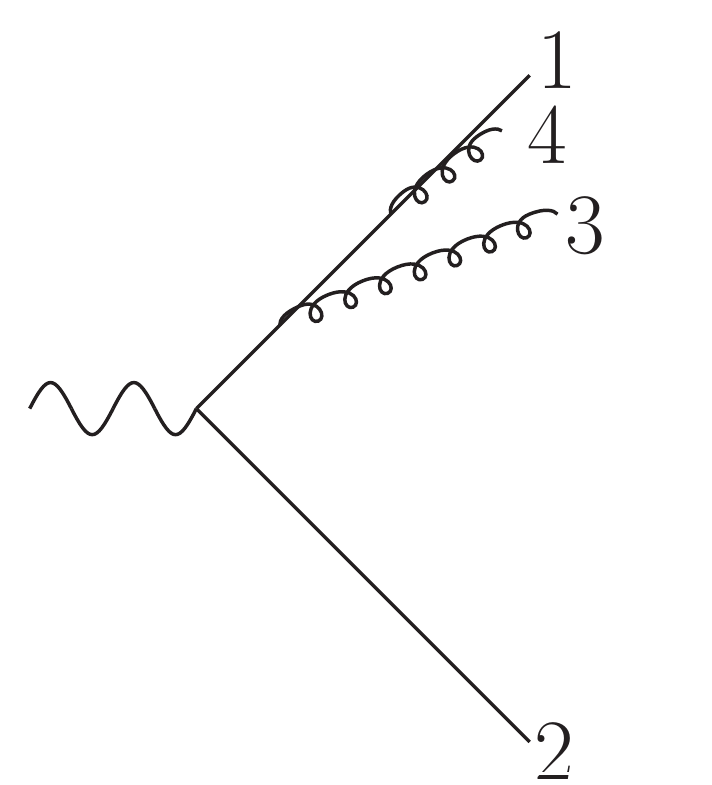}}
	\subfigure[$\propto P_{31}^{[3]}$]{\includegraphics[width=0.2\textwidth]{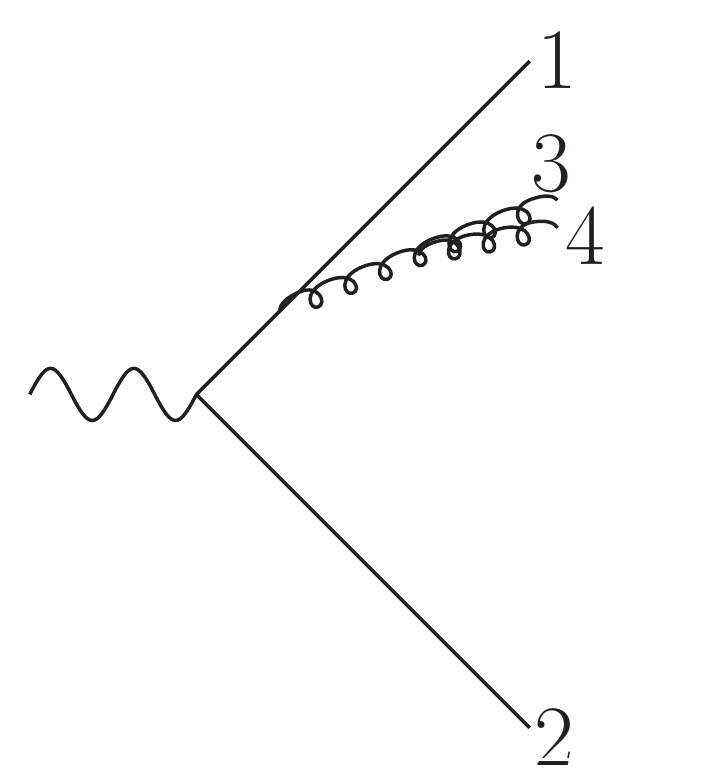}} 	
	\subfigure[$\propto P_{21}^{[2]} $]{\includegraphics[width=0.2\textwidth]{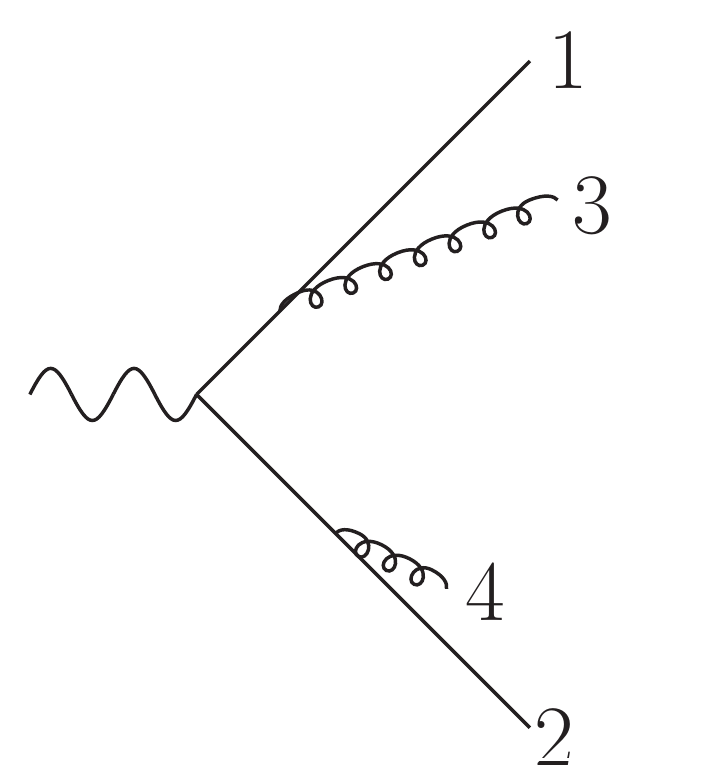}}
	\subfigure[$\propto \tilde{P}_{(1+3)2}^{[1+3]}(\theta_{13})$]{\includegraphics[width=0.2\textwidth]{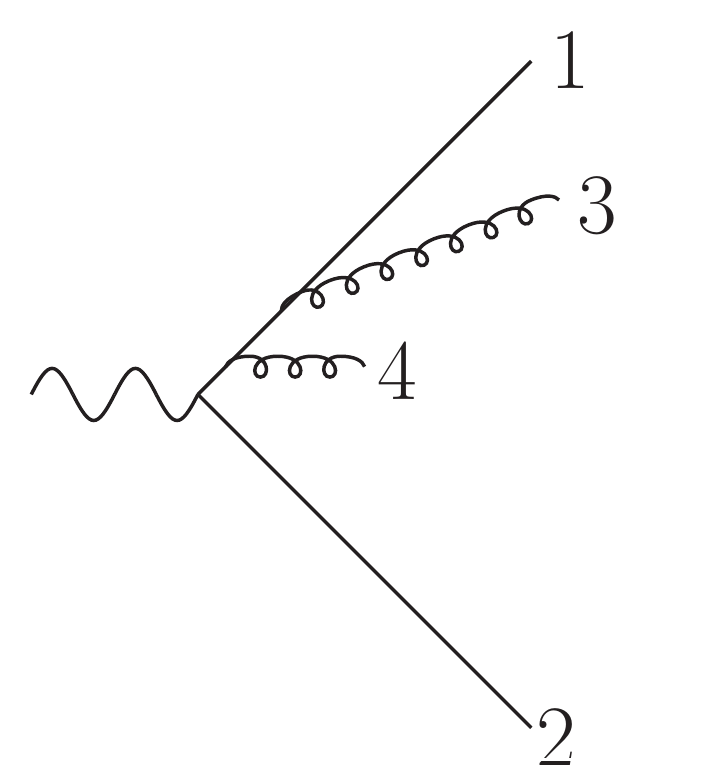}}
	\caption{Diagrams illustrating the angular ordered interpretation of $\gamma^{*} \rightarrow q^{(1)} \bar{q}^{(2)} g^{(3)} g^{(4)}$. The relative lengths of lines depict the relative energies of the particles. Likewise, the angles between lines are indicative. These terms are computable from $\v{A}_{4}$ in the limit that particle $3$ is collinear to particle $1$, and particle $4$ is soft. As the amplitude density matrix is diagonal, these diagrams can be considered representative of contributions at cross-section level.}
	\label{fig:intuitionforcoherence}
\end{figure}

The lowest order amplitude density matrix that displays coherence is $\v{A}_{4}$, which could represent a process $e^{+}e^{-}\rightarrow q^{(1)} \bar{q}^{(2)} g^{(3)}g^{(4)}$ (see Figure \ref{fig:matrixlement}). We consider the limit that particles $1$ and $2$ define two back-to-back hard jets, particle $3$ is collinear to particle $1$, and particle $4$ is soft ($E_{4} \ll E_{1},E_{2},E_{3}$). $\v{A}_{4}$, expressed in terms of strings, can be read off from Eq.~\eqref{eq:amplitudeinringsandstrings}:
\begin{align}
   \v{A}_{4} = \frac{2\As}{\pi} \big(\, {}^{(3)}\TT{S}^{2,3}_{1}[1\bcdot  1] + \, {}^{(3)}\TT{S}^{1,3}_{2}[2\bcdot  2]  + \, {}^{(3)}\TT{S}^{1,2}_{3}[3\bcdot 3] \big), \label{eq:A4}
\end{align}
where $[i\bcdot j]=\v{T}_{i} \lkl M_{3}  \> \< M_{3}  \rkl \v{T}^{\dagger}_{j}$. To derive QCD coherence, one usually makes the following manipulations. Firstly one defines a function
\begin{align} \label{eq:Pdef}
2P_{ij} =  w^{(n)}_{ij} + \frac{E_{i}}{E_{n} \, q_{i}\cdot q_{n}} - \frac{E_{j}}{E_{n} \, q_{j}\cdot q_{n}}.
\end{align}
This function has the property that it azimuthally averages to give a step function in the emission angle ($\theta_{in}$) and the dipole opening angle ($\theta_{ij}$):
\begin{align}
E^{2}_{n}\int^{2\pi}_{0} \frac{\td \phi^{(i)}_{q}}{2\pi} P_{ij} = \frac{1}{1 - \cos \theta_{in}} \Theta(\theta_{in} < \theta_{ij}),
\end{align}
where $q_{i} \cdot q_{j}= E_{i}E_{j}(1 - \cos\theta_{ij})$ and $q_{i} \cdot q_{n}= E_{i}E_{n}(1 - \cos\theta_{in})$. In turn, this is used to define
\begin{align}
P_{ij}^{[i]} = -\frac{(\td\cos\theta_{in}) \; \td \phi^{(i)}_{q}}{4\pi} \; \int^{2\pi}_{0} \frac{\td \phi^{(i)}_{q}}{2\pi} P_{ij}.
\end{align}
After azimuthal averaging, the string, $\, {}^{(n-1)}\TT{S}^{j,k}_{i}$, can be written\footnote{A word on notation. In this paper there are two different approximations which we will discuss: azimuthal averaging and the collinear limit. Generally speaking we will signpost a collinear limit with either a restriction (e.g. ``$|_{n_{4}\rightarrow n_{1}}$'') or by including an explicit limit function. The combination $\frac{\td \Omega_{n}}{4\pi} \approx$ implies that what follows has been azimuthally averaged.}
\begin{align}
    \, {}^{(n-1)}\TT{S}^{j,k}_{i} \frac{\td \Omega_{n}}{4\pi} \approx \frac{1}{2}\left(P^{[i]}_{ij} + P^{[i]}_{ik}\right) +  \frac{1}{2}\left(P^{[j]}_{ji} - P^{[j]}_{jk}\right) + \frac{1}{2}\left(P^{[k]}_{ki} - P^{[k]}_{kj}\right).
\end{align}
Now consider the limit that two of the three particles ($i,j,k$) are collinear and the third is anti-collinear. We will neglect any terms that do not have a collinear pole in $\theta_{in}$ and vanish in the exact collinear limit, as is usual when deriving QCD coherence. To aid us, we also define $\tilde{P}^{[j]}_{jk} (\theta_{ij})= P^{[j]}_{jk} \Theta(\theta_{jn} > \theta_{ij})$. In the collinear limit, $P^{[k]}_{ki} \approx P^{[k]}_{kj}$ and $\tilde{P}^{[j]}_{jk} (\theta_{ij}) \approx \tilde{P}^{[i]}_{ik} (\theta_{ij})$, and so 
\begin{align}
    &\, {}^{(n-1)}\TT{S}^{j,k}_{i} \frac{\td \Omega_{n}}{4\pi} \bigg|_{n_i \rightarrow n_j} \approx P^{[i]}_{ij}, \qquad \, {}^{(n-1)}\TT{S}^{j,k}_{i} \frac{\td \Omega_{n}}{4\pi} \bigg|_{n_i \rightarrow n_k} \approx P^{[i]}_{ik}, \nonumber \\ & \, {}^{(n-1)}\TT{S}^{j,k}_{i} \frac{\td \Omega_{n}}{4\pi} \bigg|_{n_j \rightarrow n_k} \approx P^{[i]}_{ij} + \tilde{P}^{[j]}_{ji}(\theta_{jk})\approx P^{[i]}_{ik} + \tilde{P}^{[k]}_{ki}(\theta_{jk}), \label{eq:stringsavaregedcoll}
\end{align} 
where $n_{i} = q_{i}/E_{i}$. Here we see the structures of coherence emerging: strings split apart into terms that can probe a collinear dipole and wide-angle terms that cannot. A two-jet coherence pattern emerges immediately from $\v{A}_{4}$ when we azimuthally average the strings in the limit $n_{1} \rightarrow n_{3}$, as is given in Eqs.~\eqref{eq:stringsavaregedcoll}:
\begin{align}
   \v{A}_{4} \frac{\td \Omega_{n}}{4\pi}  \bigg|_{n_1 \rightarrow n_3} & \approx \frac{2\As}{\pi} \big(P^{[1]}_{13}[1\bcdot 1] + (P^{[2]}_{21} + \tilde{P}^{[1+3]}_{(1+3)2}(\theta_{13}))[2\bcdot 2]  + P^{[3]}_{31}[3\bcdot 3] \big) \nonumber \\
   & = \frac{2\As}{\pi} \big(P^{[1]}_{13}[1\bcdot 1] + P^{[2]}_{21}[2\bcdot 2] + \tilde{P}^{[1+3]}_{(1+3)2}(\theta_{13})[1+3\bcdot 1+3]  + P^{[3]}_{31}[3\bcdot 3] \big), \label{eq:2-jetcoherenceaveraged}
\end{align}
where we have used colour conservation to arrive at the third term and where $q_{1+3} =q_{1}+q_{3}$. This is the usual coherence result, which can be given the angular-ordered interpretation in Figure \ref{fig:intuitionforcoherence}.
 
\subsection{Strings and rings in the collinear limit}
\label{sec:collinearlimit}

We have just demonstrated that a coherent radiation pattern can be obtained from an amplitude density matrix dependent only on strings. We wish to generalise this radiation pattern to include hard processes beyond the two-jet limit. However, first we must discuss a few subtleties regarding the collinear limit.

To motivate the following discussion, let us consider the ring $\, {}^{(n-1)}\TT{R}^{i,j}_{k,l}$ in the limit that $n_i \rightarrow n_j$ where $n_{i,j} = q_{i,j}/E_{i,j}$. It is easy to convince oneself that $\, {}^{(n-1)}\TT{R}^{i,j}_{k,l} \rightarrow -2\, {}^{(n-1)}\TT{S}^{k,l}_{i}$, assuming all other momenta are well spaced. We refer to this result as the naive limit. Now let us compare the naive limit with the same limit as computed on the azimuthally averaged ring:
\begin{align}
    \, {}^{(n-1)}\TT{R}^{i,j}_{k,l}\frac{\td \Omega_{n}}{4\pi} \approx \left(P^{[i]}_{ij} - P^{[i]}_{ik}\right) +  \left(P^{[j]}_{ji} - P^{[j]}_{jl}\right) + \left(P^{[k]}_{kl} - P^{[k]}_{ki}\right) + \left(P^{[l]}_{lk} - P^{[l]}_{lj}\right).
\end{align}
We can regroup terms so that the cancellation of collinear poles is apparent:
\begin{align}
    \, {}^{(n-1)}\TT{R}^{i,j}_{k,l}\frac{\td \Omega_{n}}{4\pi} \approx & \tilde{P}^{[i]}_{ij} (\theta_{ik}) + \tilde{P}^{[j]}_{ji} (\theta_{jl}) + \tilde{P}^{[k]}_{kl} (\theta_{ik}) + \tilde{P}^{[l]}_{lk} (\theta_{jl}) \nonumber \\ 
    &-\tilde{P}^{[i]}_{ik} (\theta_{ij}) - \tilde{P}^{[j]}_{jl} (\theta_{ij}) - \tilde{P}^{[k]}_{ik} (\theta_{kl}) - \tilde{P}^{[l]}_{lj} (\theta_{kl}),
\end{align}
where each $\tilde{P}^{[j]}_{jk} (\theta_{ij})= P^{[j]}_{jk} \Theta(\theta_{jn} > \theta_{ij})$ has no collinear pole. Upon taking the limit $n_i \rightarrow n_j$ we find that
\begin{align}
    \, {}^{(n-1)}\TT{R}^{i,j}_{k,l}\frac{\td \Omega_{n}}{4\pi} \rightarrow -2\, {}^{(n-1)}\TT{S}^{k,l}_{i}\Theta(\theta_{in} > \theta_{ij}) \frac{\td \Omega_{n}}{4\pi}. \label{eq:averagedapproxring}
\end{align}
The naive limit and the limit of the azimuthally averaged ring are consistent when $\theta_{ij}=0$ and $\theta_{in}>0$, however they are otherwise inconsistent. The naive limit misses the important region where gluon $n$ is soft and collinear, $\theta_{in} < \theta_{ij}$, by assuming all momenta other than $q_{i}$ and $q_{j}$ are well spaced. In the $\theta_{in} < \theta_{ij}$ region the ring does not diverge but the naive limit loses track of this fact and thus loses all dependence on $j$. The $\theta_{in} < \theta_{ij}$ region is handled by the azimuthally averaged ring as the step functions take care of the $\theta_{in} < \theta_{ij}$ region and so the well spaced assumption is not necessary.

The naive limit failed because the limits $q_{n}/E_{n} \rightarrow n_{i}$ and $n_{i} \rightarrow n_{j}$ do not commute. As an illustrative example 
$$\lim_{q_{n}/E_{n} \rightarrow n_{i}}\lim_{n_{i} \rightarrow n_{j}} \omega_{ij}(q_{n}) \rightarrow 0,$$
whilst
\begin{align}
    \lim_{n_{i} \rightarrow n_{j}} \lim_{q_{n}/E_{n} \rightarrow n_{i}}\omega_{ij}(q_{n}) \rightarrow \infty .
\end{align}
The limits $q_{n}/E_{n} \rightarrow n_{i}$ and $n_{i} \rightarrow n_{j}$ are not smoothly connected. There is a discontinuity between the regions where $\theta_{ni} \ll \theta_{ij} \ll 1$ and $\theta_{ij} \ll \theta_{ni} \ll 1$.\footnote{The region of phase-space where $q_{n}/E_{n} \rightarrow n_{i}$ and $n_{j} \rightarrow n_{i}$ at the same rate contains the region where $q_{n}/E_{n}= n_{j}$ and so also diverges (though non-trivially as it depends on the directions which $q_{n}/E_{n}$ and $n_{j}$ approach $n_{i}$). The presence of this pole signals that the discontinuous transition between the $q_{n}/E_{n} \rightarrow n_{i}$ and $n_{i} \rightarrow n_{j}$ limiting behaviours occurs at $\theta_{ni} = \Lambda \theta_{ij}$ for $\Lambda=1$.} This feature leads to the limiting properties of strings and rings being non-trivial, and is what azimuthally averages to generate the step function. 

Let us look more carefully at the limits $q_{n}/E_{n} \rightarrow n_{i}$ and $n_{i} \rightarrow n_{j}$ of a ring ${}^{(n-1)}\TT{R}^{i,j}_{k,l}$. We find that
$$\lim_{q_{n}/E_{n} \rightarrow n_{i}}\lim_{n_{i} \rightarrow n_{j}} {}^{(n-1)}\TT{R}^{i,j}_{k,l} \rightarrow \lim_{q_{n}/E_{n} \rightarrow n_{i}} -2 {}^{(n-1)}\TT{S}^{k,l}_{i} \rightarrow -\infty,$$
whilst
\begin{align}
    \lim_{n_{i} \rightarrow n_{j}} \lim_{q_{n}/E_{n} \rightarrow n_{i}}{}^{(n-1)}\TT{R}^{i,j}_{k,l} \rightarrow \omega_{kl} + \mathcal{O}(\theta^{0}_{in},\theta^{0}_{ij}).
\end{align}
If we wish to study the limit $n_{i} \rightarrow n_{j}$ independently of the angle at which gluon $n$ is emitted (as we do in the collinear limit relevant to coherence) we must deal with these non-commuting limits carefully. This requires that we separate regions as
\begin{align}
    {}^{(n-1)}\TT{R}^{i,j}_{k,l} = {}^{(n-1)}\TT{R}^{i,j}_{k,l} \Theta(\theta_{in} > \theta_{ij}) + {}^{(n-1)}\TT{R}^{i,j}_{k,l} \Theta(\theta_{in} < \theta_{ij}), \label{eq:divideuptheregions}
\end{align}
where the first term contains divergences in $\theta_{in}$ as $n_{i} \rightarrow n_{j}$ and the second is finite. The boundary of the two regions ($\theta_{in} = \theta_{ij}$) is determined by the presence of an integrable pole in ${}^{(n-1)}\TT{R}^{i,j}_{k,l}$ which emerges when $q_{n}$ lies in the $ij$ plane with $\theta_{in} = \theta_{ij}$ (this integrable pole is what azimuthally averages to generate the usual angular-ordered step function in Eq.~\eqref{eq:averagedapproxring}).\footnote{ More  formally, to find the $n_{i} \rightarrow n_{j}$ limiting behavior of $\v{A}_{n}$ at an arbitrary point in the $n$-parrticle phase-space we must partition the phase-space into multiple regions such that in each region the density matrix is a smooth function upon which the limit is well defined. These regions are necessarily bounded by the presence of poles (integrable or otherwise). Once the phase-space is suitably partitioned, we can compute the Laurent series around the $n_{i} \rightarrow n_{j}$ pole in each region. ${}^{(n-1)}\TT{R}^{i,j}_{k,l}$ has four integrable poles, which appear at $\theta_{in} = 0,\theta_{ij},\theta_{ik},\theta_{il}$ with particular values of the azimuth $\phi_{n}$ (for simplicity we here assume that $0,\theta_{ij},\theta_{ik},\theta_{il}$ are ordered in size). Therefore, strictly speaking, the phase-space should be divided into four regions $0<\theta_{in} < \theta_{ij}$, $\theta_{ij}<\theta_{in} < \theta_{ik}$, $\theta_{ik}<\theta_{in} < \theta_{il}$, and $\theta_{il}<\theta_{in} < \pi$. In each region the $n_{i} \rightarrow n_{j}$ limit should be independently analysed. However, the latter three regions trivially all share the same limiting behavior and so have been grouped together in Eq.~\eqref{eq:divideuptheregions}.} 

We wish to approximate $\, {}^{(n)}\TT{R}^{i,j}_{k,l}$ in the collinear limit $n_{i} \rightarrow n_{j}$ at the same level of accuracy with which we approximated the functions $P_{ij}^{[i]}$ when deriving QCD coherence. Specifically, we keep all terms with non-integrable collinear poles in the emission angle of gluon $n$ ($\theta_{in}$ and $\theta_{jn}$)\footnote{Integrable poles do not produce logarithms. } and terms which remain finite as $\theta_{ij} \rightarrow 0$. $\, {}^{(n-1)}\TT{R}^{i,j}_{k,l}\Theta(\theta_{in} < \theta_{ij})$ both does not have a collinear pole and vanishes in the limit $\theta_{ij} \rightarrow 0$ and so we set it to zero. Thus, in this \textit{quasi-collinear} limit, $\theta_{ij} \ll 1$\footnote{Hereafter we will use quasi-collinear to specify the limit $\theta \ll 1$ whilst $\theta \neq 0$.}, the ring is
\begin{align}
   & {}^{(n-1)}\TT{R}^{i,j}_{k,l} \bigg|_{n_i \rightarrow n_j} \approx -2\, {}^{(n-1)}\TT{S}^{k,l}_{i} \Theta(\theta_{in} > \theta_{ij}) \approx -2\, {}^{(n-1)}\TT{S}^{k,l}_{j}\Theta(\theta_{jn} > \theta_{ij}). \label{eq:ringlimit}
\end{align}
Importantly, the quasi-collinear limit is consistent with the collinear limit of the azimuthally averaged ring, Eq.~\eqref{eq:averagedapproxring}, for all values of $\theta_{ij}$ and $\theta_{in}$. We can also consider the quasi-collinear limits $\theta_{ik} \ll 1$ and $\theta_{il} \ll 1$:
\begin{align}
   & \, {}^{(n-1)}\TT{R}^{i,j}_{k,l} \bigg|_{n_i \rightarrow n_k} \approx 2\, {}^{(n-1)}\TT{S}^{j,l}_{i}\Theta(\theta_{in} > \theta_{ik}) \approx 2\, {}^{(n-1)}\TT{S}^{j,l}_{k} \Theta(\theta_{kn} > \theta_{ik}), \\
   &\, {}^{(n-1)}\TT{R}^{i,j}_{k,l} \bigg|_{n_i \rightarrow n_l} \approx 0. \label{eq:ringlimit2}
\end{align}

The same level of care must also be taken with strings in the quasi-collinear limit. We can write 
\begin{align}
    \, {}^{(n-1)}\TT{S}^{j,k}_{i} = \, {}^{(n-1)}\TT{S}^{j,k}_{i} (\Theta(\theta_{in} < \theta_{ij}) + \Theta(\theta_{in} > \theta_{ij})),
\end{align}
and note that
\begin{align}
     (\omega_{ij}(q_{n}) + \omega_{ik}(q_{n}) - \omega_{jk}(q_{n}))\Theta(\theta_{in} > \theta_{ij})\big|_{n_{j}\rightarrow n_{i}} \approx 0,
\end{align}
as it has no collinear pole and vanishes as $n_{j} \rightarrow n_{i}$. Consequently, 
\begin{align}
    \, {}^{(n-1)}\TT{S}^{j,k}_{i}\big|_{n_{j}\rightarrow n_{i}} \approx \, {}^{(n-1)}\TT{S}^{j,k}_{i} \Theta(\theta_{in} < \theta_{ij}). \label{eq:stringlimit}
\end{align} 

Already, we can see that coherence is encoded in the behaviour of strings and rings even without azimuthal averaging. Just as in the averaged case, in the quasi-collinear limit rings and strings split into small-angle terms which probe the colour structure of the density matrix in the neighbourhood of a collinear pole and wide-angle terms that cannot probe the colour structure in the neighbourhood of a collinear pole. In what follows, we will use the identities in Eqs.~\eqref{eq:ringlimit}, \eqref{eq:ringlimit2}, and \eqref{eq:stringlimit} to derive coherence properties of amplitude density matrices without azimuthal averaging.

\subsection{Revisiting two-jet coherence without azimuthal averaging}
\label{sec:revisit}

In the previous discussion we demonstrated that strings and rings display coherence-like properties in the quasi-collinear limit without azimuthal averaging. In the following sections we look to exploit these properties to study coherence in processes without an azimuthal symmetry (namely three-jet topologies or double soft emission), which are usually considered beyond the scope of the traditional angular-ordered approach of Eq.~\eqref{eq:2-jetcoherenceaveraged}. 

First, let us return to Eq.~\eqref{eq:2-jetcoherenceaveraged} and derive it without azimuthal averaging. To do so, we must compute $\v{A}_{4}$ in the limit that particles $1$ and $2$ define two hard jets, particle $3$ is collinear to particle $1$, and particle $4$ is soft. $\v{A}_{4}$ is given in terms of strings in Eq.~\eqref{eq:A4}. We can use \eqref{eq:stringlimit} and colour conservation to find the $n_{1} \rightarrow n_{3}$ quasi-collinear limit:
\begin{align}
   \v{A}_{4}  \big|_{n_1 \rightarrow n_3} &\approx \frac{2\As}{\pi} \bigg[ \, {}^{(3)}\TT{S}^{2,3}_{1} \Theta(\theta_{14} < \theta_{13})  [1\bcdot 1]  \nonumber  \\ & \qqquad + \, {}^{(3)}\TT{S}^{1,3}_{2} \Theta(\theta_{(1+3)4} > \theta_{13}) [2\bcdot 2] + \, {}^{(3)}\TT{S}^{1,2}_{3} \Theta (\theta_{34} < \theta_{13}) [3\bcdot 3]  \bigg], \nonumber \\
   &\approx \frac{2\As}{\pi} \bigg[ \, {}^{(3)}\TT{S}^{2,3}_{1} (\theta_{14} < \theta_{13})  [1\bcdot 1] + P_{(1+3)2} \Theta(\theta_{(1+3)4} > \theta_{13}) [1+3\bcdot 1+3] \nonumber \\& \qqquad + P_{2(1+3)} [2\bcdot 2] + \, {}^{(3)}\TT{S}^{1,2}_{3} (\theta_{34} < \theta_{13}) [3\bcdot 3]  \bigg]. \label{eq:A4notaveraged}
\end{align}
This has the same angular-ordered interpretation as Eq.~\eqref{eq:2-jetcoherenceaveraged}, shown diagrammatically in Figure \ref{fig:intuitionforcoherence}. However, by replacing the azimuthally averaged kernels with their unaveraged counterparts, Eq.~\eqref{eq:A4notaveraged} retains the azimuthal correlations which are lost in Eq.~\eqref{eq:2-jetcoherenceaveraged}. The azimuthal correlations in Eq.~\eqref{eq:A4notaveraged} actually vanish in the quasi-collinear, two-jet limit due to the axial symmetry. This is not the case when we come to consider the three-jet limit in the next section. The use of the functions $P_{ij}$ (defined in Eq.~\eqref{eq:Pdef}) is not unique. $P_{ij}$ could be replaced by any function $g_{ij}$ that is finite as $\theta_{4j} \rightarrow 0$ and which satisfies $\omega^{(4)}_{ij} = g_{ij} + g_{ji}$. It is easy to check that, when particles $1$ and $2$ are back-to-back, averaging Eq.~\eqref{eq:A4notaveraged} returns Eq.~\eqref{eq:2-jetcoherenceaveraged}.

We can generalise Eq.~\eqref{eq:A4notaveraged} to an all-orders treatment by replacing particles $1$, $2$ and $3$ with bunches of exactly collinear partons defining jets 1, 2 and 3, i.e. we let the invariant mass of any sub-clusters of particles inside each jet go to zero, $q_{i}^{2}/(2q_{i}\cdot q_{j}) \approx 0$ where $q_{i}$ ($q_{j}$) is the total momentum of the jet $i=1,2,3$ ($j=1,2,3$ and $i\neq j$). The amplitude density matrix for a soft emission from the three perfectly collimated jets is
\begin{align}
   \v{A}_{3 \, \TT{jet}+(g)} = - \frac{\As}{\pi} \left( \omega^{(4)}_{12}(|1\bcdot 2|_{3} + |2\bcdot 1|_{3})+ \omega^{(4)}_{13}(|1\bcdot 3|_{3} + |3\bcdot 1|_{3}) + \omega^{(4)}_{23}(|2\bcdot 3|_{3} + |3\bcdot 2|_{3})\right),
\end{align}
where $|i\bcdot j|_{3} = \v{T}_{i} \lkl M_{3 \, \TT{jet}}  \> \< M_{3 \, \TT{jet}}  \rkl \v{T}^{\dagger}_{j}$ given $\v{T}_{i} = \sum_{m \in \, \TT{jet} \, i} \v{T}_{m}$, and where $q_{i} = \sum_{m \in \, \TT{jet} \, i} q_{m}$ for $i \in \{1,2,3\}$. Here $\lkl M_{3 \, \TT{jet}}  \> = \lkl M_{n-1}  \> |_{\TT{3\, jet}}$ where $|_{\TT{3\, jet}}$ requires that each of the $n-1$ particles are exactly collinear to one of the 3 jet directions. We use that $\omega^{(n)}_{ij} = \, {}^{(n-1)}\TT{S}^{j,k}_{i} +  \, {}^{(n-1)}\TT{S}^{i,k}_{j}$ to write 
\begin{align}
   \v{A}_{3 \, \TT{jet}+(g)} = \frac{2\As}{\pi} \big(\, {}^{(3)}\TT{S}^{2,3}_{1}|1\bcdot 1|_{3} + \, {}^{(3)}\TT{S}^{1,3}_{2}|2\bcdot 2|_{3}  + \, {}^{(3)}\TT{S}^{1,2}_{3}|3\bcdot 3|_{3} \big). \label{eq:exact3jet}
\end{align}
As expected, we see that each string multiplies a diagonal colour structure only dependent on the combined colour charge of the jet. As per $\v{A}_{4}$, we can apply Eq.~\eqref{eq:stringlimit} in the limit that jets 1 and 3 are quasi-collinear (which we label $1||3$) to see two-jet coherence emerge. The outcome is
\begin{align}
\v{A}_{3 \, \TT{jet} + (g)}\big|_{1||3} \approx \frac{2\As}{\pi} \bigg[& \, {}^{(3)}\TT{S}^{2,3}_{1} (\theta_{14} < \theta_{13})  |1\bcdot 1|_{3} + \, {}^{(3)}\TT{S}^{1,3}_{2}  |2\bcdot 2|_{3} + \, {}^{(3)}\TT{S}^{1,2}_{3} (\theta_{34} < \theta_{13}) |3\bcdot 3|_{3}  \bigg] . \label{eq:2-jetazimthalcorrlations}
\end{align}
In the limit that jets $1$ and $2$ are back-to-back, this azimuthally averages to the familiar result \cite{CATANI1991635}.

\subsection{Coherence and angular-ordering in the three-jet limit}
\label{sec:coherencegenerally}

We will now use strings to extend coherence to the three-jet limit, including azimuthal correlations.

\begin{figure}[t]
	\centering
	\subfigure[A term from $\lkl M_{3 \, \TT{jet}+(g)}\>$ where the soft gluon is emitted from jet 2.\label{fig:3jet}] {\includegraphics[width=0.35\textwidth]{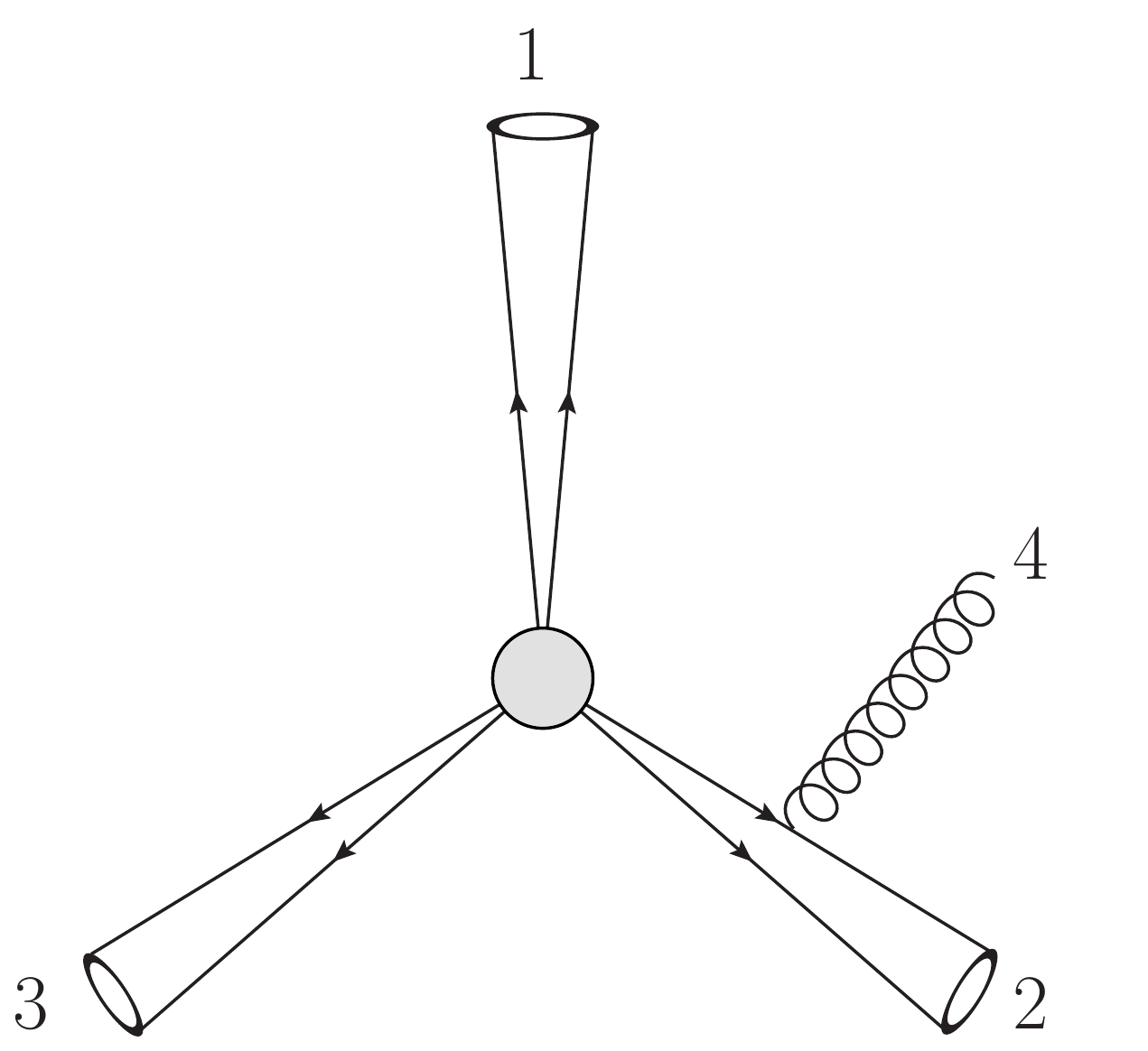}} 	~~~~~~~
	\subfigure[A term from $\lkl M_{4 \, \TT{jet}+(g)}\>\big|_{1||4}$ where the soft gluon is emitted from jet 2.\label{fig:4jet}] {\includegraphics[width=0.35\textwidth]{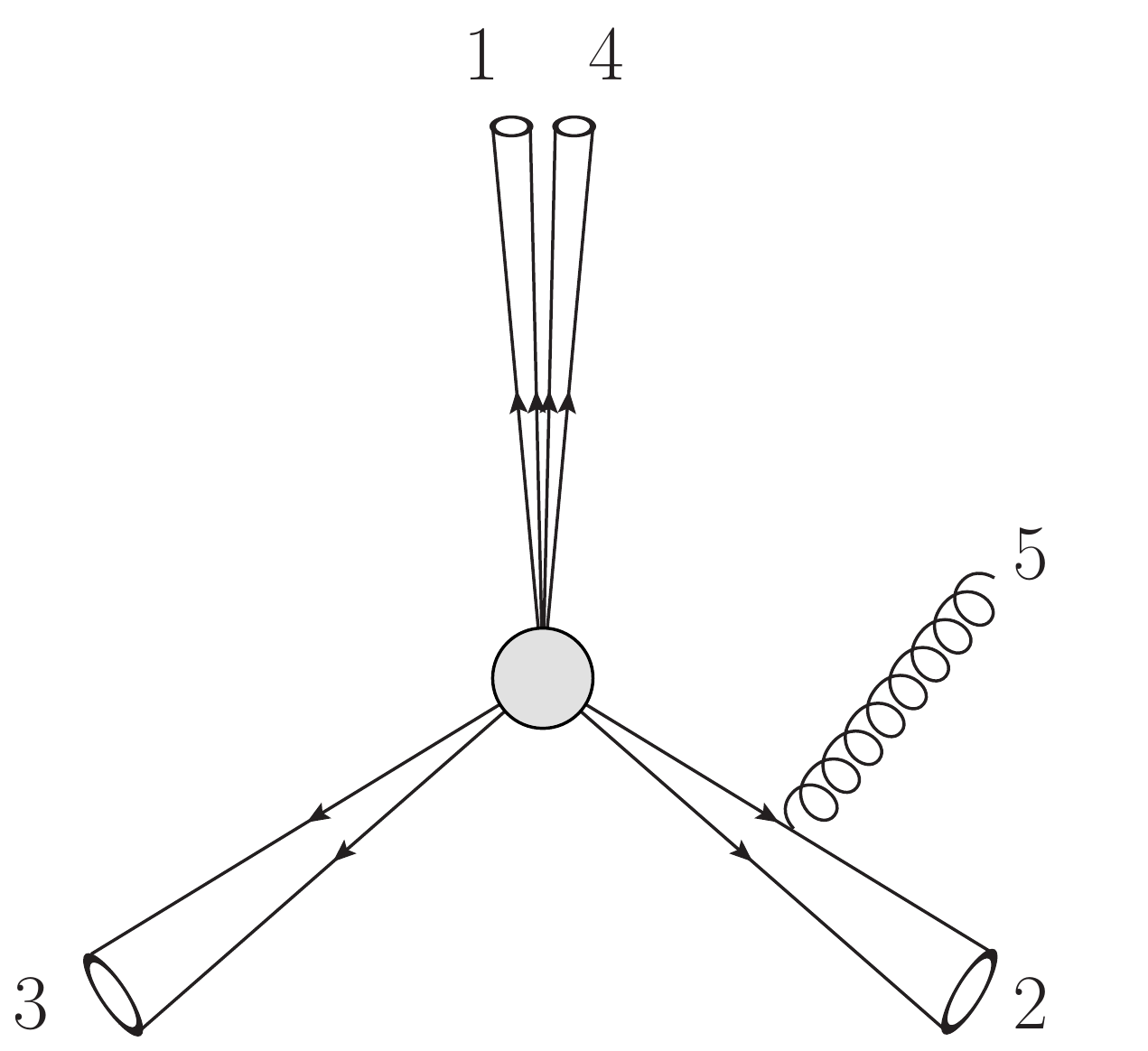}}
	\caption{Contributions to $\lkl M_{3 \, \TT{jet}+(g)}\>$ and $\lkl M_{4 \, \TT{jet}+(g)}\>\big|_{1||4}$. Cones represent jets of collinear particles and the grey circle represents the hard process. The difference between $\lkl M_{4 \, \TT{jet}+(g)}\>\big|_{1||4}$ and $\lkl M_{3 \, \TT{jet}+(g)}\>$ is interpreted as being due to substructure in jet 1.}
\end{figure}

Consider a soft gluon emitted from a three-jet system (jets $1,2,3$), Figure \ref{fig:3jet}. In Appendix \ref{sec:fullderivation} we study this system by analysing Eq.~\eqref{eq:amplitudeinringsandstrings} in the limit that particles $1,2,3$ define the three jet directions (jets $1,2,3$ respectively) and that particles $4,\dots, n-1$ are quasi-collinear with a strong hierarchy in angles between the particle and their parent jet. In this section we will present a simpler approach to computing the all-orders coherence properties of a three-jet system. We will assume that every particle in a jet is an exactly collinear parton and compute the amplitude density matrix. Then we will consider perturbing the three-jet system by splitting one of the jets into two subjets (each built of exactly collinear particles), see Figure \ref{fig:4jet}. This simpler approach captures all the essential features of coherence in the three-jet limit and can be applied iteratively to reconstruct the picture of coherent radiation one finds from the `brute-force' analysis in Appendix \ref{sec:fullderivation}.

The amplitude density matrix of interest is $\v{A}_{4 \, \TT{jet} + (g)} = \lkl M_{4 \, \TT{jet} + (g)}  \> \< M_{4 \, \TT{jet} + (g)}  \rkl$ where $q_{5}$ is a soft gluon, and where jets 1 and 4 become quasi-collinear so as to form a single jet. The difference between the $\v{A}_{4 \, \TT{jet} + (g)}\big|_{1||4}$ and $\v{A}_{3 \, \TT{jet}+(g)}$ is to be considered a perturbation to $\v{A}_{3 \, \TT{jet}+(g)}$ due to substructure in jet $1$ that vanishes when $\theta_{14} = 0$ (given $\theta_{15} \neq 0$). Using the same basis as in Eq.~\eqref{eq:amplitudeinringsandstrings}:
\begin{align}
\v{A}_{4 \, \TT{jet} + (g)} = & \frac{2\As}{\pi} \bigg[\, {}^{(4)}\TT{S}^{2,3}_{1}  |1\bcdot 1|_{4} + \, {}^{(4)}\TT{S}^{1,3}_{2}  |2\bcdot 2|_{4} + \, {}^{(4)}\TT{S}^{1,2}_{3}  |3\bcdot 3|_{4} + \, {}^{(4)}\TT{S}^{1,2}_{4}  |4\bcdot 4|_{4} \bigg] \nonumber \\
& - \frac{\As}{2\pi} \bigg[ \, {}^{(4)}\TT{R}^{3,4}_{1,2}(|1\bcdot 4|_{4} + |3\bcdot 4|_{4} - |2\bcdot 4|_{4} + |4\bcdot 1|_{4} + |4\bcdot 3|_{4} - |4\bcdot 2|_{4}) \nonumber \\
& \qquad  + \, {}^{(4)}\TT{R}^{4,3}_{1,2}(|2\bcdot 4|_{4} + |3\bcdot 4|_{4} - |1\bcdot 4|_{4} + |4\bcdot 2|_{4} + |4\bcdot 3|_{4} - |4\bcdot 1|_{4}) \bigg].
\end{align}
As per the three-jet case, $\lkl M_{4 \, \TT{jet}}  \> \equiv \lkl M_{n-1}  \> |_{4 \TT{\, jet}}$ where $|_{4 \TT{\, jet}}$ indicates that each of the $n-1$ particles are exactly collinear to one of the four jet axes, so that $q_{i}^{2}/(2q_{i}\cdot q_{j}) \approx 0$ when $i\neq j$\footnote{In what follows $\lkl M_{p \, \TT{jet}}  \> \equiv \lkl M_{n-1}  \> |_{p \TT{\, jet}}$ for $|_{p \TT{\, jet}}$ requires that each of the $n-1$ particles are exactly collinear to one of the $p$ jet axes. Following this definition, four-jet colour charge structures, $|i\bcdot j|_{4}$, are naturally generalised to $p$-jet colour structures, $|i\bcdot j|_{p}$.}.

We now enforce the quasi-collinear ($1||4$) limit on the rings, setting 
$$\, {}^{(4)}\TT{R}^{3,4}_{1,2} \approx 0,$$ 
and 
$$\, {}^{(4)}\TT{R}^{4,3}_{1,2} \approx 2\, {}^{(4)}\TT{S}^{2,3}_{1} \Theta(\theta_{15} > \theta_{14}).$$ 
After this
\begin{align}
&\v{A}_{4 \, \TT{jet} + (g)}\big|_{1||4} \approx  \frac{2\As}{\pi} \bigg[\, {}^{(4)}\TT{S}^{2,3}_{1} \Theta(\theta_{15} < \theta_{14})  |1\bcdot 1|_{4} + \, {}^{(4)}\TT{S}^{1,3}_{2}  |2\bcdot 2|_{4} + \, {}^{(4)}\TT{S}^{1,2}_{3}  |3\bcdot 3|_{4} + \, {}^{(4)}\TT{S}^{1,2}_{4}  |4\bcdot 4|_{4} \bigg] \nonumber \\
& - \frac{\As}{\pi} \, {}^{(4)}\TT{S}^{2,3}_{1} \Theta(\theta_{15} > \theta_{14})(|2\bcdot 4|_{4} + |3\bcdot 4|_{4} - |1\bcdot 4|_{4} + |4\bcdot 2|_{4} + |4\bcdot 3|_{4} - |4\bcdot 1|_{4} - 2|1\bcdot 1|_{4}).
\end{align}
Colour conservation can be used on the final bracket so that 
\begin{align}
\v{A}_{4 \, \TT{jet} + (g)}\big|_{1||4} \approx \frac{2\As}{\pi} \bigg[& \, {}^{(4)}\TT{S}^{2,3}_{1} \Theta(\theta_{15} < \theta_{14})  |1\bcdot 1|_{4} + \, {}^{(4)}\TT{S}^{1,3}_{2}  |2\bcdot 2|_{4} + \, {}^{(4)}\TT{S}^{1,2}_{3}  |3\bcdot 3|_{4} + \nonumber \\ & {}^{(4)}\TT{S}^{1,2}_{4}  |4\bcdot 4|_{4} 
 + \, {}^{(4)}\TT{S}^{2,3}_{1} \Theta(\theta_{15} > \theta_{14})|1+4\bcdot 1+4|_{4}\bigg] . \label{eq:3-jet1storder}
\end{align}
In the quasi-collinear limit
\begin{align}
    \, {}^{(4)}\TT{S}^{1,2}_{4}\big|_{n_{4}\rightarrow n_{1}} \approx \, {}^{(4)}\TT{S}^{1,2}_{4} \Theta(\theta_{45} < \theta_{14}) \approx \, {}^{(4)}\TT{S}^{2,3}_{4} \Theta(\theta_{45} < \theta_{14}),
\end{align} 
and consequently the difference between $\v{A}_{4 \, \TT{jet} + (g)}\big|_{1||4}$ and $\v{A}_{3 \, \TT{jet}+(g)}$ is confined to an angular cone around jet $1$ of size $\theta_{14}$.
Eq.~\eqref{eq:3-jet1storder} is a key result. It corresponds to a three-jet coherence pattern, just as in the two-jet case, and azimuthal averaging was not used; only the quasi-collinear approximation.
It is straightforward enough to iterate this to generate an angular-ordered cascade of emissions corresponding to multiple subjets that are quasi-collinear with one of the three primary jets. This is done in Appendix \ref{sec:fullderivation}.

To highlight some interesting features of the three-jet limit, we can once again consider the azimuthally averaged strings: 
\begin{align}
    \, {}^{(3)}\TT{S}^{j,k}_{i}\frac{\td \Omega_{n}}{4\pi} \approx \frac{1}{2}\left(P^{[i]}_{ij} + P^{[i]}_{ik}\right) +  \frac{1}{2}\left(\tilde{P}^{[j]}_{ji}(\theta_{jk}) - \tilde{P}^{[j]}_{jk}(\theta_{ij})\right) + \frac{1}{2}\left(\tilde{P}^{[k]}_{ki}(\theta_{jk}) - \tilde{P}^{[k]}_{kj}(\theta_{ik})\right).
\end{align}
When $\theta_{ij},\theta_{ik}>\theta_{jk}$ the averaged string generates a purely positive ``semi-classical" radiation pattern, the limiting behaviour $\theta_{ij},\theta_{ik}\gg \theta_{jk}$ capturing two-jet coherence. When $\theta_{jk}$ grows larger than either $\theta_{ij}$ or $\theta_{ik}$, negative terms enter. When $\theta_{ij}=\theta_{ik}=\theta_{jk}$ negative contributions exactly vanish, as does much of the ``semi-classical" radiation pattern in the direction anti-collinear to $i$. Specifically, in the limit that the jets lie in a plane and are well-spaced (so that $\theta_{ij}\approx\theta_{ik}\approx\theta_{jk}\approx 2\pi/3$), we find that
\begin{align}
    \, {}^{(3)}\TT{S}^{j,k}_{i}\frac{\td \Omega_{n}}{4\pi}\bigg|_{\theta_{ij}\approx\theta_{ik}\approx\theta_{jk}\approx 2\pi/3} \approx \frac{1}{2}\left(P^{[i]}_{ij} + P^{[i]}_{ik}\right) \approx P^{[i]}_{ij}.
\end{align}
This is the naive two-jet, angular-ordered emission kernel one would use if we simply ignored interference with jet $k$. For the azimuthally averaged amplitude density matrix, in the planar, well-spaced limit of the three-jets, we find
\begin{align}
\v{A}_{4 \, \TT{jet} + (g)}\big|_{1||4} \frac{\td \Omega_{n}}{4\pi} \approx \frac{2\As}{\pi} \bigg[& P^{[1]}_{14}  |1\bcdot 1|_{4} + \tfrac{1}{2}\left(P^{[2]}_{21} + P^{[2]}_{23}\right) |2\bcdot 2|_{4} + \tfrac{1}{2}\left(P^{[3]}_{31} + P^{[3]}_{32}\right)  |3\bcdot 3|_{4} \nonumber \\
& + P^{[4]}_{41}  |4\bcdot 4|_{4} + \tfrac{1}{2}\left(\tilde{P}^{[1]}_{12} (\theta_{14}) + \tilde{P}^{[1]}_{13}(\theta_{14})\right)|1+4\bcdot 1+4|_{4} \bigg] . \label{eq:3-jet1storderaveraged}
\end{align}
This particular limit of the three-jet coherence pattern is  equivalent to what one would find by naively applying two-jet angular ordering. 

At present, relatively few precision analytic resummations of three (or more)-jet observables have been performed \cite{Banfi:2000si,Arpino:2019ozn,Stewart:2010tn} (requiring the exponentiation of soft single logarithms). The prototypical three-jet observables \cite{Banfi:2000si,Arpino:2019ozn} resum logarithms that measure the aplanarity of a process. These logarithms diverge when the jets lie exactly in a plane, analogous to thrust logarithms diverging for back-to-back, pencil-like jets\footnote{This analogy can be made concrete. One can define the planar limit as $T \sim T_{M} \gg T_{m}$ where $T$ is the thrust, $T_{M}$ is the thrust-major and $T_{m}$ is the thrust-minor \cite{Banfi:2000si}. Logarithms in the aplanarity can be related to logarithms of the small quantity $T_{m}/T_{M}$.}. In these resummations, matrix elements for the emission of a soft gluon from three coloured particles are needed to compute wide-angle, soft logarithms at NLL accuracy, and strings unavoidably appear. For example, they appear in the correlated emission matrix elements in Eq.~(2.23) in \cite{Banfi:2000si}, which contains a string that accounts for coherent radiation from a gluon jet. It is interesting to note that, for the $q \bar{q} g$ topology, \cite{Banfi:2000si} argue that the string-like structure persists beyond one loop (see Eq.~(2.26) in \cite{Banfi:2000si}). In \cite{Arpino:2019ozn}, soft physics dressing three planar jets is exponentiated, at NNLL accuracy, using webs \cite{Falcioni:2014pka,Gardi:2010rn}. This goes beyond the accuracy of the present work. The link between the analysis we present and the approach taken in \cite{Arpino:2019ozn} is not immediately clear and worthy of further investigation.

Finally, a word on full colour coherence with more than three jets. Below Eq.~\eqref{eq:key} we summarised how one can use the three-jet result derived above to achieve LC accuracy with more than three jets, however one could be more ambitious and attempt to repeat the analysis above for an arbitrary number of jets. The key outcome of our three-jet analysis is that a soft gluon emission can be entirely accounted for using angular ordering, a single kinematic function (the string), and Casimir colour factors. This is the exact structure one needs to construct a typical cross-section level parton shower. Whilst the above approach can be repeated for any multiplicity of jets, the amplitude density matrix will not reduce to being diagonal and so the computation of a soft gluon emission with full-colour accuracy cannot be achieved by only considering colour Casimirs. Similarly, with more than three-jets the emission of a soft gluon will always depend on terms containing rings that do not reduce to single strings (beyond the trivial ${}^{(n)}\TT{R}^{jl}_{ik} = {}^{(n)}\TT{S}^{ij}_{l} - {}^{(n)}\TT{S}^{ik}_{l}$). Thus, the treatment of more than three-jets will necessarily remain akin to how $2\rightarrow 2$ processes have historically been handled in the resummation literature \cite{Kidonakis:1998nf,Oderda:1998en,Oderda:1999kr,Banfi:2004yd,Banfi:2010xy}; requiring a degree of amplitude-level full-colour computation to include coherent soft gluons separately from factorised collinear physics.

\subsubsection{Hard-collinear physics, the CMW coupling and unitarity}
\label{sec:extensions}

In this section we will provide a short discussion on the inclusion of hard-collinear physics, the running coupling and virtual corrections. We will base our discussions around the simpler, all-orders result in Eq.~\eqref{eq:3-jet1storder} with the knowledge that our discussions can be applied directly to the `brute-force' result in Eq.~\eqref{eq:3-jetallorder} without much work.

We can extend Eq.~\eqref{eq:3-jet1storder} to include hard-collinear physics by noting that $E^{2}_{n}  \,{}^{(n-1)}\TT{S}^{y,z}_{x}$ has no energy dependence (ignoring momentum conservation, on which more at the end of this section)\footnote{We are ignoring $g\rightarrow q \bar{q}$ transitions which are without a soft pole and so are beyond the scope of this paper.}. Hard-collinear physics can therefore be included as a perturbation to the energy measure:
\begin{align}
    \frac{\td E_{q}}{E_{q}} \mapsto \frac{\td E_{q}}{E_{q}} (1+\text{hard-collinear}) \equiv P(z) \, \td z ,
\end{align}
where $z$ is an energy fraction and $P$ is an appropriate collinear splitting function, identified by its colour factor. Consequently, Eq.~\eqref{eq:3-jet1storder} becomes
\begin{align}
&\v{A}_{4 \, \TT{jet} + (g)} \frac{\td^{3-2\epsilon}\vec{q}_{5}}{2 E_{5}} \Bigg|_{1||4} \approx \frac{\As}{\pi} \bigg[ \, {}^{(4)}\TT{S}^{2,3}_{1} P_{1\rightarrow 1g} \left(\tfrac{E_{5}}{E_{1}},\epsilon\right) \frac{E_{5}^{2}\td E_{5}}{E_{1}} \Theta(\theta_{15} < \theta_{14})  |1\bcdot 1|_{4} \nonumber \\
&+ \, {}^{(4)}\TT{S}^{1,3}_{2} P_{2\rightarrow 2g}\left(\tfrac{E_{5}}{E_{2}},\epsilon\right) \frac{E_{5}^{2}\td E_{5}}{E_{2}} |2\bcdot 2|_{4}  + \, {}^{(4)}\TT{S}^{1,2}_{3} P_{3\rightarrow 3g}\left(\tfrac{E_{5}}{E_{3}},\epsilon\right) \frac{E_{5}^{2}\td E_{5}}{E_{3}} |3\bcdot 3|_{4} \nonumber \\
& + \, {}^{(4)}\TT{S}^{2,3}_{4} P_{4\rightarrow 4g}\left(\tfrac{E_{5}}{E_{4}},\epsilon\right) \frac{E_{5}^{2}\td E_{5}}{E_{1}} \Theta(\theta_{45} < \theta_{14}) |4\bcdot 4|_{4} \nonumber \\
& + \, {}^{(4)}\TT{S}^{2,3}_{1} P_{(1+4)\rightarrow (1+4)g}\left(\tfrac{E_{5}}{E_{1}+E_{4}},\epsilon\right) \frac{E_{5}^{2}\td E_{5}}{E_{1}+E_{4}} \Theta(\theta_{15} > \theta_{14})|1+4\bcdot 1+4|_{4}\bigg] \td^{2-2\epsilon}\Omega_{5}. \label{eq:3-jet1storderhard}
\end{align}
where when $i$ is a quark jet, $\Tr|i\bcdot i|_{4} = C_{F}|M_{4 \, \TT{jet}}|^{2}\big|_{1||4}$,
\begin{align}
     P_{i\rightarrow ig} \left(1-z,\epsilon\right) = (1-z)^{-\epsilon} \left( \frac{1+z^{2}}{1-z} - \epsilon (1-z) \right),
\end{align}
and when $i$ is a gluon jet, $\Tr|i\bcdot i|_{4} = C_{A}|M_{4 \, \TT{jet}}|^{2}\big|_{1||4}$,
\begin{align}
     P_{i\rightarrow ig} \left(1-z,\epsilon\right) = 2 (1-z)^{-\epsilon} \left( \frac{z}{1-z} + \frac{1-z}{z} + z(1-z) \right).
\end{align}
These are the usual unregularised, collinear splitting functions into which we have absorbed the factor $(1-z)^{-\epsilon}$ from the measure.

The CMW running coupling \cite{CATANI1991635} is a modification to the $\overline{\TT{MS}}$ coupling which incorporates the possibility that an emitted soft gluon may further branch into unresolved soft and collinear partons. The unresolved branchings generate universal terms, starting at $\sim \As^{2} \ln^{2} E_{g}\theta_{g}$, which suppress the emission probability of a soft gluon. The unresolved branchings preserve the overall colour structure of the density matrix. Therefore, the CMW coupling can be included in our previous discussions with the simple substitution $\As \mapsto \As^{\TT{CMW}}$. More precisely, in Eq.~\eqref{eq:3-jet1storderhard} we should replace
\begin{align}
    \As |i \bcdot i| \mapsto \As^{\TT{CMW}}|i\bcdot i| = \As(E^{2}_{5}\theta^{2}_{i5})\left(1 + \frac{\As(E^{2}_{5}\theta^{2}_{i5})}{2\pi}\left(\left(\frac{67}{18} - \frac{\pi^{2}}{6}\right)C_{A} -\frac{5}{9}n_{f} \right)\right)|i\bcdot i|,
\end{align} 
where $\As(\mu^{2})$ is the $\overline{\TT{MS}}$ coupling. 

Finally, the colour diagonal structure of Eq.~\eqref{eq:3-jet1storderhard} means that the $\mathcal{O}(\As)$ virtual correction to $\v{A}_{4 \, \TT{jet}}\big|_{1||4}$ (computed in the same collinear approximation as $\v{A}_{4 \, \TT{jet} + (g)}\big|_{1||4}$) can be found by a simple application of cross-section-level unitarity and exponentiated to give a Sudakov factor
\begin{align}
&\ln V = - \int_{\TT{P.S.}} \frac{\Tr \v{A}_{4 \, \TT{jet} + (g)}}{\Tr \v{A}_{4 \, \TT{jet}}} \Bigg|_{1||4} \frac{\td^{3-2\epsilon}\vec{q}_{5}}{2 E_{5}}. \label{eq:Sudakov}
\end{align}

Including hard-collinear physics requires one to consider the conservation of momentum. Longitudinal momentum conservation can be achieved using a local DGLAP prescription \cite{APSplitting}. However, the complete conservation of momentum in directions transverse and anti-collinear to the collinear direction is also important \cite{Herwig_shower,Bewick:2019rbu,Hamilton:2020rcu}, particularly for parton showers. The correct mechanism for momentum conservation in the three-jet limit is not clear from our analysis, which has been based around the recoil-free, soft limit. It is also not immediately clear whether the complications of momentum conservation in the three-jet limit will amount to a negligible effect. We consider a broad study into three-jet momentum conservation beyond the scope of this paper and direct the reader to the current literature for discussions on the conservation of momentum in angular-ordered showers \cite{Bewick:2019rbu} and the three-jet limit \cite{Luisoni:2020efy}. That said, we will give an overview of how the problem could be tackled and the issues one might run into in the following section. 

\subsubsection{Angular-ordered parton showers}
\label{sec:partonshower}

\begin{figure}[b]
	\centering
	\subfigure[\label{fig:subfiga}]{\includegraphics[width=0.4\textwidth]{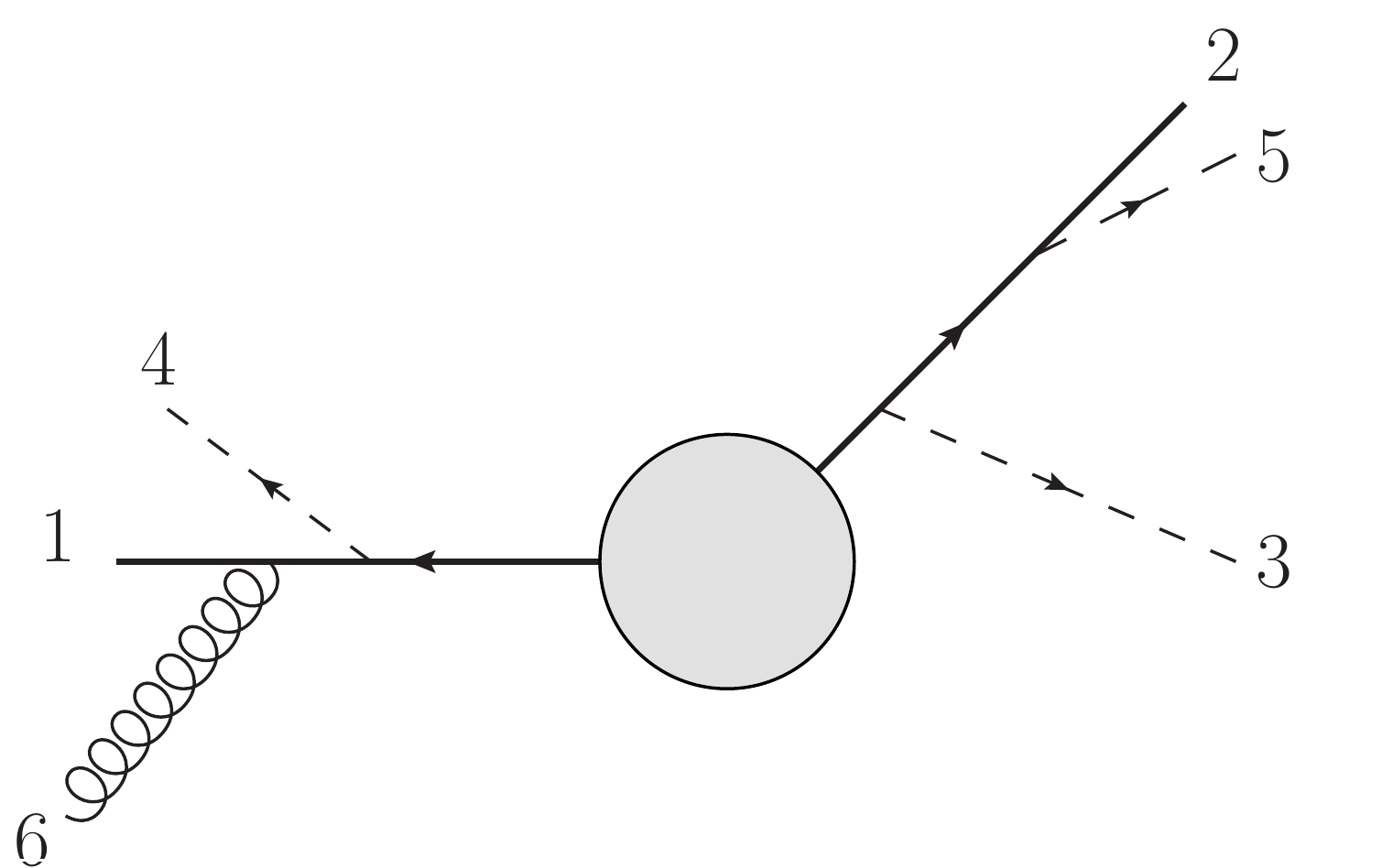}} ~~~~
	\subfigure[\label{fig:subfigb}]{\includegraphics[width=0.4\textwidth]{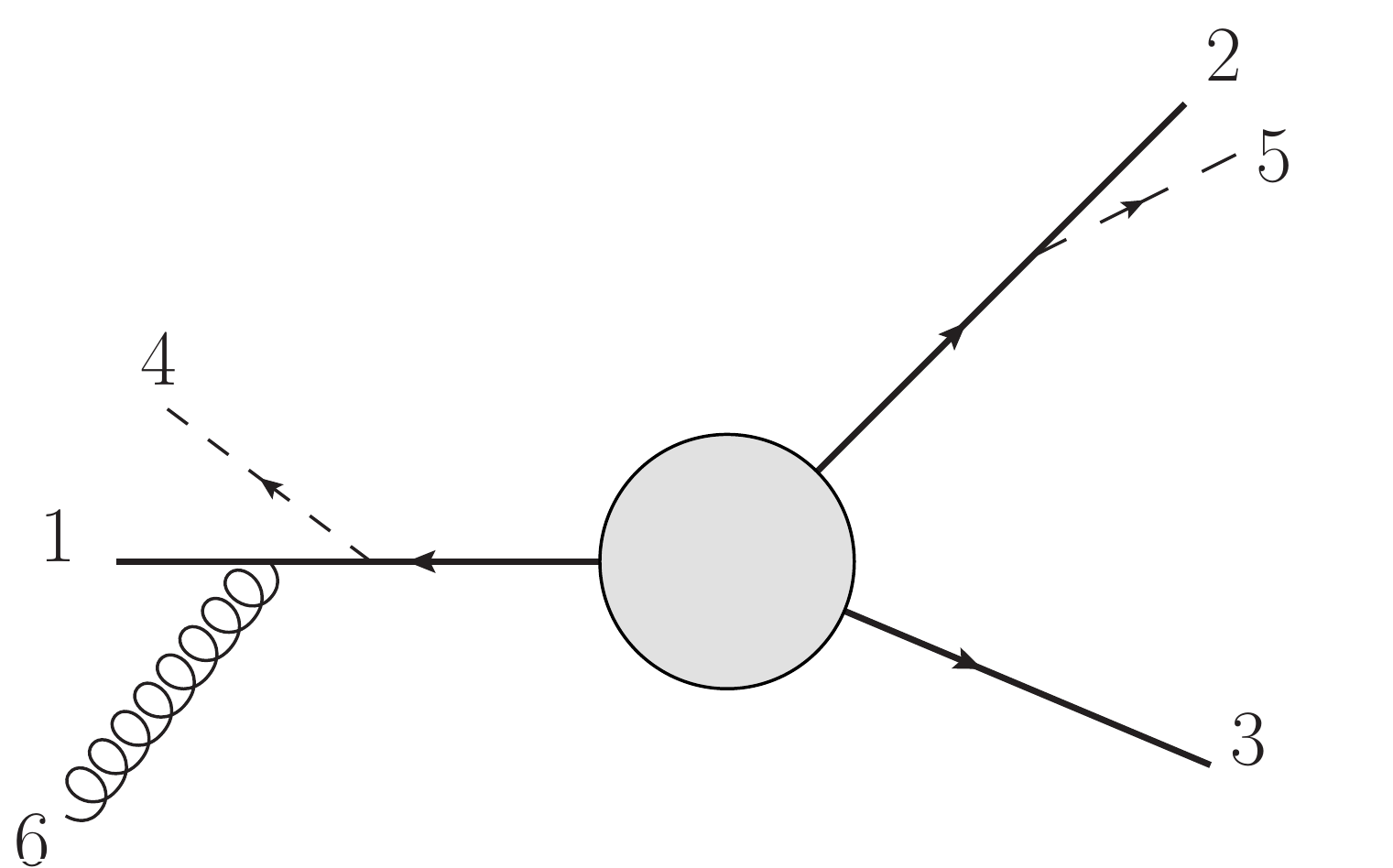}}
	\caption{Diagrams illustrating substitution \eqref{eq:substitution} in the case of six coloured particles. These particular contributions correspond to a term $\sim {}^{(5)}\TT{S}^{2,3}_{1}\, E^{2}_{6} \,\Theta(\theta_{16} < \theta_{25})\, \td(\cos \theta_{16} )$. Solid lines are coloured particles from the hard process and dashed lines are showered partons, ordered in angle. Diagram (a) has two hard partons (a two-jet process), whilst (b) has three hard partons (a three-jet process). The emission probability is the same for both, modulo momentum conservation and recoil against parton $3$, which is not necessarily hard in (a). For a discussion on momentum conservation see the text surrounding Eq.~\eqref{eq:2-jetlikemap}. This is in contrast to an angular-ordered shower, where (a) and (b) are handled differently from each other \cite{Herwig_shower}.}
	\label{fig:partonshower}
\end{figure}

It therefore follows that an angular-ordered shower can be systematically improved to include some of the azimuthal correlations\footnote{Specifically, the correlation between the three primary jets and the widest-angle soft gluon will be correct.}
from full-colour, soft physics by making a simple substitution in the emission kernels, for particle multiplicities $\geq 3$,
\begin{align}
    \frac{\td(\cos \theta_{in} )\td \phi_{n}}{1 - \cos \theta_{in}} \Theta(\theta_{in} < \theta)  \mapsto  \, {}^{(n-1)}\TT{S}^{j,k}_{i} \, E^{2}_{n} \, \Theta(\theta_{in} < \theta) \, \td(\cos \theta_{in} )\td \phi_{n}, \label{eq:substitution}
\end{align}
where $\theta$ is the previous angular scale in the shower ($\theta=\pi$ for the first emission so that the complete wide-angle-soft gluon spectrum is recovered) and $i$ is the parent of the emitted gluon $n$. The definitions of $j$ and $k$ are contextual, see Figures \ref{fig:subfiga} and \ref{fig:subfigb} which illustrate the substitution for two- and three-jet processes. For a three-coloured-particle hard-process $q_{j,k}$ are the momenta of the two jets from which $i$ does not originate. For a two-coloured-particle hard-process, the situation is identical to the three jet case, where the third jet is the first (widest angle) emission. 

For hard processes with $>3$ coloured particles, colour-flows from the hard process can be used to reduce a process to a superposition of individual two- or three- jet systems with leading colour accuracy (as is usual in existing angular ordered showers \cite{Herwig_shower}).
In this case, a shower using \eqref{eq:substitution} would compute squared matrix elements (in angular-ordered regions of the $n$-parton phase-space) which correspond to the first term in a perturbative expansion of Eq.~\eqref{eq:amplitudeinringsandstrings} around $\, ^{(n-1)}R^{ij}_{kl} = 0$. This expansion is not controlled by a small parameter and so does not generate a formal improvement in accuracy. However, an expansion around $\, ^{(n-1)}R^{ij}_{kl} = 0$ may prove to be more powerful than we can currently justify. It has been observed in the literature that, for some observables, subleading $\Nc$ terms which correspond to when $\, ^{(n-1)}R^{ij}_{kl} \neq 0$ appear to vanish \cite{Forshaw:2020wrq,Hatta:2020wre} upon resummation. However, this effect is not yet understood.

Using \eqref{eq:substitution} will lead to some ambiguity in how a parton shower should conserve momentum in three-jet processes. It is typical in a parton shower to assign a ``parent'' particle/jet \cite{Herwig_shower,Herwig_dipole_shower,Pythia8,Dasgupta:2020fwr,Forshaw:2020wrq}, whose momentum is adjusted to ensure momentum conservation, i.e. applying this to Eq.~\eqref{eq:3-jet1storderhard}, longitudinal momentum should be conserved between $q_{5}$ and jet 1 in the term which is proportional to $|1 \bcdot 1|$ and has $1/\theta_{15}$ a collinear pole. For instance, if we parameterise $q_{5}$ as
\begin{align}
    q_{5} = \alpha q_{1} + k_{\bot} + \mathcal{O}(k^{2}_{\bot}),
\end{align}
where $q_{1} \cdot k_{\bot}=0$, to conserve momentum it is necessary to transform jet 1's momentum after emitting particle $q_{5}$ as
\begin{align}
    q_{1} \mapsto (1-\alpha) q_{1} + \mathcal{O}(k_{\bot}),
\end{align}
conserving momentum longitudinal to the jet. This reasoning is applicable to terms in our three-jet coherence pattern where the phase-space of $q_{5}$ is restricted to an angular cone around a given jet (which can be chosen to be the parent). For these terms, momentum conservation can be treated in the usual way \cite{Bewick:2019rbu,Forshaw:2020wrq,Dasgupta:2020fwr}. However, the terms that allow $q_{5}$ to be emitted at an arbitrarily wide angle are less clear (such as those in lines 2 and 4 of Eq.~\eqref{eq:3-jet1storderhard}). For example, let us look at the term proportional to $\, {}^{(4)}\TT{S}^{1,3}_{2}  |2 \bcdot 2|$ in Eq.~\eqref{eq:3-jet1storderhard}. In the limit that jets 1 and 3 become collinear to each other, after azimuthal averaging, Eq.~\eqref{eq:3-jet1storderhard} reduces to a two-jet coherence pattern akin to that in Eq.~\eqref{eq:2-jetcoherenceaveraged}. In this limit, $\, {}^{(4)}\TT{S}^{1,3}_{2}  |2 \bcdot 2|$ generates both the term where gluon $5$ is emitted from jet 2 but also the term where gluon $5$ is emitted coherently from the combined jet $1+3$. Therefore, consistency with the two-jet coherence limit requires that we can express $q_{5}$, in the limits $n_{1} \cdot n_{3} \rightarrow 0$ and $\theta_{15}>\theta_{13},\theta_{14}$, as
\begin{align}
    q_{5} = \alpha q_{2}+ \beta(q_{1}+q_{3}+q_{4}) + k_{\bot},
\end{align}
where $q_{1} \cdot k_{\bot}=(q_{1}+q_{3}+q_{4})\cdot k_{\bot}=0$ and where after emitting particle $5$ the jet momenta should be transformed as
\begin{align}
    q_{1} \mapsto (1-\beta) q_{1} + \mathcal{O}(k_{\bot}), \qquad q_{2} \mapsto (1-\alpha) q_{2} + \mathcal{O}(k_{\bot}), \nonumber \\  q_{3} \mapsto (1-\beta) q_{3} + \mathcal{O}(k_{\bot}), \qquad q_{4} \mapsto (1-\beta) q_{4} + \mathcal{O}(k_{\bot}). \label{eq:2-jetlikemap}
\end{align}
How this extends beyond the two-jet limit is unclear. We can consider using Eq.~\eqref{eq:2-jetlikemap} outside the two-jet limit, however this has the undesirable feature that when particle $5$ becomes collinear to jet $1$ large amounts of momentum are conserved against jets 2 and 3 (and vice versa). This is against the intuition guiding the parton shower approach where momentum should be mostly conserved between particle $5$ and the jet with which it is collinear. Alternatively, we could handle this term by expressing $q_{5}$ in a three-jet decomposition,
\begin{align}
    q_{5} = \alpha q_{2}+ \beta q_{1}+\gamma(q_{3}+q_{4}) + k_{\bot},
\end{align}
where $q_{1} \cdot k_{\bot}=q_{2} \cdot k_{\bot}=(q_{3}+q_{4})\cdot k_{\bot}=0$ and conserving momentum via 
\begin{align}
    q_{1} \mapsto (1-\beta) q_{1} + \mathcal{O}(k_{\bot}), \qquad q_{2} \mapsto (1-\alpha) q_{2} + \mathcal{O}(k_{\bot}), \nonumber \\  q_{3} \mapsto (1-\gamma) q_{3} + \mathcal{O}(k_{\bot}), \qquad q_{4} \mapsto (1-\gamma) q_{4} + \mathcal{O}(k_{\bot}). \label{eq:3-jetlikemap}
\end{align}
This methodology preserves the parton shower intuition. However, this decomposition is ill-defined in the two-jet limit where $k_{\bot}$ goes from having $1$ degree of freedom to $2$, and where $\beta$ and $\gamma$ become degenerate. Furthermore, the decomposition allows ``unpleasant'' phase-space boundaries since, with the usual parton shower boundary conditions (either $\alpha,\beta,\gamma <1$ or $0<\alpha +\beta +\gamma <1$), it is possible for one of $\alpha, \beta, \gamma$ to become negative\footnote{Simply requiring $\alpha,\beta,\gamma >0$ does not fix the problem since it prevents the population of the low $k_{\bot}$ phase-space region. This problem is visualised by considering a frame where all three jets lie in the same hemisphere. It is impossible to populate the opposing hemisphere if $\alpha,\beta,\gamma >0$ is required. This problem is not present in the two-jet decomposition as the on-shell requirement, $q_{5}^{2}=0$, is sufficient to ensure $\alpha,\beta>0$.}. At its heart, the problem is that strings at wide angles describe genuine $3 \rightarrow 4$ particle transitions where none of the particles can be singled out as a spectator. A work around could be to express each string in terms of antenna functions and use one of the established mechanisms for conserving momentum using antenna functions (such as those used in modern dipole showers \cite{Dasgupta:2020fwr,Forshaw:2020wrq}). However, this too could cause problems as strings involve both positive and negative antenna functions, and negative weights in parton showers can inhibit numerical convergence, though there exist proposals to mitigate the effect of weighted parton shower algorithms, e.g. \cite{Platzer:2011dq,Olsson:2019wvr}. 

\section{Conclusions}

The colour structures and collinear poles occurring in QCD amplitudes of soft gluons lead naturally to the notion of strings, which have collinear poles, and rings, which do not. Colour off-diagonal elements in amplitude density matrices are without collinear poles and so can be efficiently expressed in terms of rings. Rings emerge naturally in the colour-flow basis, where their presence facilitates flips in colour flows which give rise to $\Nc$ suppressed terms. Expressing sub-amplitudes for colour flows in terms of rings helps to improve the statistical convergence of the \texttt{CVolver} amplitude-evolution code.

Expressing squared amplitudes in terms of rings and strings separates the wide-angle contributions from terms with collinear poles. Therefore QCD coherence, which leads to the factorisation of terms with collinear poles from wide-angle terms, is naturally described in terms of rings and strings. This description has allowed us to extend the notion of angular ordering to include three-jet processes where we are able to include azimuthal correlations. This is accomplished with a simple replacement of the usual emission kernels:
\begin{align}
    \frac{\td(\cos \theta_{in} )\td \phi_{n}}{1 - \cos \theta_{in}} \Theta(\theta_{in} < \theta)  \mapsto  \, {}^{(n-1)}\TT{S}^{j,k}_{i} \, E^{2}_{n} \, \Theta(\theta_{in} < \theta) \, \td(\cos \theta_{in} )\td \phi_{n}. \label{eq:subconclusions}
\end{align}

\section*{Acknowledgements}

This work is supported in part by the GLUODYNAMICS project funded by the ``P2IO LabEx (ANR-10-LABX-0038)'' in the framework ``Investissements d’Avenir'' (ANR-11-IDEX-0003-01) managed by the Agence Nationale de la Recherche (ANR), France. JH thanks the UK Science and Technology Facilities Council for the award of a postgraduate studentship. SP
is grateful to the Erwin Schr\"odinger Institute Vienna for
hospitality and support while significant parts of this work have been
achieved within the Research in Teams programme 
``Amplitude Level Evolution I: Initial State Evolution.'' (RIT0421). 

\appendix

\section{Supplementary material on rings and strings}
\label{app:rings}

In this appendix we discuss some representations of rings and strings and more of their properties. We first introduce two new representations of the antenna functions. Firstly, the pole representation:
\begin{align}
    \omega_{ij}(q_{n}) \equiv \begin{pmatrix} \delta_{1,i} + \delta_{1,j} \\
     \delta_{2,i} + \delta_{2,j} \\
     \vdots \\
      \delta_{n-1,i} + \delta_{n-1,j}
    \end{pmatrix} = \vec{\beta}_{ij}, 
\end{align}
i.e. for a 4-particle amplitude of which particle 4 is soft
\begin{align}
    \omega_{12}(q_{4}) \equiv \begin{pmatrix} 1 \\
     1 \\
     0 
    \end{pmatrix},  \qquad 
    \omega_{13}(q_{4}) \equiv \begin{pmatrix} 1 \\
     0 \\
     1 
    \end{pmatrix},  \qquad 
    \omega_{23}(q_{4}) \equiv \begin{pmatrix} 0 \\
     1 \\
     1 
    \end{pmatrix}.
\end{align}
Secondly, the combination representation:
\begin{align}
    \omega_{ij}(q_{n}) \equiv \begin{pmatrix} \delta_{\{i,j\},\sigma_{1}} \\
     \delta_{\{i,j\},\sigma_{2}} \\
     \vdots \\
      \delta_{\{i,j\},\sigma_{_{(n-1)}C_{2}}}
    \end{pmatrix}  = \vec{\gamma}_{ij}, 
\end{align}
where $\sigma_{i}$ is the $i$th element in a set of unique combinations of two numbers from $1$ to $n-1$. Once again, for a four-particle amplitude of which particle 4 is soft, a combination basis representation is
\begin{align}
    \omega_{12}(q_{4})\equiv \begin{pmatrix} 1 \\
     0 \\
     0 
    \end{pmatrix},  \qquad 
    \omega_{13}(q_{4}) \equiv \begin{pmatrix} 0 \\
     1 \\
     0 
    \end{pmatrix},  \qquad 
    \omega_{23}(q_{4}) \equiv \begin{pmatrix} 0 \\
     0 \\
     1 
    \end{pmatrix}.
\end{align}
For $i,j \in \{1,\dots,n-1\}$ there are $ _{(n-1)}C_{2}$ antenna functions. The dimension of the combination representation is $ _{(n-1)}C_{2}$ and the antenna functions (trivially) form an orthonormal basis spanning this representation. However, the pole basis has dimension $n-1$. There are $_{(n-1)}C_{2} - (n-1)$ linearly independent combinations of antenna functions which are degenerate in the pole representation (i.e. in this representation they differ by the null vector). A linear combination of antenna functions without collinear poles is represented by a null vector in the pole representation. Thus, the $_{(n-1)}C_{2} - (n-1)$ linear combinations of antenna functions form a basis over rings. Non-zero vectors in the pole representation are strings. Vectors that form a basis over the pole representation are what we refer to as `basis strings'. However, note that `basis strings' form a basis over collinear poles not over all possible strings, since adding a null vector (equivalent to a ring) to a string returns another string. Also note that in the combination representation, if a ring has the representation $\vec{\gamma}^{\TT{R}}$ then $\sum_{i}(\vec{\gamma}^{\TT{R}})_{i} = 0$.

For amplitude density matrices $\v{A}_{n}$ with $n<5$ there are no rings, as the smallest ring requires four particles, and so there is no off-diagonal colour. At $n=5$ there are two linearly independent rings $\vec{\gamma}^{\TT{R}^{1}}$ and $\vec{\gamma}^{\TT{R}^{2}}$, i.e.
\begin{align}
\TT{For} \; 
\vec{\gamma}_{ij}  = \begin{pmatrix} \delta_{\{i,j\},\{1,2\}} \\
     \delta_{\{i,j\},\{1,3\}} \\
     \delta_{\{i,j\},\{1,4\}} \\
     \delta_{\{i,j\},\{2,3\}} \\
     \delta_{\{i,j\},\{2,4\}} \\
     \delta_{\{i,j\},\{3,4\}}
    \end{pmatrix}, \qquad 
    \vec{\gamma}^{\TT{R}^{1}} = \begin{pmatrix} 1\\
     -1 \\
     0 \\
     0 \\
     -1 \\
     1
    \end{pmatrix}, \qquad \TT{and} \quad \vec{\gamma}^{\TT{R}^{2}} = \begin{pmatrix} 1\\
     0 \\
     -1 \\
     -1 \\
     0 \\
     1
    \end{pmatrix}.
\end{align}
In terms of antenna functions these are $\, {}^{(4)}\TT{R}^{1,2}_{3,4}$ and $\, {}^{(4)}\TT{R}^{1,2}_{4,3}$ respectively. All other soft functions corresponding to four hard particles and no collinear poles can be expressed as a linear combination of these two rings, e.g. 
$$ 
2\omega_{12} - \omega_{13} - \omega_{14} + 2\omega_{34} - \omega_{23} - \omega_{24} = \, {}^{(4)}\TT{R}^{1,2}_{3,4} + \, {}^{(4)}\TT{R}^{1,2}_{4,3}.
$$
Let us look at the symmetries of $\, {}^{(4)}\TT{R}^{i,j}_{k,l}$. Under the exchange of indices it has the following symmetries and anti-symmetries
\begin{align}
    \, {}^{(4)}\TT{R}^{i,j}_{k,l} = \, {}^{(4)}\TT{R}^{j,i}_{l,k} \, , \qquad \, {}^{(4)}\TT{R}^{i,j}_{k,l} = -\, {}^{(4)}\TT{R}^{l,j}_{k,i} \, , \qquad \, {}^{(4)}\TT{R}^{i,j}_{k,l} = -\, {}^{(4)}\TT{R}^{i,k}_{j,l} \, .
\end{align}
$\, {}^{(4)}\TT{R}^{i,j}_{k,l}$ can be viewed as a tensor with 24 degrees of freedom (found by considering all permutations of the indices). Each of these symmetries/anti-symmetries constrains a further 6 degrees of freedom. Finally,
\begin{align}
\, {}^{(4)}\TT{R}^{i,j}_{k,l} = \, {}^{(4)}\TT{R}^{i,j}_{l,k} - \, {}^{(4)}\TT{R}^{i,k}_{l,j}, \label{eq:4particleconstraint}
\end{align}
which, taken in conjunction with the symmetries/anti-symmetries of $\, {}^{(4)}\TT{R}^{i,j}_{k,l}$, constrains a further four degrees of freedom. Thus $\, {}^{(4)}\TT{R}^{i,j}_{k,l}$ only has two linearly independent elements.

Now consider $n=6$, there are $_{5}C_{2}-5=5$ linearly independent rings in this case. Once again we can count the degrees of freedom that $\, {}^{(5)}\TT{R}^{i,j}_{k,l}$ has, this time for $i,j,k,l \in \{1,\dots,5\}$. With the previously given four-particle constraints, the degrees of freedom for $\, {}^{(n-1)}\TT{R}^{i,j}_{k,l}$ is equal to number of degrees of freedom for $\, {}^{(4)}\TT{R}^{i,j}_{k,l}$ multiplied by $_{n-1}C_{4}$. That is $\, {}^{(5)}\TT{R}^{i,j}_{k,l}$ is constrained to 10 degrees of freedom. However, $\, {}^{(5)}\TT{R}^{i,j}_{k,l}$ also has another five-particle constraint:
\begin{align}
    \, {}^{(5)}\TT{R}^{i,j}_{k,l} = \, {}^{(5)}\TT{R}^{i,j}_{k,m}\, -  {}^{(5)}\TT{R}^{l,j}_{k,m}. \label{eq:5particleconstraint}
\end{align}
This removes a further five degrees of freedom. Thus $\, {}^{(5)}\TT{R}^{i,j}_{k,l}$ has five independent degrees of freedom which can be used to form a basis over rings, e.g. $\, {}^{(5)}\TT{R}^{1,2}_{3,4}$, $\, {}^{(5)}\TT{R}^{1,2}_{4,3}$, $\, {}^{(5)}\TT{R}^{1,2}_{3,5}$, $\, {}^{(5)}\TT{R}^{1,2}_{5,3}$, and $\, {}^{(5)}\TT{R}^{1,2}_{4,5}$. The linear independence of these functions is easily seen in the combination representation:
\begin{align}
\vec{\gamma}_{ij}  = \begin{pmatrix} 
    \delta_{\{i,j\},\{1,2\}} \\
     \delta_{\{i,j\},\{1,3\}} \\
     \delta_{\{i,j\},\{1,4\}} \\
     \delta_{\{i,j\},\{1,5\}} \\
     \delta_{\{i,j\},\{2,3\}} \\
     \delta_{\{i,j\},\{2,4\}} \\
     \delta_{\{i,j\},\{2,5\}} \\
     \delta_{\{i,j\},\{3,4\}} \\
     \delta_{\{i,j\},\{3,5\}} \\
     \delta_{\{i,j\},\{4,5\}} 
    \end{pmatrix}, \; 
    \vec{\gamma}^{\TT{R}^{1}} = \begin{pmatrix} 
    1\\
     -1 \\
     0 \\
     0 \\
     0 \\
     -1 \\
     0 \\
     1 \\
     0 \\
     0 
    \end{pmatrix}, \; \vec{\gamma}^{\TT{R}^{2}} = \begin{pmatrix} 
    1\\
     0 \\
     -1 \\
     0 \\
     -1 \\
     0 \\
     0 \\
     1 \\
     0 \\
     0 \\
    \end{pmatrix}, \; \vec{\gamma}^{\TT{R}^{3}} = \begin{pmatrix} 
    1\\
     -1 \\
     0 \\
     0 \\
     0 \\
     0 \\
     -1 \\
     0 \\
     1 \\
     0 \\
    \end{pmatrix}\; \vec{\gamma}^{\TT{R}^{4}} = \begin{pmatrix} 
    1\\
     0 \\
     0 \\
     -1 \\
     -1 \\
     0 \\
     0 \\
     0 \\
     1 \\
     0 \\
    \end{pmatrix}\; \vec{\gamma}^{\TT{R}^{5}} = \begin{pmatrix} 
    1\\
     0 \\
     -1 \\
     0 \\
     0 \\
     0 \\
     -1 \\
     0 \\
     0 \\
     1 \\
    \end{pmatrix}.
\end{align}

For arbitrary $n$, $\, {}^{(n-1)}\TT{R}^{i,j}_{k,l}$ has $_{(n-1)}C_{2} - (n-1)$ degrees of freedom. This is most easily shown by considering the cutting rule, Eq.~\eqref{eq:cutrule}, and the repeated edge rule, Eq.~\eqref{eq:repeatedgerule}. The logic is as follows. We know from our previous arguments that there are $_{(n-1)}C_{2} - (n-1)$ unique ring polygons. Rules Eq.~\eqref{eq:cutrule} and Eq.~\eqref{eq:repeatedgerule} ensure that a polygon can always be used to reduced to linear combinations of independent squares. Thus there must be $_{(n-1)}C_{2} - (n-1)$ independent squares. A complete set of independent squares necessarily forms a basis over which we can express the tensor $\, {}^{(n-1)}\TT{R}^{i,j}_{k,l}$. Thus, it must be the case that $\, {}^{(n-1)}\TT{R}^{i,j}_{k,l}$ also has $_{(n-1)}C_{2} - (n-1)$ degrees of freedom. It is therefore trivial that a basis for the degrees of freedom in $\, {}^{(n-1)}\TT{R}^{i,j}_{k,l}$ necessarily also forms a basis over all rings. One such basis is given in Eq.~\eqref{eq:basis}.

We can find a functional form of the rings, which is explicitly without collinear poles, by applying Catani-Seymour (CS) dipole factorisation \cite{Catani:1996vz} and then using partial fractions. After CS factorisation a basis ring can be written as
\begin{align}
   \, {}^{(n-1)}\TT{R}^{i,j}_{k,l} = \frac{s_{ij}}{s_{i} \, s_{i+j}} - \frac{s_{ik}}{s_{i} \, s_{i+k}} + \frac{s_{ij}}{s_{j} \, s_{i+j}} - \frac{s_{jl}}{s_{j} \, s_{j+l}} + \frac{s_{kl}}{s_{k} \, s_{k+l}} - \frac{s_{ik}}{s_{k} \, s_{i+k}} + \frac{s_{kl}}{s_{l} \, s_{k+l}} - \frac{s_{jl}}{s_{l} \, s_{j+l}},
\end{align}
where $s_{ij} = q_{i}\cdot q_{j}$, $s_{i} = q_{i}\cdot q_{n}$, and $s_{i+j} = q_{n}\cdot (q_{i}+q_{j})$. Partial fractions can be used so that
\begin{align}
   \frac{s_{ij}}{s_{i} \, s_{i+j}} - \frac{s_{ik}}{s_{i} \, s_{i+k}} = \frac{s_{ik}-s_{ij}(s_{ik}-s_{ij}-s_{jk})}{s_{i+j}\, s_{ij}(s_{ik}-s_{ij}-s_{jk})} - \frac{s_{ij}-s_{ik}(s_{ij}-s_{ik}-s_{jk})}{s_{i+k} \, s_{ik}(s_{ij}-s_{ik}-s_{jk})},
\end{align}
which has no non-integrable collinear divergences. Therefore
\begin{align}
   \, {}^{(n-1)}\TT{R}^{i,j}_{k,l} =& \frac{s_{ik}-s_{ij}(s_{ik}-s_{ij}-s_{jk})}{s_{i+j}\, s_{ij}(s_{ik}-s_{ij}-s_{jk})} - \frac{s_{ij}-s_{ik}(s_{ij}-s_{ik}-s_{jk})}{s_{i+k} \, s_{ik}(s_{ij}-s_{ik}-s_{jk})} \nonumber \\ 
   &+ \frac{s_{jl}-s_{ji}(s_{jl}-s_{ji}-s_{il})}{s_{j+i}\, s_{ji}(s_{jl}-s_{ji}-s_{il})} - \frac{s_{ji}-s_{jl}(s_{ji}-s_{jl}-s_{il})}{s_{j+l} \, s_{jl}(s_{ji}-s_{jl}-s_{il})} \nonumber \\  
   &+ \frac{s_{ki}-s_{kl}(s_{ki}-s_{kl}-s_{li})}{s_{k+l}\, s_{kl}(s_{ki}-s_{kl}-s_{li})} - \frac{s_{kl}-s_{ki}(s_{kl}-s_{ki}-s_{li})}{s_{k+i} \, s_{ki}(s_{kl}-s_{ki}-s_{li})} \nonumber \\  
   &+ \frac{s_{lj}-s_{lk}(s_{lj}-s_{lk}-s_{kj})}{s_{l+k}\, s_{lk}(s_{lj}-s_{lk}-s_{kj})} - \frac{s_{lk}-s_{lj}(s_{lk}-s_{lj}-s_{kj})}{s_{l+j} \, s_{lj}(s_{lk}-s_{lj}-s_{kj})},
\end{align}
which also has no non-integrable collinear divergences.

Now we turn our attention to applying the same techniques to basis strings. This is comparatively easy: there are many simple sets of basis strings we could construct. The simplest approach is to use the pole representation, i.e. for $n+1$ particles
\begin{align}
\vec{\beta}^{\, \TT{S}}_{i} = \begin{pmatrix} \delta_{1,i} \\
     \delta_{2,i}  \\
     \vdots \\
      \delta_{n,i}
    \end{pmatrix}.
\end{align}
In terms of antenna functions, $\vec{\beta}^{\, \TT{S}}_{i}$ is equivalent to
\begin{align}
\vec{\beta}^{\, \TT{S}}_{i} \equiv \, {}^{(n)}\TT{S}^{j,k}_{i} = \frac{1}{2}\left(\omega_{ij} + \omega_{ik} - \omega_{jk} \right),
\end{align}
for arbitrary $j,k$. For example, one could pick $j= (i + 1) \, (\TT{mod} \, n)$ and $k= (i - 1) \, (\TT{mod} \, n)$. An antenna function can be recovered from these strings by $\, {}^{(n)}\TT{S}^{j,k}_{i} + \, {}^{(n)}\TT{S}^{i,k}_{j} = \omega_{ij}(q_{n+1})$.\footnote{Note that a complete basis of $n$ strings can only ever be used to reconstruct at most $n$ antenna functions, not the complete set of $_{n}C_{2}$ antenna functions. For $n>3$, reconstructing all $_{n}C_{2}$ antenna functions requires both rings and strings.}

\section{Linearly independent strings and the large $\Nc$ expansion}

\label{app:LargeNc}

In this appendix we will provide more details on how strings capture colour diagonal physics. Leading colour (LC) soft physics can be computed in its entirety from the dipole formalism, where we exchange
\begin{align}
    \TT{LC}:\frac{\As}{\pi} \sum_{i \neq j} \omega_{ij}(q_{n})\v{T}_{i} \lkl M_{n-1}  \> \< M_{n-1}  \rkl \v{T}^{\dagger}_{j} \; \mapsto \; \frac{\As \Nc}{2\pi}\sum_{\sigma} \sum_{(i,j) \, \TT{c.c.} \, \sigma} \omega_{ij}(q_{n}) |M^{(\sigma)}_{n-1}|^{2}, \label{eq:B.1}
\end{align}
where $\sigma$ is a given $n-1$ particle planar (leading $\Nc$) colour flow and the sum over ``$(i,j) \, \TT{c.c.} \, \sigma$'' means that we sum over the dipoles colour connected in $\sigma$. LC physics is colour diagonal and only positive antenna functions contribute. Therefore, there are $n-1$ collinear poles in Eq.~\eqref{eq:B.1}. There are $n-1 - \tfrac{1}{2}N_{q}$ colour connected dipoles in $\sigma$ where $N_{q}$ is the number of quarks in the process (each gluon is connected to two other particles whilst a quark to only one). It is tempting to think that we can build a basis for strings that fully encapsulates the LC limit and consequently ensures that rings always appear subleading in colour. To this end, we can construct a set of antenna functions $\{\,{}^{(n-1)}\TT{S}^{(\sigma)}_{i}\} = \{\omega^{(n)}_{ij} | (i,j) \, \TT{c.c.} \, \sigma\} \cup \{\omega^{(n)}\}$ which we can try to use as a string basis, where $\{\omega^{(n)}\}$ is a set of $\tfrac{1}{2}N_{q}$ antenna functions whose dipoles are not colour connected in $\sigma$. These are needed so that the dimension of $\{\,{}^{(n-1)}\TT{S}^{(\sigma)}_{i}\}$ matches the number of collinear poles. By construction, functions from the set $\{\,{}^{(n-1)}\TT{S}^{(\sigma)}_{i}\}$ can be used to express the LC matrix elements. However, constructing a string basis from $\{\,{}^{(n-1)}\TT{S}^{(\sigma)}_{i}\}$ has a problem: for each colour singlet pair of gluons in $\sigma$ from which an even number of colour lines originate $\{ \,{}^{(n-1)}\TT{S}^{(\sigma)}_{i} \}$ contains a ring. Consequently the set of strings built from $\{ \,{}^{(n-1)}\TT{S}^{(\sigma)}_{i} \}$ is not a string basis (the strings are not linearly independent). For example, consider a purely gluonic matrix element $M^{\sigma}_{4}$ where in $\sigma$ the following dipoles are colour connected
\begin{align}
    (1,2) \; (2,3) \; (3,4) \; (4,1),
\end{align}
therefore
\begin{align}
    \{\,{}^{(4)}\TT{S}^{\sigma}_{i}\} = \{ \omega^{(5)}_{12}, \omega^{(5)}_{23}, \omega^{(5)}_{34}, \omega^{(5)}_{14} \}.
\end{align}
However,
\begin{align}
    \Pole \left(\omega^{(5)}_{12} -\omega^{(5)}_{23} +\omega^{(5)}_{34}\right) = \Pole\left( \omega^{(5)}_{14} \right).
\end{align}
Consequently, $\{\,{}^{(n-1)}\TT{S}^{(\sigma)}_{i}\}$ contains a ring and cannot be used to form a basis over the collinear poles. This is a general feature, the set $\{\,{}^{(n-1)}\TT{S}^{(\sigma)}_{i}\}$ forms a string basis, ensuring rings are subleading in colour, for any amplitude that does not contain a colour singlet pair of gluons. However, for each colour singlet pair of gluons, with an even number of daughter gluons, one ring function must be present at leading colour. For instance $e^+e^- \rightarrow q \bar{q} (ggg)$ does not suffer a LC singlet gluon pair and so can be expressed in terms of the string basis sets $\{\,{}^{(3)}\TT{S}^{(\sigma_{0})}_{i}\}$ and $\{\,{}^{(4)}\TT{S}^{(\sigma_{1,2})}_{i}\}$:
\begin{align}
    \frac{\td \Sigma^{e^+e^- \rightarrow q \bar{q} (ggg)}_{3}}{\td \Phi_{3}} = -2 C_{\TT{F}} \Nc^{2} \omega^{(3)}_{12} \bigg( & \, {}^{(3)}\TT{S}^{\sigma_{0}}_{1}\left(\,{}^{(4)}\TT{S}^{\sigma_{1}}_{1}+\,{}^{(4)}\TT{S}^{\sigma_{1}}_{2}+\,{}^{(4)}\TT{S}^{\sigma_{1}}_{3}\right) \nonumber \\ &+ \, {}^{(3)}\TT{S}^{\sigma_{0}}_{2}\left(\,{}^{(4)}\TT{S}^{\sigma_{2}}_{1}+\,{}^{(4)}\TT{S}^{\sigma_{2}}_{2}+\,{}^{(4)}\TT{S}^{\sigma_{2}}_{3}\right)\bigg) + \mathcal{O}(\Nc),
\end{align}
for $e^+e^- \rightarrow q_1 \bar{q}_2 (g_3 g_4 g_5)$ with colour connected pairs $(1,2)(2,3)$ in $\sigma_{0}$, $(1,4)(2,3)(3,4)$ in $\sigma_{1}$ and $(1,3)(2,4)(3,4)$ in $\sigma_{2}$.

\section{The three-jet limit of $\v{A}_{n}$}
\label{sec:fullderivation}

\begin{figure}[t]
	\centering
	\subfigure[] {\includegraphics[width=0.3\textwidth]{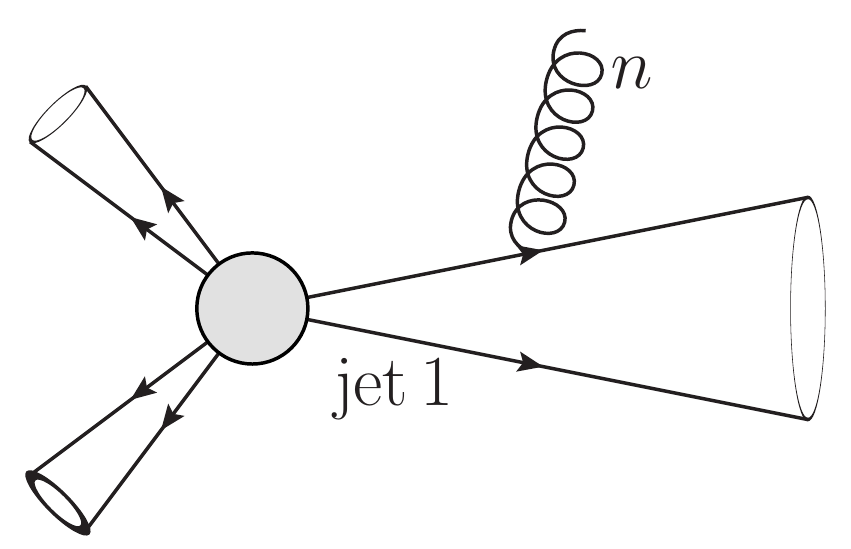}} \\
	\subfigure[] {\includegraphics[width=0.7\textwidth]{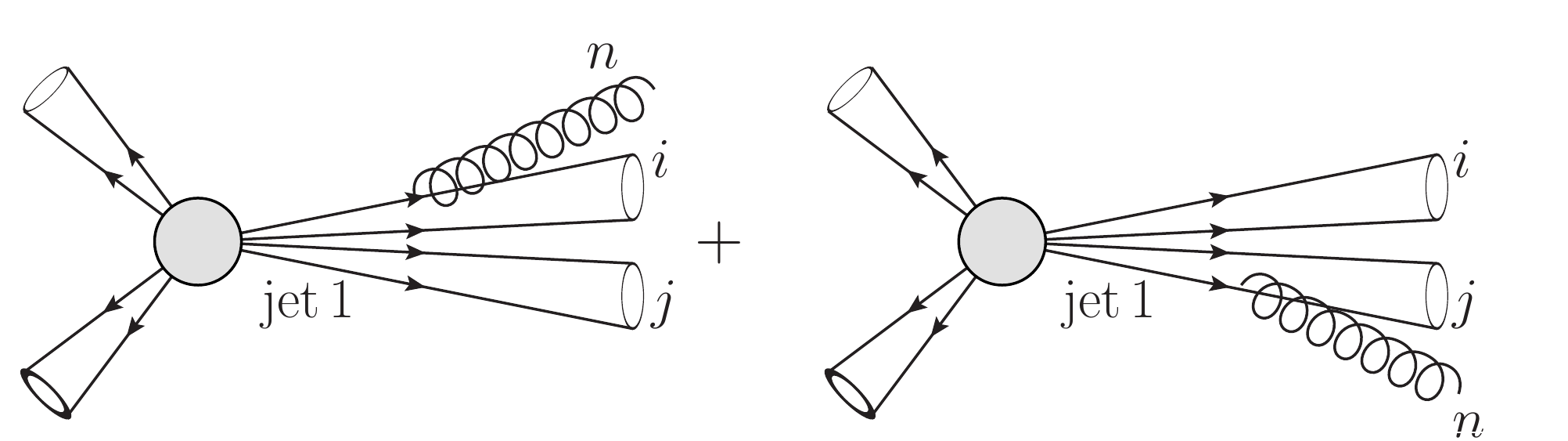}}
	\subfigure[] {\includegraphics[width=\textwidth]{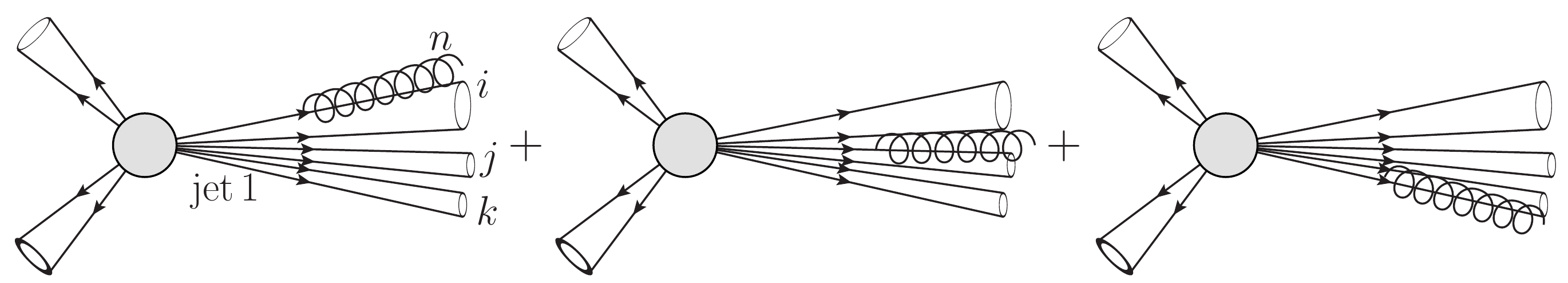}}
	\caption{Figures (a), (b) and (c) show a gluon $n$ emitted from jet $1$ (in a three-jet process) with collinear substructure at progressively finer angular scales. Cones represent jets of collinear particles and the grey circle represents the hard process.  \label{fig:3jetallorder}}
\end{figure}

In Section \ref{sec:coherencegenerally} we viewed the difference between $\v{A}_{4 \, \TT{jet} + (g)}\big|_{1||4}$ and $\v{A}_{3 \, \TT{jet}+(g)}$ as a perturbation to the exact three-jet limit in Eq.~\eqref{eq:exact3jet}, due to jet substructure at a finer resolution scale. This perturbation can be applied recursively at finer and finer resolution scales to build up a complete description of the collinear physics in the full three-jet system (see Figure \ref{fig:3jetallorder}). We will now show that this interpretation is correct by studying Eq.~\eqref{eq:amplitudeinringsandstrings} in the limit that particles $1,2,3$ define jet axes for collimated jets (jet 1, jet 2 and jet 3), and that particle $n$ is soft. We start by re-arranging Eq.~\eqref{eq:amplitudeinringsandstrings}, and applying the three-jet limit to rings,
\begin{align}
    \v{A}_{n}\big|_{\TT{3\, jet \, limit}} = & \frac{2\As}{\pi} \Bigg[\, {}^{(n-1)}\TT{S}^{2,3}_{1}  [1\bcdot 1] + \, {}^{(n-1)}\TT{S}^{1,3}_{2}  [2\bcdot 2] + \, {}^{(n-1)}\TT{S}^{1,2}_{3}  [3\bcdot 3] + \nonumber \\ 
    & \qquad  + \sum_{j  \in \, \TT{jet} \, 1} \, {}^{(n-1)}\TT{S}^{1,2}_{j} [j\bcdot j] + \sum_{j  \in \, \TT{jet} \, 2} \, {}^{(n-1)}\TT{S}^{1,2}_{j} [j\bcdot j] + \sum_{j  \in \, \TT{jet} \, 3} \, {}^{(n-1)}\TT{S}^{1,2}_{j}  [j\bcdot j]\Bigg] \nonumber \\ 
    & + \frac{\As}{\pi} \Bigg[\sum_{j  \in \, \TT{jet} \, 1} \, {}^{(n-1)}\TT{S}^{2,3}_{j} \; \Theta(\theta_{jn} > \theta_{j1}) \; ([j\bcdot 1-2] + [1-2\bcdot j]) \nonumber \\
    & \qquad - \sum_{j  \in \, \TT{jet} \, 2} \, {}^{(n-1)}\TT{S}^{1,3}_{j} \; \Theta(\theta_{jn} > \theta_{j2}) \; ([j\bcdot 1-2] + [1-2\bcdot j]) \nonumber \\
    & \qquad - \frac{1}{2} \bigg( \sum_{i \in \, \TT{jet} \, 1}\sum_{k \in \, \TT{jet} \, 2} + \sum_{i \in \, \TT{jet} \, 1}\sum_{k \in \, \TT{jet} \, 3} + \sum_{i \in \, \TT{jet} \, 3}\sum_{k \in \, \TT{jet} \, 2} \nonumber \\ 
    &\qqquad + \sum_{i \in \, \TT{jet} \, 3}\sum_{k \in \, \TT{jet} \, 3} \bigg) \, {}^{(n-1)}\TT{R}^{i,k}_{1,2} ([i\bcdot k] + [k\bcdot i]) \Bigg],
\end{align}
where $$[i\bcdot j]=\v{T}_{i} \; \v{A}_{n-1}\big|_{\TT{3\, jet \, limit}} \; \v{T}^{\dagger}_{j},$$ and where in the sums $j \in \{4,\dots , n-1\}$, $i \in \{3,\dots , n-1\}$, $k \in \{3,\dots , n-1\}\setminus i$. The first line is akin what we found in our heuristic calculation, Eq.~\eqref{eq:exact3jet}, whilst the rest can be considered perturbations to this result upon expanding in the collinear limit. The first angular bracket came from strings whilst the second originates from rings that reduce to strings in the three-jet limit. As usual, we have only kept terms with collinear divergences and which do not vanish in exact collinear limits. The term in large parentheses can be evaluated in the three-jet limit: 
\begin{align}
   &\bigg( \sum_{i \in \, \TT{jet} \, 1}\sum_{k \in \, \TT{jet} \, 2} + \sum_{i \in \, \TT{jet} \, 1}\sum_{k \in \, \TT{jet} \, 3} + \sum_{i \in \, \TT{jet} \, 3}\sum_{k \in \, \TT{jet} \, 2} + \sum_{i \in \, \TT{jet} \, 3}\sum_{k \in \, \TT{jet} \, 3} \bigg) \, {}^{(n-1)}\TT{R}^{i,k}_{1,2} ([i\bcdot k] + [k\bcdot i])\Big|_{\TT{3 \, jet \, limit}}  \nonumber \\
   &= \sum_{i \in \, \TT{jet} \, 1}\sum_{k \in \, \TT{jet} \, 2} \Big( \, {}^{(n-1)}\TT{S}^{k,2}_{i}\; \Theta(\theta_{in} > \theta_{i1})\;\Theta(\theta_{kn} > \theta_{k2}) \nonumber \\
   & \qqquad \qqquad + \, {}^{(n-1)}\TT{S}^{i,1}_{k}\; \Theta(\theta_{kn} > \theta_{k2})\;\Theta(\theta_{in} > \theta_{i1}) \Big)([i\bcdot k] + [k\bcdot i])  \nonumber \\
   & \qquad+ 2 \Bigg( \sum_{i \in \, \TT{jet} \, 1} \, {}^{(n-1)}\TT{S}^{2,3}_{i} \; \Theta(\theta_{in} > \theta_{i1}) + \sum_{i \in \, \TT{jet} \, 2}\, {}^{(n-1)}\TT{S}^{1,3}_{i} \; \Theta(\theta_{in} > \theta_{i2}) \Bigg) ([i\bcdot J_{3}] + [J_{3}\bcdot i]) \nonumber \\
   & \qquad - 2 \sum_{i \in \, \TT{jet} \, 3} \, {}^{(n-1)}\TT{S}^{1,2}_{i} \sum_{k \in \, \TT{jet} \, 3}\Theta(\theta_{in}>\theta_{ik}) ([i\bcdot k] + [k\bcdot i]),
\end{align}
where we have approximated the directions of particles without their nearest jet axis in terms with a collinear pole and 
\begin{align}
    \v{T}_{J_{3}} = \sum_{i \in \, \TT{jet} \, 3} \v{T}_{i}.
\end{align} 
The second and third lines have been symmetrised and written in this form to make the commutativity of the limits $n_{i} \rightarrow n_{1}$ and $n_{k} \rightarrow n_{2}$ obvious (note that $\, {}^{(n-1)}\TT{S}^{2,2}_{1} = \, {}^{(n-1)}\TT{S}^{1,1}_{2}$). This required the presence of the additional step function. Grouping terms
\begin{align}
    \v{A}_{n}\big|_{\TT{3\, jet \, limit}} = & \frac{2\As}{\pi} \Bigg[\, {}^{(n-1)}\TT{S}^{2,3}_{1}  [1\bcdot 1] + \, {}^{(n-1)}\TT{S}^{1,3}_{2}  [2\bcdot 2] + \sum_{j  \in \, \TT{jet} \, 1} \, {}^{(n-1)}\TT{S}^{1,2}_{j} [j\bcdot j]  + \sum_{j  \in \, \TT{jet} \, 2} \, {}^{(n-1)}\TT{S}^{1,2}_{j} [j\bcdot j] \Bigg] \nonumber \\ 
    & + \frac{\As}{\pi} \Bigg[\sum_{j  \in \, \TT{jet} \, 1} \, {}^{(n-1)}\TT{S}^{2,3}_{j} \; \Theta(\theta_{jn} > \theta_{j1}) \; ([j\bcdot 1-2-J_{3}] + [1-2-J_{3}\bcdot j] ) \nonumber \\
    & \qquad + \sum_{j  \in \, \TT{jet} \, 2} \, {}^{(n-1)}\TT{S}^{1,3}_{j} \; \Theta(\theta_{jn} > \theta_{j2}) \; ([j\bcdot 2-1-J_{3}] + [2-1-J_{3}\bcdot j] ) \nonumber \\
    & \qquad + \frac{1}{2}\sum_{i \in \, \TT{jet} \, 1}\sum_{k \in \, \TT{jet} \, 2} \Big( \, {}^{(n-1)}\TT{S}^{k,2}_{i}\; \Theta(\theta_{in} > \theta_{i1})\;\Theta(\theta_{kn} > \theta_{k2}) \nonumber \\
    & \qqquad \qqquad + \, {}^{(n-1)}\TT{S}^{i,1}_{k}\; \Theta(\theta_{kn} > \theta_{k2})\;\Theta(\theta_{in} > \theta_{i1}) \Big)([i\bcdot k] + [k\bcdot i])  \nonumber \\
    & \qquad + \sum_{i \in \, \TT{jet} \, 3} \, {}^{(n-1)}\TT{S}^{1,2}_{i} \sum_{i' \in \, \TT{jet} \, 3}\Theta(\theta_{in}>\theta_{ii'}) ([i\bcdot i'] + [i'\bcdot i]) \Bigg],
\end{align}
where $i' \in \{3,\dots,n-1\}$. Up to now, our indexing of particles has been allowed to be arbitrary (other than specifying that particles 1, 2 and 3 define the jet directions). To make further headway, we will assume particles are indexed forming an angular hierarchy. In particular, we index particles so that $\theta_{l} < \theta_{(l-1)}$ where $\theta_{l}$ is the angle between particle $l$ and the axis of the jet in which it is constituent. We take $\theta_{1} = \theta_{2} = \theta_{3} = \pi$ and $\theta_{n} = 0$ as boundary conditions\footnote{This is inverted compared to what one might expect, as the angle between particle 3 and jet axis 3 is zero. However, enforcing $\theta_{3} = \pi$ as a boundary condition ensures $\theta_{l} < \theta_{(l-1)}$ always holds. }. Next we split each $\, {}^{(n-1)}\TT{S}^{y,z}_{x}$ into regions $\theta_{l}> \theta_{xn} > \theta_{m}$ so that
\begin{align}
    \,{}^{(n-1)}\TT{S}^{y,z}_{x} = \sum^{n}_{l=4} {}^{(n-1)}\TT{S}^{y,z}_{x} \; \Theta(\theta_{l-1}> \theta_{xn} > \theta_{l}),
\end{align}
and
\begin{align}
    \,{}^{(n-1)}\TT{S}^{y,z}_{x} \; \Theta(\theta_{xn} > \theta_{xa})= \sum_{l=4}^{x} {}^{(n-1)}\TT{S}^{y,z}_{x} \; \Theta(\theta_{l-1}> \theta_{xn} > \theta_{l})
\end{align}
for $x$ in jet $a$. The labelling of jets $1,2,3$ was arbitrary and $\v{A}_{n}\big|_{\TT{3\, jet \, limit}}$ is symmetric under the exchange of each jet. Therefore we will now just look at the terms associated with collinear physics in jet $3$ (we picked our basis to simplify the algebra in this limit). We do this in the knowledge that terms describing jets $1$ and $2$ can be derived by the re-labelling the jet $3$ terms. We find
\begin{align}
    \v{A}_{n}\big|_{\TT{3\, jet \, limit}} \supset \frac{\As}{\pi} \sum_{i \in \, \TT{jet} \, 3} \sum_{l=4}^{n} & \, {}^{(n-1)}\TT{S}^{1,2}_{i} \; \Theta(\theta_{l-1}> \theta_{in} > \theta_{l}) \nonumber \\ 
    & \times \sum_{i' \in \, \TT{jet} \, 3}\Theta(\theta_{in}>\theta_{ii'}) ([i\bcdot i'] + [i'\bcdot i]).
\end{align}
Now note that 
$$\TT{S}^{1,2}_{i} \; \Theta(\theta_{l-1}> \theta_{in} > \theta_{l})\big|_{\TT{3\, jet \, limit}} \approx \TT{S}^{1,2}_{m} \; \Theta(\theta_{l-1}> \theta_{mn} > \theta_{l}),$$ 
for all $m >l+1$ and with $\theta_{l+1} > \theta_{im}$. Consequently, we can write 
\begin{align}
    \v{A}_{n}\big|_{\TT{3\, jet \, limit}} \supset \frac{2\As}{\pi}   \sum_{l \in \, \TT{jet} \, 3} ~ \sum_{i\leq l |i \in \, \TT{jet} \, 3} \,{}^{(n-1)} \, \TT{S}^{1,2}_{i} \; \Theta(\theta_{l}> \theta_{in} > \theta_{l+1}) [J^{(i,l)}_{3}\bcdot J^{(i,l)}_{3}],
\end{align}
where $\v{T}_{J^{(i,l)}_{3}}$ is the combined colour charge of every particle in the neighbourhood of $i$ unresolved at an angular scale in the range $\theta_{l+1}$ to $\theta_{l}$:
$$\v{T}_{J^{(i,l)}_{3}} = \sum_{i' \in \, \TT{jet} 3} \v{T}_{i'}\Theta(\theta_{l+1} > \theta_{ii'} ).$$ 
A similarly careful treatment of terms generating jets 1 and 2 returns the same result under the appropriate exchange of indices. The complete result is
\begin{align}
    \v{A}_{n}\big|_{\TT{3\, jet \, limit}} = \frac{2\As}{\pi} \sum_{a = 1,2,3} ~ \sum_{l \in \, \TT{jet} \, a} ~ \sum_{i\leq l |i \in \, \TT{jet} \, a}& \,{}^{(n-1)} \, \TT{S}^{(a - 1) \, (\TT{mod}\, 3),(a + 1) \, (\TT{mod}\, 3)}_{i} \nonumber \\
    & \times  \Theta(\theta_{l}> \theta_{in} > \theta_{l+1}) \; [J^{(i,l)}_{a}\bcdot J^{(i,l)}_{a}]. \label{eq:3-jetallorder}
\end{align}

Eq.~\eqref{eq:3-jetallorder} is equivalent to the radiation pattern built up by iterating perturbations due to quasi-collinear physics with each iteration resolving yet smaller angular substructure in the jets. For instance, we can recreate the substructure given in Figure \ref{fig:3jetallorder}, which represents a possible set of perturbations to jet 1, if we equate 
\begin{align}
    & [J^{(1,1)}_{1}\bcdot J^{(1,1§)}_{1}] = | 1 \bcdot 1|_{3}, \nonumber \\
    &[J^{(i,i)}_{1}\bcdot J^{(i,i)}_{1}] = | i \bcdot i|_{4},\quad [J^{(j,i)}_{1}\bcdot J^{(j,i)}_{1}] = | j \bcdot j|_{4}, \nonumber \\ & [J^{(i,j)}_{1}\bcdot J^{(i,j)}_{1}] = | i \bcdot i|_{5},\quad [J^{(j,j)}_{1}\bcdot J^{(j,j)}_{1}] = | j \bcdot j|_{5},  \quad
    [J^{(k,j)}_{1}\bcdot J^{(k,j)}_{1}] = | k \bcdot k|_{5}, \nonumber \\ &\TT{and} \quad \theta_{i} = \theta_{ij},\quad \theta_{j} = \theta_{jk},
\end{align}
for a given configuration of particles where $i<j,k$ and all other angular scales are less than $\theta_{jk}$.

We can re-organise Eq.~\eqref{eq:3-jetallorder} into a simple angular-ordered recurrence relation by assuming particles $4, \dots , n$ are gluons that are soft relative to the hard-process particles ($1,2,3$), which define the 3-jet axis. We re-label the particles in the order of their emission angles so that $4$ is the widest angle gluon and $n$ is the smallest angle. The recurrence relation is 
\begin{align}
    \v{A}_{n}\big|_{\TT{3\, jet \, limit}} = \frac{2\As}{\pi} \sum_{a = 1,2,3} ~ \sum_{i \in \, \TT{jet} \, a} & \,{}^{(n-1)} \, \TT{S}^{(a - 1) \, (\TT{mod}\, 3),(a + 1) \, (\TT{mod}\, 3)}_{i}   \Theta(\theta_{n-1}> \theta_{in}) \; [i\bcdot i],
\end{align}
with the initial condition that $\v{A}_{3} = \v{H}(Q)$, the hard-process density matrix at a hard scale $Q$, and $\theta_{3}=\pi$.

\bibliographystyle{JHEP}
\bibliography{stringsandrings}

\end{document}